\documentclass[aps,showpacs,amsmath,amssymb]{revtex4}

\usepackage{latexsym}
\usepackage{graphicx}
\usepackage{dcolumn}
\usepackage{bm}
\usepackage{hyperref}

\newcommand{\beq}{\begin{equation}}
\newcommand{\eeq}{\end{equation}}
\newcommand{\beqa}{\begin{eqnarray}}
\newcommand{\eeqa}{\end{eqnarray}}
\newcommand{\beqar}{\begin{eqnarray*}}
\newcommand{\eeqar}{\end{eqnarray*}}
\newcommand{\bra}[1]{\mbox{$\left\langle{#1}\right|$}}
\newcommand{\ket}[1]{\mbox{$\left|{#1}\right\rangle$}}
\newcommand{\diracsp}[2]{\mbox{$\langle{#1}|{#2}\rangle$}}

\def\I{{\rm i}}
\def\d{{\rm d}}
\def\e{{\rm e}}

\newcounter{saveeqn}

\begin{document}

\title{Quantum mechanics in general quantum systems (II): \\ perturbation theory}
\author{An Min Wang}\email{anmwang@ustc.edu.cn}
\affiliation{Quantum Theory Group, Department of Modern Physics,
University of Science and Technology of China, Hefei, 230026,
P.R.China}

\begin{abstract}

We propose an improved scheme of perturbation theory based on our
exact solution [An Min Wang, quant-ph/0602055] in general quantum
systems independent of time. Our elementary start-point is to
introduce the perturbing parameter as late as possible. Our main
skills are Hamiltonian redivision so as to overcome a flaw of the
usual perturbation theory, and the perturbing Hamiltonian matrix
product decomposition in order to separate the contraction and
anti-contraction terms. Our calculational technology is the limit
process for eliminating apparent divergences. Our central idea is
``dynamical rearrangement and summation" for the sake of the partial
contributions from the high order even all order approximations
absorbed in our perturbed solution. Consequently, we obtain the
improved forms of the zeroth, first, second and third order
perturbed solutions absorbing the partial contributions from the
high order even all order approximations of perturbation. Then we
deduce the improved transition probability. In special, we propose
the revised Fermi's golden rule. Moreover, we apply our scheme to
obtain the improved forms of perturbed energy and perturbed state.
In addition, we study an easy understanding example of two-state
system to illustrate our scheme and show its advantages. All of this
implies the physical reasons and evidences why our improved scheme
of perturbation theory are actually calculable, operationally
efficient, conclusively more accurate. Our improved scheme is the
further development and interesting application of our exact
solution, and it has been successfully used to study on open system
dynamics [An Min Wang, quant-ph/0601051].

\end{abstract}


\maketitle

\section{Introduction}\label{sec1}

The known perturbation theory \cite{seq,Diracpqm} is an extremely
important tool for describing real quantum systems, as it turns out
to be very difficult to find exact solutions to the Schr\"dinger
equation for Hamiltonians of even moderate complexity. Recently, we
see the dawn to overcome this difficulty because we obtained the
exact solution of the Schr\"oding equation \cite{seq} in general
quantum systems independent of times \cite{My1}. However, this does
not mean that the perturbation theory is unnecessary because our
exact solution is still an infinite power series of perturbation.
Our solution is called ``exact one" in the sense including all order
approximations of perturbation. In practice, if we do not intend to
apply our exact solution to investigations of the formal theory of
quantum mechanics, we often need to cut off our exact solution
series to some given order approximation in the calculations of
concrete problems. Perhaps, one argues that our exact solution will
back to the usual perturbation theory, and it is, at most, an
explicit form that can bring out the efficiency amelioration.
Nevertheless, the case is not so. Such a view, in fact, ignores the
significance of the general term in an infinite series, and forgets
the technologies to deal with an infinite series in the present
mathematics and physics. From our point of view, since the general
term is known, we can systematically and reasonably absorb the
partial contributions from some high order even all order
approximations to the lower order approximations just like one has
done in quantum field theory via summation over a series of
different order but similar feature Feynman figures. In this paper,
based on such a method we develop our ``dynamical arrangement and
summation" idea, and then propose an improved scheme of perturbation
theory via introducing several useful skills and methods.

It is very interesting that we find a flaw in the usual perturbation
theory, that is, the perturbing parameter is introduced too early so
that the contributions from the high order even all order
approximations of the diagonal and off-diagonal elements of the
perturbing Hamiltonian matrix are, respectively, inappropriately
dropped and prematurely cut off. For some systems, the influences on
the calculation precision from this flaw can be not neglectable with
the evolution time increasing. This motivates us to set our
start-point to introduce the perturbing parameter as late as
possible in order to guarantee the generality and precision. It is
natural from a mathematics view if we think the perturbing parameter
in a general perturbation theory is a formal multiplier. Based on
this start point, we propose Hamiltonian redivision skill and
further methods so as to overcome the above flaw in the usual
perturbation theory, viz. the Hamiltonian redivision makes the
contributions from all order approximations of the diagonal elements
of the perturbing Hamiltonian matrix can be absorbed in our improved
form of perturbed solution. Hence, this skill advances the
calculation precision in theory, extends the application range for
the perturbation theory and can remove degeneracies in some systems.

Since our exact solution series has apparent divergences, we provide
the methods of perturbing Hamiltonian matrix product decomposition
in order to separate the contraction terms with apparent divergences
and anti-contraction terms without apparent divergences. Here,
``apparent" refers to an unture thing, that is, the apparent
divergences are not real singularities and they can be eliminated by
mathematical and/or physical methods, while the ``perturbing
Hamiltonian matrix" refers to the representation matrix of the
perturbing Hamiltonian in the unperturbed Hamiltonian
representation. Then, by the limit process we can eliminate these
apparent divergences in the contraction terms. Furthermore we apply
``dynamical rearrangement and summation" idea for the sake of the
partial contributions from the high order even all order
approximations absorbed in our perturbed solution. In terms of these
useful ideas, skills and methods we build an improved scheme of
perturbation theory. Without any doubt, they are given definitely
dependent on our exact solution. In fact, our exact solution
inherits the distinguished feature in a $c$-number function form
just like the Feynman \cite{Feynman} path integral expression and
keeps the advantage in Dyson series \cite{Dyson} that is a power
series of perturbation. At the same time, our exact solution is so
explicit that when applying it to a concrete quantum system, all we
need to do is only the calculations of perturbing Hamiltonian matrix
and the limitations of primary functions.

As well known, a key idea of the existed perturbation theory to
research the time evolution of system is to split the Hamiltonian of
system into two parts, that is \beq H=H_0+H_1,\eeq where the
eigenvalue problem of so called unperturbed Hamiltonian $H_0$ is
solvable, and so-call perturbing Hamiltonian $H_1$ is the rest part
of the Hamiltonian. In other words, this splitting is chosen in such
a manner that the solutions of $H_0$ are known as \beq\label{h0eeq}
H_0\ket{\Phi^\gamma}=E_\gamma \ket{\Phi^\gamma}, \eeq where
$\ket{\Phi^\gamma}$ is the eigenvector of $H_0$ and $E_\gamma$ is
the corresponding eigenvalue. Whole $\ket{\Phi^\gamma}$, in which
$\gamma$ takes over all possible values, form a representation of
the unperturbed Hamiltonian. It must be emphasized that the
principle of Hamiltonian split is not just the best solvability
mentioned above in more general cases. If there are degeneracies,
the Hamiltonian split is also restricted by the condition that the
degeneracies can be completely removed via the usual diagonalization
procedure of the degenerate subspaces and the Hamiltonian redivision
proposed in this paper, or specially, if the remained degeneracies
are allowed, it requires that the off-diagonal element of the
perturbing Hamiltonian matrix between any two degenerate levels are
always vanishing so as to let our improved scheme of perturbation
theory work well. As an example, it has been discussed in our serial
study \cite{My3}. In addition, if the cut-off approximation of
perturbation is necessary, it requires that for our improved scheme
of perturbation theory, the off-diagonal elements of $H_1$ matrix is
small enough compared with the diagonal element of $H=H_0+H_1$
matrix in the unperturbed representation.

Nevertheless, there are some known shortcomings in the existed
perturbation theory, for example, when $H_1$ is not so small
compared with $H_0$ that the high order approximations should be
considered, and/or when the partial contributions from the higher
order approximations become relatively important to the studied
problems, and/or the evolution time is long enough, the usual
perturbation theory might be difficult to calculate to an enough
precision in an effective way, even not feasible practically since
the lower approximation might break the physical symmetries and/or
constraints. In order to overcome these difficulties and problems,
we recently study and obtain the exact solution in general quantum
systems via explicitly expressing the time evolution operator as a
$c$-number function and a power series of perturbation including all
order approximations \cite{My1}. In this paper, our purpose is to
build an improved scheme of perturbation theory based on our exact
solution \cite{My1} so that the physical problems are more
accurately and effectively calculated. For simplicity, we focus on
the cases of Schr\"odinger dynamics \cite{seq}. It is direct to
extend our improved scheme to the cases of the von Neumann dynamics
\cite{vonneumann}.

Just well-known, quantum dynamics and its perturbation theory have
been sufficiently studied and have many successful applications.
Many famous physicists created their nice formulism and obtained
some marvelous results. An attempt to improve its part content or
increase some new content as well as some new methods must be very
difficult in their realizations. However, our endeavors have
obtained their returns, for examples, our exact solution \cite{My1},
perturbation theory and open system dynamics \cite{My3} in general
quantum systems independent of time.

In this paper, we start from proposing our ideas, skills and
methods. We expressly obtain the improved forms of the zeroth,
first, second and third order approximations of perturbed solution
absorbing partial contributions from the high order even all order
approximations, finding the improved transition probability,
specially, the revised Fermi's golden rule, and providing an
operational scheme to calculate the perturbed energy and perturbed
state. Furthermore, by studying a concrete example of two state
system, we illustrate clearly that our solution is more efficient
and more accurate than the usual perturbative method. In short, our
exact solution and perturbative scheme are formally explicit,
actually calculable, operationally efficient, conclusively more
accurate (to the needed precision).

This paper is organized as the following: in Sec. \ref{sec2} we find
a flaw of the usual perturbation theory and introduce Hamiltonian
redivision to overcome it. Then, we propose the skill of the
perturbing Hamiltonian matrix product decomposition in order to
separate the contraction terms with apparent divergences and
anti-contraction terms without apparent divergences. By the limit
process we can eliminate these apparent divergences. More
importantly, we use the ``dynamical rearrangement and summation"
idea so that the partial contributions from the high order even all
order approximations are absorbed in our perturbed solution and the
above flaw is further overcome; in Sec. \ref{sec3} we obtain the
improved forms of the zeroth, first, second and third order
perturbed solutions of dynamics absorbing partial contributions from
the high order even all order approximations; in Sec. \ref{sec4} we
deduce the improved transition probability, specially, the revised
Fermi's golden rule. In Sec. \ref{sec5} we provide a scheme to
calculate the perturbed energy and the perturbed state; in Sec.
\ref{sec6} we study an example of two state system in order to
concretely illustrate our solution to be more effective and more
accurate than the usual method; in Sec. \ref{sec7} we summarize our
conclusions and give some discussions. Finally, we write an appendix
as well as a supplementary where some expressions are calculated in
order to derive out the improved forms of perturbed solutions.

\section{Skills and methods in the improved scheme of perturbation
theory}\label{sec2}

In our recent work \cite{My1}, by splitting a Hamiltonian into two
parts, using the solvability of eigenvalue problem of one part of
the Hamiltonian, proving an useful identity and deducing an
expansion formula of operator binomials power, we obtain an explicit
and general form of the time evolution operator in the
representation of solvable part (unperturbed part) of the
Hamiltonian. Then we find out an exact solution of the
Schr\"{o}dinger equation in general quantum systems independent of
time \beqa \label{ouress} \ket{\Psi(t)}&=&\sum_{l=0}^\infty
\mathcal{A}_l(t)\ket{\Psi(0)}=\sum_{l=0}^\infty\sum_{\gamma,\gamma^\prime}
A_l^{\gamma\gamma^\prime}(t)
\left[\diracsp{\Phi^{\gamma^\prime}}{\Psi(0)}\right]\ket{\Phi^\gamma},
\eeqa where \beqa \mathcal{A}_l(t)&=&\sum_{\gamma,\gamma^\prime}
A_l^{\gamma\gamma^\prime}(t)\ket{{\Phi}^{\gamma}}\bra{{\Phi}^{\gamma^\prime}},\\
A_0^{\gamma\gamma^\prime}(t)&=&\sum_{\gamma,\gamma^\prime}\e^{-\I
E_{\gamma}
t}\delta_{\gamma\gamma^\prime}, \\
\label{Aldefinition}
A_l^{\gamma\gamma^\prime}(t)&=&\sum_{\gamma_1,\cdots,\gamma_{l+1}}\left[
\sum_{i=1}^{l+1}(-1)^{i-1}\frac{\e^{-\I E_{\gamma_i}
t}}{d_i(E[\gamma,l])}\right]\left[
\prod_{j=1}^{l}H_1^{\gamma_j\gamma_{j+1}}\right]
\delta_{\gamma_1\gamma}\delta_{\gamma_{l+1}\gamma^\prime}.\eeqa and
all
$H_1^{\gamma_j\gamma_{j+1}}=\bra{\Phi^{\gamma_j}}H_1\ket{\Phi^{\gamma_{j+1}}}$
form so-called ``perturbing Hamiltonian matrix", that is, the
representation matrix of the perturbing Hamiltonian in the
unperturbed Hamiltonian representation. While \beqa
d_1(E[\gamma,l])&=&\prod_{i=1}^{l}\left(E_{\gamma_{1}}
-E_{\gamma_{i+1}}\right),\\
 d_i(E[\gamma,l])&=&
\prod_{j=1}^{i-1}\left(E_{\gamma_{j}}
-E_{\gamma_{i}}\right)\!\!\!\prod_{k=i+1}^{l+1}\left(E_{\gamma_{i}}
-E_{\gamma_{k}}\right),\\[-3pt] d_{l+1}(E[\gamma,l])
&=&\prod_{i=1}^{l}\left(E_{\gamma_{i}}-E_{\gamma_{l+1}}\right),\eeqa
here $2\leq i \leq l$.

It is clear that there are apparent divergences in the above
solution. For example \beqa
A_1^{\gamma\gamma^\prime}(t)&=&\left[\frac{\e^{-\I E_{\gamma}
t}}{E_{\gamma}-E_{\gamma^\prime}}-\frac{\e^{-\I E_{\gamma^\prime}
t}}{E_{\gamma}-E_{\gamma^\prime}}\right]H_1^{\gamma\gamma^\prime}.\eeqa
when $E_{\gamma}=E_{\gamma^\prime}$ (which can appear in the
summation or degeneracy cases), it is $\infty-\infty=-\I
H_1^{\gamma\gamma^\prime} t$. As pointed out in our paper
\cite{My1}, we need to understand $A_k^{\gamma\gamma^\prime}(t)$ in
the sense of limitation. Moreover, in practice, we should present
how to calculate their limitation in order to eliminate the apparent
divergences.

Now, the key problems are how and when to introduce the cut-off
approximation in order to obtain the perturbed solution. For
studying and solving them, we first need to propose some skills and
methods in this section. These skills and methods profit from the
fact that the general term is clearly known and explicitly expressed
in our exact solution. By using them we can derive out the improved
forms of perturbed solution absorbing the partial contributions from
the high order even all order approximations of perturbation. It
will be seen that all the steps are well-regulated and only
calculational technology is to find the limitation of primary
functions. In other words, our exact solution and perturbation
theory are easily calculative and operational, and they have better
precision and higher efficiency. Frankly speaking, before we know
our exact solution, we are puzzled by too many irregular terms and
very trouble dependence on previous calculation steps. Moreover, we
are often anxious about the precision of the results in such some
calculations because those terms proportional to $t^a \e^{-\I
E_{\gamma_i} t}$ $(a=1,2,\cdots)$ in the high order approximations
might not be ignorable with time increasing. Considering the
contributions of these terms can obviously improve the precision.
However, the known perturbation theory does not give the general
term, considering this task to absorb reasonably the high order
approximations is impossible.

Since our exact solution has given the explicit form of any order
approximation, that is a general term of an arbitrary order
perturbed solution, and their forms are simply the summations of a
power series of the perturbing Hamiltonian. Just enlightened by this
general term of arbitrary order perturbed solution, we use two
skills and ``dynamical rearrangement and summation" method to build
an improved scheme of perturbation theory, which are respectively
expressed in the following two subsections.

\subsection{Hamiltonian Redivision}

The first skill starts from the decomposition of the perturbing
Hamiltonian matrix, that is the matrix elements of $H_1$ in the
representation of $H_0$, into diagonal part and off-diagonal part:
\beq\label{H1d2}
H_1^{\gamma_j\gamma_{j+1}}=h_1^{\gamma_j}\delta_{\gamma_j\gamma_{j+1}}
+g_1^{\gamma_j\gamma_{j+1}},\eeq so as to distinguish them because
the diagonal and off-diagonal elements can be dealed with in the
different way. In addition, it makes the concrete expression of a
given order approximation can be easily calculated. Note that
$h_1^{\gamma_j}$ has  been chosen as its diagonal elements and then
$g_1^{\gamma_j\gamma_{j+1}}$ has been set as its off-diagonal
elements: \beq g_1^{\gamma_j\gamma_{j+1}}=g_1^{\gamma_j\gamma_{j+1}}
(1-\delta_{\gamma_j\gamma_{j+1}}).\eeq

As examples, for the first order approximation, it is easy to
calculate that \beq
\label{A1h}A_1^{\gamma\gamma^\prime}(h)=\sum_{\gamma_1,\gamma_{2}}\left[
\sum_{i=1}^{2}(-1)^{i-1}\frac{\e^{-\I E_{\gamma_i}
t}}{d_i(E[\gamma,l])}\right]
\left(h_1^{\gamma_1}\delta_{\gamma_1\gamma_{2}}\right)
\delta_{\gamma\gamma_1}\delta_{\gamma^\prime\gamma_{2}}=\frac{(-\I
h_1^{\gamma} t)}{1!}\e^{-\I E_{\gamma}
t}\delta_{\gamma\gamma^\prime},\eeq \beqa\label{A1g}
A_1^{\gamma\gamma^\prime}(g)&=&\sum_{\gamma_1,\gamma_{2}}\left[
\sum_{i=1}^{2}(-1)^{i-1}\frac{\e^{-\I E_{\gamma_i}
t}}{d_i(E[\gamma,l])}\right]
g_1^{\gamma_1\gamma_2}(1-\delta_{\gamma_1\gamma_2})
\delta_{\gamma\gamma_1}\delta_{\gamma^\prime\gamma_{2}}
=\left[\frac{\e^{-\I E_{\gamma}
t}}{E_{\gamma}-E_{\gamma^\prime}}-\frac{\e^{-\I E_{\gamma^\prime}
t}}{E_{\gamma}-E_{\gamma^\prime}}\right]g_1^{\gamma\gamma^\prime}.\eeqa
Note that here and after we use the symbol
$A_i^{\gamma\gamma^\prime}$ denoting the contribution from the $i$th
order approximation, which is defined by (\ref{Aldefinition}), while
its argument indicates the product form of perturbing Hamiltonian
matrix. However, for the second order approximation, since \beqa
\label{H12deq}\prod_{j=1}^{2}H_1^{\gamma_j\gamma_{j+1}}&=&
\left(h_1^{\gamma_1}\right)^2\delta_{\gamma_1\gamma_{2}}\delta_{\gamma_2\gamma_{3}}
+h_1^{\gamma_1}g_1^{\gamma_2\gamma_{3}}\delta_{\gamma_1\gamma_2} +
g_1^{\gamma_1\gamma_{2}}h_1^{\gamma_2}\delta_{\gamma_2\gamma_3}
+g_1^{\gamma_1\gamma_{2}}g_1^{\gamma_2\gamma_{3}}.\eeqa we need to
calculate the mixed product of diagonal and off-diagonal elements of
perturbing Hamiltonian matrix. Obviously, we have \beqa \label{A2hh}
A_2^{\gamma\gamma^\prime}(hh)&=&\sum_{\gamma_1,\gamma_{2},\gamma_3}\left[
\sum_{i=1}^{3}(-1)^{i-1}\frac{\e^{-\I E_{\gamma_i}
t}}{d_i(E[\gamma,2])}\right]
\left(h_1^{\gamma_1}\delta_{\gamma_1\gamma_{2}}h_1^{\gamma_2}\delta_{\gamma_2\gamma_{3}}\right)
\delta_{\gamma\gamma_1}\delta_{\gamma^\prime\gamma_{3}} =\frac{(-\I
h_1^{\gamma} t)^2}{2!}\e^{-\I E_{\gamma}
t}\delta_{\gamma\gamma^\prime},\eeqa \beqa \label{A2hg}
A_2^{\gamma\gamma^\prime}(hg)&=&\sum_{\gamma_1,\gamma_2,\gamma_{3}}\left[
\sum_{i=1}^{3}(-1)^{i-1}\frac{\e^{-\I E_{\gamma_i}
t}}{d_i(E[\gamma,2])}\right] h_1^{\gamma_1}g_1^{\gamma_2\gamma_{3}}
\delta_{\gamma_1\gamma_2}\delta_{\gamma\gamma_1}\delta_{\gamma^\prime\gamma_3}\nonumber\\
& =&\left[-\frac{\e^{-\I E_{\gamma}
t}}{(E_{\gamma}-E_{\gamma^\prime})^2}+\frac{\e^{-\I
E_{\gamma^\prime} t}}{(E_{\gamma}-E_{\gamma^\prime})^2}+(-\I
t)\frac{\e^{-\I E_{\gamma}
t}}{E_{\gamma}-E_{\gamma^\prime}}\right]h_1^{\gamma}g_1^{\gamma\gamma^\prime},\eeqa
\beqa
\label{A2gh}A_2^{\gamma\gamma^\prime}(gh)&=&\sum_{\gamma_1,\gamma_2,\gamma_{3}}\left[
\sum_{i=1}^{3}(-1)^{i-1}\frac{\e^{-\I E_{\gamma_i}
t}}{d_i(E[\gamma,2])}\right]
h_1^{\gamma_2}g_1^{\gamma_1\gamma_{2}}\delta_{\gamma_2\gamma_3}
\delta_{\gamma\gamma_1}\delta_{\gamma^\prime\gamma_3}\nonumber\\
& =&\left[\frac{\e^{-\I E_{\gamma}
t}}{(E_{\gamma}-E_{\gamma^\prime})^2}-\frac{\e^{-\I
E_{\gamma^\prime} t}}{(E_{\gamma}-E_{\gamma^\prime})^2}-(-\I
t)\frac{\e^{-\I E_{\gamma^\prime}
t}}{E_{\gamma}-E_{\gamma^\prime}}\right]g_1^{\gamma\gamma^\prime}h_1^{\gamma^\prime},\eeqa
\beqa
\label{A2gg}A_2^{\gamma\gamma^\prime}(gg)&=&\sum_{\gamma_1,\gamma_2,\gamma_{3}}\left[
\sum_{i=1}^{3}(-1)^{i-1}\frac{\e^{-\I E_{\gamma_i}
t}}{d_i(E[\gamma,2])}\right]
g_1^{\gamma_1\gamma_2}g_1^{\gamma_2\gamma_{3}}
\delta_{\gamma\gamma_1}\delta_{\gamma^\prime\gamma_3}\nonumber\\
&=&\sum_{\gamma_1}\left[\frac{\e^{-\I
E_{\gamma}t}}{(E_{\gamma}-E_{\gamma_1})(E_{\gamma}-E_{\gamma^\prime})}-\frac{\e^{-\I
E_{\gamma_1}t}}{(E_{\gamma}-E_{\gamma_1})(E_{\gamma_1}-E_{\gamma^\prime})}+\frac{\e^{-\I
E_{\gamma^\prime}t}}{(E_{\gamma}-E_{\gamma^\prime})(E_{\gamma_1}-E_{\gamma^\prime})}\right]
g_1^{\gamma\gamma_1}g_1^{\gamma_1\gamma^\prime}. \eeqa In the usual
time-dependent perturbation theory, the zeroth order approximation
of time evolution of quantum state keeps its original form \beq
\ket{\Psi^{(0)}(t)}=\e^{-\I E_\gamma t}\ket{\Phi^\gamma},\eeq where
we have set the initial state as $\ket{\Phi^\gamma}$ for simplicity.
By using our solution, we easily calculate out the contributions of
all order approximations from the product of completely diagonal
elements $h$ of the perturbing Hamiltonian matrix to this zeroth
order approximation \beqa \sum_{\gamma_1,\cdots,\gamma_{l+1}}\left[
\sum_{i=1}^{l+1}(-1)^{i-1}\frac{\e^{-\I E_{\gamma_i}
t}}{d_i(E[\gamma,l])}\right]
\left(\prod_{j=1}^{l}h_1^{\gamma_j}\delta_{\gamma_j\gamma_{j+1}}\right)
\delta_{\gamma\gamma_1}\delta_{\gamma^\prime\gamma_{l+1}}=\frac{(-\I
h_1^{\gamma} t)^l}{l!}\e^{-\I E_{\gamma}
t}\delta_{\gamma\gamma^\prime}.\eeqa Therefore, we can absorb the
contributions of all order approximation parts from the product of
completely diagonal elements $h$ of the perturbing Hamiltonian
matrix into this zeroth order approximation to obtain \beq
\label{0thawithde} \ket{{\Psi^\prime}^{(0)}(t)}=\e^{-\I
\left(E_\gamma+h_1^\gamma\right) t}\ket{\Phi^\gamma}.\eeq Similarly,
by calculation, we can deduce that up to the second approximation,
the perturbed solution has the following form \beqa
\label{2thawithde}
\ket{\Psi^\prime(t)}&=&\sum_{\gamma,\gamma^\prime}\left\{\e^{-\I
\left(E_{\gamma}+h_1^\gamma\right)t}\delta_{\gamma\gamma^\prime} +
\left[\frac{\e^{-\I \left(E_{\gamma}+h_1^\gamma\right) t}-\e^{-\I
\left(E_{\gamma^\prime}+h_1^{\gamma^\prime}\right)
t}}{\left(E_{\gamma}+h_1^{\gamma}\right)
-\left(E_{\gamma^\prime}+h_1^{\gamma^\prime}\right)}\right]
g_1^{\gamma\gamma^\prime}\right.\nonumber\\
& & +\sum_{\gamma_1}\left[\frac{\e^{-\I
\left(E_{\gamma}+h_1^\gamma\right)t}}{\left[\left(E_{\gamma}+h_1^\gamma\right)
-\left(E_{\gamma_1}+h_1^{\gamma_1}\right)\right]\left[\left(E_{\gamma}+h_1^\gamma\right)
-\left(E_{\gamma^\prime}+h_1^{\gamma^\prime}\right)\right]}\right.\nonumber\\
& & -\frac{\e^{-\I
\left(E_{\gamma_1}+h_1^{\gamma_1}\right)t}}{\left[\left(E_{\gamma}+h_1^\gamma\right)
-\left(E_{\gamma_1}+h_1^{\gamma_1}\right)\right]\left[\left(E_{\gamma_1}+h_1^{\gamma_1}\right)
-\left(E_{\gamma^\prime}+h_1^{\gamma^\prime}\right)\right]}\nonumber\\
& &\left.\left.+\frac{\e^{-\I\left(
E_{\gamma^\prime}+h_1^{\gamma^\prime}\right)t}}{\left[\left(E_{\gamma}+h_1^\gamma\right)
-\left(E_{\gamma^\prime}+h_1^{\gamma^\prime}\right)\right]\left[\left(E_{\gamma_1}+h_1^{\gamma_1}\right)
-\left(E_{\gamma^\prime}+h_1^{\gamma^\prime}\right)\right]}\right]
g_1^{\gamma\gamma_1}g_1^{\gamma_1\gamma^\prime}\right\}\nonumber\\
& &
\left[\diracsp{\Phi^{\gamma^\prime}}{\Psi(0)}\right]\ket{\Phi^\gamma}+\mathcal{O}(H_1^3).
\eeqa However, for the higher order approximation, the corresponding
calculation is heavy. In fact, it is unnecessary to calculate the
contributions from those terms with the diagonal elements of $H_1$
since introducing the following skill. This is a reason why we omit
the relevant calculation details. Here we mention it only for
verification of the correctness of our exact solution in this way.

The results (\ref{0thawithde}) and (\ref{2thawithde}) are not
surprised because of the fact that the Hamiltonian is re-divisible.
Actually, we can furthermore use a trick of redivision of the
Hamiltonian so that the new $H_0$ contains the diagonal part of
$H_1$, that is, \beqa
H_0^\prime&=&H_0+\sum_{\gamma}h_1^\gamma\ket{\Phi^\gamma}\bra{\Phi^\gamma},\\
H_1^\prime&=&H_1-\sum_{\gamma}h_1^\gamma\ket{\Phi^\gamma}\bra{\Phi^\gamma}
=\sum_{\gamma,\gamma^\prime}g_1^{\gamma\gamma^\prime}\ket{\Phi^\gamma}\bra{\Phi^{\gamma\prime}}.\eeqa
In other words, without loss of generality, we always can choose
that $H_1^\prime$ has only the off-diagonal elements in the
$H_0^\prime$ (or $H_0$) representation and \beq
H_0^\prime\ket{\Phi^\gamma}=\left(E_\gamma+h_1^\gamma\right)\ket{\Phi^\gamma}
=E_{\gamma}^\prime\ket{\Phi^\gamma}.\eeq It is clear that this
redivision does not change the representation of the unperturbed
Hamiltonian, but can change the corresponding eigenvalues. In spite
that our skill is so simple, it seems not be sufficiently transpired
and understood from the fact that the recent some textbooks of
quantum mechanics still remain the contributions from the diagonal
elements of the perturbing Hamiltonian matrix in the expression of
the second order perturbed state. It is clear that the directly
cut-off approximation in the usual perturbation theory drops the
contributions from all higher order approximations of the diagonal
element of the perturbing Hamiltonian matrix. From our point of
view, the usual perturbation theory introduces the perturbing
parameter too early so that this flaw is resulted in.

If there is degeneracy, our notation has to be changed as \beqa
E_{\gamma_i}&\rightarrow&E_{\gamma_i a_{\gamma_i}}=E_{\gamma_i},\\
 \delta_{\gamma_i\gamma_j}&\rightarrow&
\delta_{\gamma_i\gamma_j}\delta_{a_{\gamma_i} a_{\gamma_j}},\\
\eta_{\gamma_i\gamma_j}&\rightarrow& \eta_{\gamma_i\gamma_j}+
\delta_{\gamma_i\gamma_j}\eta_{a_{\gamma_i} a_{\gamma_j}}.\eeqa
Thus, we can find \beq A_1^{\gamma a_{\gamma},\gamma^\prime
a_{\gamma^\prime}}(h)=\frac{(-\I h_1^{\gamma} t)}{1!}\e^{-\I
E_{\gamma}
t}\delta_{\gamma\gamma^\prime}\delta_{a_{\gamma}a_{\gamma^\prime}},\eeq
\beqa \label{A1gd} A_1^{\gamma a_{\gamma},\gamma^\prime
a_{\gamma^\prime}}(g)&=&\left[\frac{\e^{-\I E_{\gamma a_\gamma}
t}}{E_{\gamma a_\gamma}-E_{\gamma^\prime
a_{\gamma^\prime}}}-\frac{\e^{-\I E_{\gamma^\prime
a_{\gamma^\prime}} t}}{E_{\gamma a_\gamma}-E_{\gamma^\prime
a_{\gamma^\prime}}}\right]g_1^{\gamma a_{\gamma},\gamma^\prime
a_{\gamma^\prime}}\eta_{\gamma\gamma^\prime}+\frac{(-\I g_1^{\gamma
a_{\gamma},\gamma a_{\gamma^\prime}} t)}{1!}\e^{-\I E_{\gamma
a_\gamma} t}\delta_{\gamma\gamma^\prime}.\hskip 1.0cm\eeqa This
seems to bring some complications. However, we can use the trick in
the usual degenerate perturbation theory, that is, we are free to
choose our base set of unperturbed kets $\ket{\Phi^{\gamma
a_{\gamma}}}$ in such a way that that $H_1$ is diagonalized in the
corresponding degenerate subspaces. In other words, we should find
the linear combinations of the degenerate unperturbed kets to
re-span the zero-order eigen subspace of $H_0$ so that \beqa
\bra{\Phi^{\gamma a_{\gamma}}}H_1\ket{\Phi^{\gamma b_{\gamma}}}&=&
g_1^{\gamma a_{\gamma},\gamma b_{\gamma}}=d_1^{\gamma a_\gamma}
\delta_{a_\gamma b_\gamma}.\eeqa (If there is still the same values
among all of $d_{\gamma a}$, this procedure can be repeated in
general.) This means that $g_1^{\gamma a_{\gamma},\gamma
a_{\gamma^\prime}}=0$. Then, we use our redivision skill again, that
is \beqa H_0^{\prime\prime}&=&H_0+\sum_{\gamma\notin D,\;
\gamma}h_1^\gamma\ket{\Phi^\gamma}\bra{\Phi^\gamma} +\sum_{\gamma\in
D,\;a_{\gamma}}d_1^{\gamma a_{\gamma}}
\ket{\Phi^{\gamma a_{\gamma}}}\bra{\Phi^{\gamma a_{\gamma}}},\\
H_1^{\prime\prime}&=&H_1-\sum_{\gamma\notin D,\;
\gamma}h_1^\gamma\ket{\Phi^\gamma}\bra{\Phi^\gamma}-\sum_{\gamma\in
D,\;a_{\gamma}}d_1^{\gamma a_{\gamma}} \ket{\Phi^{\gamma
a_{\gamma}}}\bra{\Phi^{\gamma a_{\gamma}}}.\eeqa where $D$ is a set
of all degenerate subspace-indexes. Thus, the last term in Eq.
(\ref{A1gd}) vanishes, \beqa \label{A1gdnew} A_1^{\gamma
a_{\gamma},\gamma^\prime a_{\gamma^\prime}}(g)&=&\left[\frac{\e^{-\I
E_{\gamma a_\gamma} t}}{E_{\gamma a_\gamma}-E_{\gamma^\prime
a_{\gamma^\prime}}}-\frac{\e^{-\I E_{\gamma^\prime
a_{\gamma^\prime}} t}}{E_{\gamma a_\gamma}-E_{\gamma^\prime
a_{\gamma^\prime}}}\right]g_1^{\gamma a_{\gamma},\gamma^\prime
a_{\gamma^\prime}}\eta_{\gamma\gamma^\prime}.\eeqa In fact, under
the preconditions of $H_1$ is diagonal in the degenerate subspaces,
we can directly do replacement \beq
g_1^{\gamma_i\gamma_j}\rightarrow g_1^{\gamma_i
a_{\gamma_i},\gamma_j a_{\gamma_j}}\eta_{\gamma_i\gamma_j}\eeq from
the non-degenerate case to the degenerate case. For simplicity, we
always assume that $H_1$ has been diagonalized in the degenerate
subspaces from now on.

It must be emphasized that the Hamiltonian redivision skill leads to
the fact that the new perturbed solution can be obtained by the
replacement \beq \label{rdenergy} E_{\gamma_i}\rightarrow
E_{\gamma_i}+h_1^{\gamma_i}\eeq in the non-degenerate perturbed
solution and its conclusions. With degeneracy present, if our method
is to work well, the degeneracy should be completely removed in the
diagonalization procedures of the degenerate subspaces and the
Hamiltonian redivision, that is, for any given degenerate subspace,
$d_1^{\gamma a}\neq d_1^{\gamma b}$ if $a\neq b$. In other words,
$E_{\gamma a}^{\prime\prime}\neq E_{\gamma b}^{\prime\prime}$ if
$a\neq b$. This means that all of eigenvalues of
$H_0^{\prime\prime}$ are no longer the same, so we can back to the
non-degenerate cases. Or specially, if we allow the remained
degeneracies, the off-diagonal element of the perturbing Hamiltonian
matrix between any two degenerate levels are always vanishing. This
implies that there is no extra contribution from the degeneracies in
the any more than the zeroth order approximations. It is important
to remember these facts. However, how must we proceed if the
degeneracies are not completely removed by the usual diagonalization
procedure and our Hamiltonian redivision as well as the special
cases with the remained degeneracies stated above are not valid. It
is known to be a challenge in the usual perturbation theory.
Although our exact solution can apply to such a kind of cases, but
the form of perturbed solution will get complicated because more
apparent divergences need to be eliminated and then some new terms
proportional to the power of evolution time will appear in general.
We will study this problem in the near future. Based on the above
reasons, we do not consider the degenerate case from now on.

From the statement above, we have seen that there are two equivalent
ways to obtain the same perturbed solution and its conclusions. One
of them is to redefine the energy level $E_{\gamma_i}$ as
$E^\prime_{\gamma_i}$ (or $E^{\prime\prime}_{\gamma_i}$), think
$E^\prime_{\gamma_i}$ (or $E^{\prime\prime}_{\gamma_i}$) to be
explicitly independent on the perturbing parameter from a redefined
view, and then use the method in the usual perturbation theory to
obtain the result from the redivided $H_1^\prime$ (or
$H_1^{\prime\prime}$). The other way is to directly derive out the
perturbed solution from the original Hamiltonian by using the
standard procedure, but the rearrangement and summation are carried
out just like what we have done above. From our point of view, this
is because the perturbing parameter is only a formal multiplier in
mathematics and it can be introduced after redefining
$E^\prime_{\gamma_i}$. It is natural although this problem seems not
be noticed for a long time. The first skill, that is, the
Hamiltonian redivision skill will be again applied to our scheme to
obtain improved forms of perturbed energy and perturbed state in
Sec. \ref{sec3}.

The Hamiltonian redivision not only overcomes the flaw of the usual
perturbation theory, but also has three obvious advantages. Firstly,
it advances the calculation precision of perturbation theory because
it makes the contributions from all order approximations of the
diagonal elements of the perturbing Hamiltonian matrix are naturally
included. Secondly, it extends the applicable range of perturbation
theory based on the same reason since the diagonal elements of the
perturbing Hamiltonian no longer is needed to be smaller. Lastly, it
can be used to remove the degeneracies, which is important for the
perturbation theory.

For simplicity, in the following, we omit the ${}^\prime$ (or
${}^{\prime\prime}$) in $H_0$, $H_1$ as well as $E_\gamma$, and
always let $H_1$ have only its off-diagonal part and let $H_0$ have
no degeneracy unless particular claiming.

\subsection{Perturbed Hamiltonian matrix product decomposition and apparent divergence elimination}

In this subsection, we present the second important skill
enlightened by our exact solution, that is, the perturbing
Hamiltonian matrix product decomposition, which is a technology to
separate the contraction terms with apparent divergences and
anti-contraction terms without apparent divergences, and then we can
eliminate these apparent divergences by the limit process. More
importantly, we can propose so-called ``dynamical rearrangement and
summation" idea in order to absorb the partial contributions from
the high order even all order approximation of perturbing
Hamiltonian into the lower order terms of our perturbation theory.
It is a key method in our improved scheme of perturbation theory.

Let us start with the second order approximation. Since we have
taken $H_1^{\gamma_j\gamma_{j+1}}$ only with the off-diagonal part
$g_1^{\gamma_j\gamma_{j+1}}$, the contribution from the second order
approximation of the perturbing Hamiltonian is only
$A_2^{\gamma\gamma^\prime}(gg)$ in eq.(\ref{A2gg}). However, we find
that in the above expression of $A_2^{\gamma\gamma^\prime}(gg)$, the
apparent divergence has not been completely eliminated or the
limitation has not been completely found out because we have not
excluded the case $E_\gamma=E_{\gamma^\prime}$ (or
$\gamma=\gamma^\prime$). This problem can be fixed by introducing a
perturbing Hamiltonian matrix product decomposition \beq
g_1^{\gamma_1\gamma_2}g_1^{\gamma_2\gamma_3}
=g_1^{\gamma_1\gamma_2}g_1^{\gamma_2\gamma_3}\delta_{\gamma_1\gamma_3}
+g_1^{\gamma_1\gamma_2}g_1^{\gamma_2\gamma_3}\eta_{\gamma_1\gamma_3},
\eeq where $\eta_{\gamma_1\gamma_3}=1-{\delta}_{\gamma_1\gamma_3}$.
Thus, the contribution from the second order approximation is made
of two terms, one so-called contraction term with the $\delta$
function factor and the other so-called anti-contraction term with
the $\eta$ function factor. Obviously, the contraction term has the
apparent divergence and anti-contraction term has no the apparent
divergence. Hence, in order to eliminate the apparent divergence in
the contraction term, we only need to find its limitation. It must
be emphasized that we only consider the non-degenerate case here and
after for simplification. When the degeneration happens, two indexes
with the same main energy level number will not have the
anti-contraction.

In terms of the above skill, we find that the contribution from the
second order approximation is made of the corresponding contraction-
and anti-contraction- terms \beq
{A}_2^{\gamma\gamma^\prime}({gg})={A}_2^{\gamma\gamma^\prime}(gg;c)
+{A}_2^{\gamma\gamma^\prime}(gg;n),\eeq where
\beqa \label{A2ggc}
{A}_2^{\gamma\gamma^\prime}(gg;c)&=&\sum_{\gamma_1,\gamma_2,\gamma_{3}}\left[
\sum_{i=1}^{3}(-1)^{i-1}\frac{\e^{-\I E_{\gamma_i}
t}}{d_i(E[\gamma,2])}\right]
g_1^{\gamma_1\gamma_2}g_1^{\gamma_2\gamma_{3}}\delta_{\gamma_1\gamma_3}
\delta_{\gamma\gamma_1}\delta_{\gamma^\prime\gamma_3}\nonumber\\
&=& \sum_{\gamma_1}\left[-\frac{\e^{-\I
E_{\gamma}t}}{\left(E_\gamma-E_{\gamma_1}\right)^2}+\frac{\e^{-\I
E_{\gamma_1}t}}{\left(E_\gamma-E_{\gamma_1}\right)^2}+(-\I
t)\frac{\e^{-\I E_{\gamma}t}}{E_\gamma-E_{\gamma_1}}\right]
\left|g_1^{\gamma\gamma_1}\right|^2\delta_{\gamma\gamma^\prime},
\eeqa
\beqa \label{A2ggn}
{A}_2^{\gamma\gamma^\prime}(gg;n)&=&\sum_{\gamma_1,\gamma_2,\gamma_{3}}\left[
\sum_{i=1}^{3}(-1)^{i-1}\frac{\e^{-\I E_{\gamma_i}
t}}{d_i(E[\gamma,2])}\right]
g_1^{\gamma_1\gamma_2}g_1^{\gamma_2\gamma_{3}}\eta_{\gamma_1\gamma_3}
\delta_{\gamma\gamma_1}\delta_{\gamma^\prime\gamma_3}\nonumber\\
&=&\sum_{\gamma_1}\left[\frac{\e^{-\I
E_{\gamma}t}}{(E_{\gamma}-E_{\gamma_1})(E_{\gamma}-E_{\gamma^\prime})}-\frac{\e^{-\I
E_{\gamma_1}t}}{(E_{\gamma}-E_{\gamma_1})(E_{\gamma_1}-E_{\gamma^\prime})}+\frac{\e^{-\I
E_{\gamma^\prime}t}}{(E_{\gamma}-E_{\gamma^\prime})(E_{\gamma_1}-E_{\gamma^\prime})}\right]
g_1^{\gamma\gamma_1}g_1^{\gamma_1\gamma^\prime}\eta_{\gamma\gamma^\prime}.
\hskip 1.0cm\eeqa

The above method can be extended to the higher order approximation
by introducing a skill of perturbing Hamiltonian matrix product
decomposition, or simply called it $g$-product decomposition when
the perturbing Hamiltonian matrix is off-diagonal. For a sequential
product of off-diagonal elements $g$ with the form $\prod_{k=1}^m
g_1^{\gamma_k\gamma_{k+1}}$ ($m\geq 2$), we define its $(m-1)$th
decomposition by \beq \label{gpd} \prod_{k=1}^m
g_1^{\gamma_k\gamma_{k+1}}=\left(\prod_{k=1}^m
g_1^{\gamma_k\gamma_{k+1}}\right)\delta_{\gamma_1\gamma_{m+1}}
+\left(\prod_{k=1}^m
g_1^{\gamma_k\gamma_{k+1}}\right)\eta_{\gamma_1\gamma_{m+1}}.\eeq
When we calculate the contributions from the $n$th order
approximation, we will first carry out $(n-1)$ the first
decompositions, that is \beq \label{rdofgp}\prod_{k=1}^n
g_1^{\gamma_k\gamma_{k+1}}=\left(\prod_{k=1}^n
g_1^{\gamma_k\gamma_{k+1}}\right)\left[\prod_{k=1}^{n-1}
\left(\delta_{\gamma_k\gamma_{k+2}}+\eta_{\gamma_k\gamma_{k+2}}\right)\right].
\eeq Obviously, from the fact that $H_1$ is usually taken as Hermit
one, it follows that \beq
g_1^{\gamma_{j}\gamma_{j+1}}g_1^{\gamma_{j+1}\gamma_{j+2}}\delta_{\gamma_j\gamma_{j+2}}
=\left|g_1^{\gamma_{j}\gamma_{j+1}}\right|^2\delta_{\gamma_j\gamma_{j+2}}.\eeq
When the contribution from a given order approximation is
considered, the summation over one of two subscripts will lead to
the contraction of $g$-production. More generally, for the
contraction of even number $g$-production \beq \left(\prod_{j=1}^m
g_1^{\gamma_j\gamma_{j+1}}
\prod_{k=1}^{m-1}\delta_{\gamma_k\gamma_{k+2}}\right)
\delta_{\gamma_1\gamma}\delta_{\gamma_{m+1}\gamma^\prime}
=\left|g_1^{\gamma\gamma_2}\right|^m\left(\prod_{k=1}^{m-1}\delta_{\gamma_k\gamma_{k+2}}\right)
\delta_{\gamma_1\gamma}\delta_{\gamma_{m+1}\gamma^\prime}\delta_{\gamma\gamma^\prime},\eeq
and for the contraction of odd number $g$-production, \beq
\left(\prod_{j=1}^m g_1^{\gamma_j\gamma_{j+1}}
\prod_{k=1}^{m-1}\delta_{\gamma_k\gamma_{k+2}}\right)
\delta_{\gamma_1\gamma}\delta_{\gamma_{m+1}\gamma^\prime}
=\left|g_1^{\gamma\gamma^\prime}\right|^{m-1}\left(\prod_{k=1}^{m-1}\delta_{\gamma_k\gamma_{k+2}}\right)
\delta_{\gamma_1\gamma}\delta_{\gamma_{m+1}\gamma^\prime}g_1^{\gamma\gamma^\prime},\eeq
where $\delta_{\gamma_1\gamma}\delta_{\gamma_{m+1}\gamma^\prime}$ is
a factor appearing in the expression of our solution.

Then, we consider, in turn, all possible the second decomposition,
the third decomposition, and up to the $(n-1)$th decomposition. It
must be emphasized that after calculating the contributions from the
terms of lower decompositions, some of terms in the higher
decompositions may be trivial because there are some symmetric and
complementary symmetric indexes in the corresponding results, that
is, the products of these results and the given
$\delta_{\gamma_k\gamma_{k^\prime}}$ or
$\eta_{\gamma_k\gamma_{k^\prime}}$ are zero. In other words, such
some higher decompositions do not need to be considered. As an
example, let us analyze the contribution from the third order
approximation. It is clear that the first decomposition of a
sequential production of three off-diagonal elements becomes \beqa
g_1^{\gamma_1\gamma_2}g_1^{\gamma_2\gamma_3}g_1^{\gamma_3\gamma_4}
&=&g_1^{\gamma_1\gamma_2}g_1^{\gamma_2\gamma_3}g_1^{\gamma_3\gamma_4}
\delta_{\gamma_1\gamma_3}\delta_{\gamma_2\gamma_4}
+g_1^{\gamma_1\gamma_2}g_1^{\gamma_2\gamma_3}g_1^{\gamma_3\gamma_4}
\delta_{\gamma_1\gamma_3}\eta_{\gamma_2\gamma_4}
\nonumber\\
&
&+g_1^{\gamma_1\gamma_2}g_1^{\gamma_2\gamma_3}g_1^{\gamma_3\gamma_4}
\eta_{\gamma_1\gamma_3}\delta_{\gamma_2\gamma_4}
+g_1^{\gamma_1\gamma_2}g_1^{\gamma_2\gamma_3}g_1^{\gamma_3\gamma_4}
\eta_{\gamma_1\gamma_3}\eta_{\gamma_2\gamma_4}.\eeqa Thus, the
related contribution is just divided into $4$ terms \beq
A_3^{\gamma\gamma^\prime}(ggg)=A_3^{\gamma\gamma^\prime}(ggg;cc)
+A_3^{\gamma\gamma^\prime}(ggg;cn)+A_3^{\gamma\gamma^\prime}(ggg;nc)
+{A}_3^{\gamma\gamma^\prime}(ggg,nn).\eeq In fact, by calculating we
know that the second decomposition of the former three terms do not
need to be considered, only the second decomposition of the last
term is nontrivial. This means that \beq
{A}_3^{\gamma\gamma^\prime}(ggg;nn)={A}_3^{\gamma\gamma^\prime}(ggg;nn,c)
+{A}_3^{\gamma\gamma^\prime}(ggg;nn,n),\eeq where we have added
$\delta_{\gamma_1\gamma_3}$ in the definition of
${A}_3^{\gamma\gamma^\prime}(ggg;nn,c)$, and
$\eta_{\gamma_1\gamma_3}$ in the definition of
${A}_3^{\gamma\gamma^\prime}(ggg;nn,n)$. Obviously, in the practical
process, this feature largely simplifies the calculations. It is
easy to see that the number of all of terms with contractions and
anti-contractions is $5$. For convenience and clearness, we call the
contributions from the different terms in the decomposition of
$g$-product as the contractions and anti-contractions of
$g$-product. Of course, the contraction and anti-contraction refer
to the meaning after summation(s) over the subscript(s) in general.
Moreover, here and after, we drop the argument $gg\cdots g$ in the
$i$th order approximation $A_i$ since its meaning has been indicated
by $i$ after the Hamiltonian is redivided. For example, the explicit
expressions of all contraction- and anti-contraction terms in the
third order approximation $A_3$ can be calculated as follows:
\beqa\label{A3cc} {A}_3^{\gamma\gamma^\prime}(cc)&=
&\sum_{\gamma_1,\cdots,\gamma_{4}}\left[
\sum_{i=1}^{4}(-1)^{i-1}\frac{\e^{-\I E_{\gamma_i}
t}}{d_i(E[\gamma,3])}\right] \left[\prod_{j=1}^3
g_1^{\gamma_j\gamma_{j+1}}\right]\left(
\prod_{k=1}^{2}\delta_{\gamma_k\gamma_{k+2}}\right)
\delta_{\gamma_1\gamma}\delta_{\gamma_{4}\gamma^\prime}\nonumber\\
& =& \left[-\frac{2\e^{-\I
E_{\gamma}t}}{\left(E_\gamma-E_{\gamma^\prime}\right)^3}+\frac{2\e^{-\I
E_{\gamma^\prime}t}}{\left(E_\gamma-E_{\gamma^\prime}\right)^3}+(-\I
t)\frac{\e^{-\I E_{\gamma}t}}
{\left(E_\gamma-E_{\gamma^\prime}\right)^2}+(-\I t)\frac{\e^{-\I
E_{\gamma^\prime}t}}{\left(E_\gamma-E_{\gamma^\prime}\right)^2}\right]
\left|g_1^{\gamma\gamma^\prime}\right|^2g_1^{\gamma\gamma^\prime},\eeqa
\beqa\label{A3cn}
{A}_3^{\gamma\gamma^\prime}(cn)&=&\sum_{\gamma_1,\cdots,\gamma_{4}}\left[
\sum_{i=1}^{4}(-1)^{i-1}\frac{\e^{-\I E_{\gamma_i}
t}}{d_i(E[\gamma,3])}\right] \left[\prod_{j=1}^3
g_1^{\gamma_j\gamma_{j+1}}\right]\delta_{\gamma_1\gamma_3}\eta_{\gamma_2\gamma_4}
\delta_{\gamma_1\gamma}\delta_{\gamma_{l+1}\gamma^\prime}\nonumber\\
& =& \sum_{\gamma_1}\left[-\frac{\e^{-\I
E_{\gamma}t}}{\left(E_\gamma-E_{\gamma_1}\right)
\left(E_\gamma-E_{\gamma^\prime}\right)^2}-\frac{\e^{-\I
E_{\gamma}t}}{\left(E_\gamma-E_{\gamma_1}\right)^2
\left(E_\gamma-E_{\gamma^\prime}\right)}+\frac{\e^{-\I
E_{\gamma_1}t}}{\left(E_\gamma-E_{\gamma_1}\right)^2
\left(E_{\gamma_1}-E_{\gamma^\prime}\right)}\right.\nonumber\\
& &\left. -\frac{\e^{-\I
E_{\gamma^\prime}t}}{\left(E_\gamma-E_{\gamma^\prime}\right)^2
\left(E_{\gamma_1}-E_{\gamma^\prime}\right)}+(-\I t)\frac{\e^{-\I
E_{\gamma}t}}
{\left(E_\gamma-E_{\gamma_1}\right)\left(E_\gamma-E_{\gamma^\prime}\right)}\right]
\left|g_1^{\gamma\gamma_1}\right|^2g_1^{\gamma\gamma^\prime}
\eta_{\gamma_1\gamma^\prime},\eeqa
\beqa \label{A3nc} {A}_3^{\gamma\gamma^\prime}(nc)&=
&\sum_{\gamma_1,\cdots,\gamma_{4}}\left[
\sum_{i=1}^{4}(-1)^{i-1}\frac{\e^{-\I E_{\gamma_i}
t}}{d_i(E[\gamma,3])}\right] \left[\prod_{j=1}^3
g_1^{\gamma_j\gamma_{j+1}}\right]\eta_{\gamma_1\gamma_3}\delta_{\gamma_2\gamma_4}
\delta_{\gamma_1\gamma}\delta_{\gamma_{l+1}\gamma^\prime}\nonumber\\
& =& \sum_{\gamma_1}\left[\frac{\e^{-\I
E_{\gamma}t}}{\left(E_\gamma-E_{\gamma_1}\right)\left(E_\gamma-E_{\gamma^\prime}\right)^2}-\frac{\e^{-\I
E_{\gamma_1}t}}{\left(E_\gamma-E_{\gamma_1}\right)
\left(E_{\gamma_1}-E_{\gamma^\prime}\right)^2}-\frac{\e^{-\I
E_{\gamma^\prime}t}}{\left(E_\gamma-E_{\gamma_1}\right)\left(E_\gamma-E_{\gamma^\prime}\right)^2}
\right.\nonumber\\
& &\left.+\frac{\e^{-\I
E_{\gamma^\prime}t}}{\left(E_\gamma-E_{\gamma_1}\right)\left(E_{\gamma_1}-E_{\gamma^\prime}\right)^2}+(-\I
t)\frac{\e^{-\I E_{\gamma^\prime}t}}
{\left(E_\gamma-E_{\gamma^\prime}\right)\left(E_{\gamma_1}-E_{\gamma^\prime}\right)}\right]
g_1^{\gamma\gamma^\prime}
\left|g_1^{\gamma_1\gamma^\prime}\right|^2\eta_{\gamma\gamma_1},\eeqa
\beqa \label{A3nn-c} {A}_3^{\gamma\gamma^\prime}(nn,c)
&=&\sum_{\gamma_1,\cdots,\gamma_{4}}\left[
\sum_{i=1}^{4}(-1)^{i-1}\frac{\e^{-\I E_{\gamma_i}
t}}{d_i(E[\gamma,3])}\right] \left[\prod_{j=1}^3
g_1^{\gamma_j\gamma_{j+1}}\right]
\delta_{\gamma_1\gamma}\delta_{\gamma_{4}\gamma^\prime}
\eta_{\gamma_1\gamma_3}
\eta_{\gamma_2\gamma_4}{\delta}_{\gamma\gamma^\prime}
\nonumber\\
&=& \sum_{\gamma_1\gamma_2}\left[-\frac{\e^{-\I
E_{\gamma}t}}{\left(E_{\gamma}-E_{\gamma_1}\right)^2
\left(E_{\gamma_1}-E_{\gamma_2}\right)}+\frac{\e^{-\I
E_{\gamma}t}}{\left(E_{\gamma}-E_{\gamma_2}\right)^2
\left(E_{\gamma_1}-E_{\gamma_2}\right)}+\frac{\e^{-\I
E_{\gamma_1}t}}{\left(E_{\gamma}-E_{\gamma_1}\right)^2
\left(E_{\gamma_1}-E_{\gamma_2}\right)}\right.\nonumber\\
& &\left. -\frac{\e^{-\I
E_{\gamma_2}t}}{\left(E_{\gamma}-E_{\gamma_2}\right)^2
\left(E_{\gamma_1}-E_{\gamma_2}\right)}+(-\I t)\frac{\e^{-\I
E_{\gamma}t}}{\left(E_{\gamma}-E_{\gamma_1}\right)
\left(E_{\gamma_1}-E_{\gamma_2}\right)}
\right]g_1^{\gamma\gamma_1}g_1^{\gamma_1\gamma_2}g_1^{\gamma_2\gamma^\prime}
\eta_{\gamma\gamma_2}{\delta}_{\gamma\gamma^\prime}, \eeqa
\beqa \label{A3nn-n} {A}_3^{\gamma\gamma^\prime}(nn,n)
&=&\sum_{\gamma_1,\cdots,\gamma_{4}}\left[
\sum_{i=1}^{4}(-1)^{i-1}\frac{\e^{-\I E_{\gamma_i}
t}}{d_i(E[\gamma,3])}\right] \left[\prod_{j=1}^3
g_1^{\gamma_j\gamma_{j+1}}\right]
\delta_{\gamma_1\gamma}\delta_{\gamma_{4}\gamma^\prime}
\eta_{\gamma_1\gamma_3}
\eta_{\gamma_2\gamma_4}\eta_{\gamma\gamma^\prime}
\nonumber\\
&=& \sum_{\gamma_1\gamma_2}\left[\frac{\e^{-\I
E_{\gamma}t}}{\left(E_{\gamma}-E_{\gamma_1}\right)
\left(E_{\gamma}-E_{\gamma_2}\right)\left(E_{\gamma}-E_{\gamma^\prime}\right)}
-\frac{\e^{-\I E_{\gamma_1}t}}{\left(E_{\gamma}-E_{\gamma_1}\right)
\left(E_{\gamma_1}-E_{\gamma_2}\right)\left(E_{\gamma_1}-E_{\gamma^\prime}\right)}\right.\nonumber\\
& & \left.+\frac{\e^{-\I
E_{\gamma_2}t}}{\left(E_{\gamma}-E_{\gamma_2}\right)
\left(E_{\gamma_1}-E_{\gamma_2}\right)\left(E_{\gamma_2}-E_{\gamma^\prime}\right)}-\frac{\e^{-\I
E_{\gamma^\prime}t}}{\left(E_{\gamma}-E_{\gamma^\prime}\right)
\left(E_{\gamma_1}-E_{\gamma^\prime}\right)\left(E_{\gamma_2}-E_{\gamma^\prime}\right)}
\right]\nonumber\\
& &\times
g_1^{\gamma\gamma_1}g_1^{\gamma_1\gamma_2}g_1^{\gamma_2\gamma^\prime}
\eta_{\gamma\gamma_2}\eta_{\gamma_1\gamma^\prime}
\eta_{\gamma\gamma^\prime}.\eeqa In the above calculations, the
uesed technologies mainly to find the limitation, dummy index
changing and summation, as well as the replacement
$g_1^{\gamma_i\gamma_j}\eta_{\gamma_i\gamma_j}=g_1^{\gamma_i\gamma_j}$
since $g_1^{\gamma_i\gamma_j}$ has been off-diagonal.

It must be emphasized that, in our notation,
$A_i^{\gamma\gamma^\prime}$ represents the contributions from the
$i$th order approximation. The other independent variables are
divided into $i-1$ groups and are arranged sequentially
corresponding to the order of $g$-product decomposition. That is,
the first variable group represents the first decomposition, the
second variable group represents the second decomposition, and so
on. Every variable group is a bit-string made of three possible
element $c,n,k$ and its length is equal to the number of the related
order of $g$-product decomposition, that is, for the $j$th
decompositions in the $i$th order approximation its length is $i-j$.
In each variable group, $c$ corresponds to a $\delta$ function, $n$
corresponds to a $\eta$ function and $k$ corresponds to $1$
(non-decomposition). Their sequence in the bit-string corresponds to
the sequence of contraction and/or anti-contraction index string.
From the above analysis and statement, the index string of the $j$th
decompositions in the $i$ order approximation is: \beq
\prod_{k=1}^{i-j}\left(\gamma_k,\gamma_{k+1+j}\right). \eeq For
example, for $A_5$, the first variable group is $cccn$, which refers
to the first decomposition in five order approximation and the terms
to include the factor
$\delta_{\gamma_1\gamma_3}\delta_{\gamma_2\gamma_4}
\delta_{\gamma_3\gamma_5}\eta_{\gamma_4\gamma_6}$ in the definition
of $A_5(cccn)$. Similarly, $cncc$ means to insert the factor
$\delta_{\gamma_1\gamma_3}\eta_{\gamma_2\gamma_4}
\delta_{\gamma_3\gamma_5}\delta_{\gamma_4\gamma_6}$ into the
definition of $A_5(cncc)$. When there are nontrivial second
contractions, for instance, two variable groups $(ccnn,kkc)$
represent that the definition of $A_5(ccnn,kkc)$ has the factor
$\left(\delta_{\gamma_1\gamma_3}\delta_{\gamma_2\gamma_4}
\eta_{\gamma_3\gamma_5}\eta_{\gamma_4\gamma_6}\right)\delta_{\gamma_3\gamma_6}$.
Since there are fully trivial contraction (the bit-string is made of
only $k$), we omit their related variable group for simplicity.

Furthermore, we pack up all the contraction- and non-contraction
terms in the following way so that we can obtain conveniently the
improved forms of perturbed solution of dynamics absorbing the
partial contributions from the high order even all order
approximations. We first decompose $A_3^{\gamma\gamma^\prime}$,
which is a summation of all above terms, into the three parts
according to $\e^{-\I E_{\gamma_i}t}, (-\I t) \e^{-\I E_{\gamma_i}t}
$ and $(-\I t)^2\e^{-\I E_{\gamma_i}t}/2$: \beq
A_3^{\gamma\gamma^\prime}=A_3^{\gamma\gamma^\prime}(\e)+A_3^{\gamma\gamma^\prime}(t\e)
+A_3^{\gamma\gamma^\prime}(t^2\e).\eeq Secondly, we decompose its
every term into three parts according to $\e^{-\I E_{\gamma}t},
\e^{-\I E_{\gamma_1}t}$ ($\sum_{\gamma_1}\e^{-\I E_{\gamma_1}t}$)
and $\e^{-\I E_{\gamma^\prime}t}$: \beqa
A_3^{\gamma\gamma^\prime}(\e)&=&A_3^{\gamma\gamma^\prime}(\e^{-\I
E_{\gamma}t})+A_3^{\gamma\gamma^\prime}(\e^{-\I
E_{\gamma_1}t})+A_3^{\gamma\gamma^\prime}(\e^{-\I E_{\gamma^\prime}t}),\\
A_3^{\gamma\gamma^\prime}(t\e)&=&A_3^{\gamma\gamma^\prime}(t\e^{-\I
E_{\gamma}t})+A_3^{\gamma\gamma^\prime}(t\e^{-\I
E_{\gamma_1}t})+A_3^{\gamma\gamma^\prime}(t\e^{-\I E_{\gamma^\prime}t}),\\
A_4^{\gamma\gamma^\prime}(t^2\e)&=&A_3^{\gamma\gamma^\prime}(t^2\e^{-\I
E_{\gamma}t})+A_3^{\gamma\gamma^\prime}(t^2\e^{-\I
E_{\gamma_1}t})+A_3^{\gamma\gamma^\prime}(t^2\e^{-\I
E_{\gamma^\prime}t}). \eeqa Finally, we again decompose every term
in the above equations into the diagonal and off-diagonal parts
about $\gamma$ and $\gamma^\prime$: \beqa
A_3^{\gamma\gamma^\prime}(\e^{-\I
E_{\gamma_i}t})&=&A_3^{\gamma\gamma^\prime}(\e^{-\I
E_{\gamma_i}t};{\rm D})+A_3^{\gamma\gamma^\prime}(\e^{-\I
E_{\gamma_i}t};{\rm N}),\\
A_3^{\gamma\gamma^\prime}(t\e^{-\I
E_{\gamma_i}t})&=&A_3^{\gamma\gamma^\prime}(t\e^{-\I
E_{\gamma_i}t};{\rm D})+A_3^{\gamma\gamma^\prime}(t\e^{-\I
E_{\gamma_i}t};{\rm N}) ,\\
A_3^{\gamma\gamma^\prime}(t^2\e^{-\I
E_{\gamma_i}t})&=&A_3^{\gamma\gamma^\prime}(t^2\e^{-\I
E_{\gamma_i}t};{\rm D})+A_3^{\gamma\gamma^\prime}(t^2\e^{-\I
E_{\gamma_i}t};{\rm N}),  \eeqa where $E_{\gamma_i}$ takes $
E_{\gamma}, E_{\gamma_1}$ and $E_{\gamma^\prime}$.

According to the above way, it is easy to obtain \beqa
A_3^{\gamma\gamma^\prime}(\e^{-\I E_{\gamma}t};{\rm D})&=&
-\sum_{\gamma_1,\gamma_2}\e^{-\I E_{\gamma}
t}\left[\frac{1}{\left(E_{\gamma}-E_{\gamma_1}\right)
\left(E_{\gamma}-E_{\gamma_2}\right)^2}+\frac{1}{\left(E_{\gamma}-E_{\gamma_1}\right)^2
\left(E_{\gamma}-E_{\gamma_2}\right)}\right]g_1^{\gamma\gamma_1}g_1^{\gamma_1\gamma_2}
g_1^{\gamma_2\gamma}\delta_{\gamma\gamma^\prime},\\
A_3^{\gamma\gamma^\prime}(\e^{-\I E_{\gamma}t};{\rm N})&=&
-\sum_{\gamma_1}\e^{-\I E_{\gamma}
t}\left[\frac{1}{\left(E_{\gamma}-E_{\gamma_1}\right)
\left(E_{\gamma}-E_{\gamma^\prime}\right)^2}+\frac{1}{\left(E_{\gamma}-E_{\gamma_1}\right)^2
\left(E_{\gamma}-E_{\gamma^\prime}\right)}\right]g_1^{\gamma\gamma_1}g_1^{\gamma_1\gamma}
g_1^{\gamma\gamma^\prime}\nonumber\\
& &+\sum_{\gamma_1,\gamma_2}\e^{-\I
E_{\gamma}t}\frac{g_1^{\gamma\gamma_1}g_1^{\gamma_1\gamma_2}
g_1^{\gamma_2\gamma^\prime}\eta_{\gamma\gamma_2}\eta_{\gamma\gamma^\prime}}
{\left(E_{\gamma}-E_{\gamma_1}\right)
\left(E_{\gamma}-E_{\gamma_2}\right)\left(E_{\gamma}-E_{\gamma^\prime}\right)},\eeqa
\beqa A_3^{\gamma\gamma^\prime}(\e^{-\I E_{\gamma_1}t};{\rm D})&=&
\sum_{\gamma_1,\gamma_2}\e^{-\I E_{\gamma_1}
t}\frac{g_1^{\gamma\gamma_1}g_1^{\gamma_1\gamma_2}
g_1^{\gamma_2\gamma}\delta_{\gamma\gamma^\prime}}{\left(E_{\gamma}-E_{\gamma_1}\right)^2
\left(E_{\gamma_1}-E_{\gamma_2}\right)},\\
A_3^{\gamma\gamma^\prime}(\e^{-\I E_{\gamma_1}t};{\rm N})&=&
-\sum_{\gamma_1,\gamma_2}\e^{-\I
E_{\gamma_1}t}\frac{g_1^{\gamma\gamma_1}g_1^{\gamma_1\gamma_2}
g_1^{\gamma_2\gamma^\prime}\eta_{\gamma_1\gamma^\prime}\eta_{\gamma\gamma^\prime}}
{\left(E_{\gamma}-E_{\gamma_1}\right)
\left(E_{\gamma_1}-E_{\gamma_2}\right)\left(E_{\gamma_1}-E_{\gamma^\prime}\right)},\eeqa
\beqa A_3^{\gamma\gamma^\prime}(\e^{-\I E_{\gamma_2}t};{\rm D})&=&
-\sum_{\gamma_1,\gamma_2}\e^{-\I E_{\gamma_2}
t}\frac{g_1^{\gamma\gamma_1}g_1^{\gamma_1\gamma_2}
g_1^{\gamma_2\gamma}\delta_{\gamma\gamma^\prime}}{\left(E_{\gamma}-E_{\gamma_2}\right)^2
\left(E_{\gamma_1}-E_{\gamma_2}\right)},\\
A_3^{\gamma\gamma^\prime}(\e^{-\I E_{\gamma_2}t};{\rm N})&=&
\sum_{\gamma_1,\gamma_2}\e^{-\I
E_{\gamma_2}t}\frac{g_1^{\gamma\gamma_1}g_1^{\gamma_1\gamma_2}
g_1^{\gamma_2\gamma^\prime}\eta_{\gamma\gamma_2}\eta_{\gamma\gamma^\prime}}
{\left(E_{\gamma}-E_{\gamma_2}\right)
\left(E_{\gamma_1}-E_{\gamma_2}\right)\left(E_{\gamma_2}-E_{\gamma^\prime}\right)},\eeqa
\beqa A_3^{\gamma\gamma^\prime}(\e^{-\I E_{\gamma^\prime}t};{\rm
D})&=&0,\\ A_3^{\gamma\gamma^\prime}(\e^{-\I
E_{\gamma^\prime}t};{\rm N})&=& \sum_{\gamma_1}\e^{-\I
E_{\gamma^\prime}
t}\left[\frac{1}{\left(E_{\gamma}-E_{\gamma^\prime}\right)
\left(E_{\gamma_1}-E_{\gamma^\prime}\right)^2}+\frac{1}{\left(E_{\gamma}-E_{\gamma^\prime}\right)^2
\left(E_{\gamma_1}-E_{\gamma^\prime}\right)}\right]g_1^{\gamma^\prime\gamma_1}g_1^{\gamma_1\gamma^\prime}
g_1^{\gamma\gamma^\prime}\nonumber\\
& &-\sum_{\gamma_1,\gamma_2}\e^{-\I
E_{\gamma^\prime}t}\frac{g_1^{\gamma\gamma_1}g_1^{\gamma_1\gamma_2}
g_1^{\gamma_2\gamma^\prime}\eta_{\gamma_1\gamma^\prime}\eta_{\gamma\gamma^\prime}}
{\left(E_{\gamma}-E_{\gamma^\prime}\right)
\left(E_{\gamma_1}-E_{\gamma^\prime}\right)\left(E_{\gamma_2}-E_{\gamma^\prime}\right)}.\eeqa

In the end of this subsection, we would like to point out that the
main purpose introducing the $g$-product decomposition and
calculating the contractions and anti-contractions of $g$-product is
to eliminate the apparent divergences and find out all the
limitations from the contributions of $g$-product contraction terms.
This is important to express the results with the physical
significance.

\section{Improved forms of perturbed solution of dynamics}\label{sec3}

In fact, the final aim using the $g$-product decomposition and then
calculating the limitation of the contraction terms is to absorb the
partial contributions from the high order approximations of
off-diagonal elements of the perturbing Hamiltonian matrix into the
lower order approximations in our improved scheme of perturbation
theory. In this section, making use of the skills and methods stated
in previous section, we can obtain the zeroth, first, second and
third order improved forms of perturbed solutions with the above
features.

In mathematics, the process to obtain the improved forms of
perturbed solutions is a kind of technology to deal with an infinite
series, that is, according to some principles and the general term
form to rearrange those terms with the same features together
forming a group, then sum all of the terms in such a particular
group that they become a compact function at a given precision,
finally this infinite series is transformed into a new series form
that directly relates to the studied problem. More concretely
speaking, since we concern the system evolution with time $t$, we
take those terms with $(-\I y_i t) \e^{-\I x_i t}$, $(-\I y_i t)^2
\e^{-\I x_i t}/2!$ and $(-\I y_i t)^3 \e^{-\I x_i t}/3!$, $\cdots$
with the same factor function $f$ together forming a group, then sum
them to obtain an exponential function
$f\exp\left[-\I\left(x_i+y_i\right)t\right]$. The physical reason to
do this is that such an exponential function represents the system
evolution in theory and it has the obvious physical significance in
the calculation of transition probability and perturbed energy.
Through rearranging and summing, those terms with factors
$t^a\e^{-\I E_{\gamma_i}t}$,  $(a=1,2,\cdots)$ in the higher order
approximations are absorbed into the improved lower approximations,
we thus can advance the precision, particularly, when the evolution
time $t$ is long enough. We can call it ``dynamical rearrangement
and summation" method.

\subsection{Improved form of the zeroth order perturbed solution of dynamics}

Let us start with the zeroth order perturbed solution of dynamics.
In the usual perturbation theory, it is well-known \beq
\ket{\Psi^{(0)}(t)}=\sum_{\gamma}\e^{-\I E_\gamma
t}\diracsp{\Phi^\gamma}{\Psi(0)}\ket{\Phi^\gamma}=\sum_{\gamma\gamma^\prime}\e^{-\I
E_\gamma
t}\delta_{\gamma\gamma^\prime}a_{\gamma^\prime}\ket{\Phi^\gamma},\eeq
where $a_{\gamma^\prime}=\diracsp{\Phi^{\gamma^\prime}}{\Psi(0)}$.
Now, we would like to improve it so that it can absorb the partial
contributions from higher order approximations. Actually, we can
find that $A_2(c)$ and $A_3(nn,c)$ have the terms proportional to
$(-\I t)$ \beqa\label{0th1} & & (-\I t)\e^{-\I E_{\gamma}
t}\left[\sum_{\gamma_1}\frac{1}{E_{\gamma}-E_{\gamma_1}}
\left|g_1^{\gamma\gamma_1}\right|^2\right]
\delta_{\gamma\gamma^\prime},\\
\label{0th2} & &(-\I t)\e^{-\I E_{\gamma}
t}\left[\sum_{\gamma_1,\gamma_2}\frac{1}{(E_{\gamma}-E_{\gamma_1})(E_{\gamma}-E_{\gamma_2})}
g_1^{\gamma\gamma_1}g_1^{\gamma_1\gamma_2}g_1^{\gamma_2\gamma^\prime}\right]
\delta_{\gamma\gamma^\prime}.\eeqa Introduce the notation \beqa
G^{(2)}_{\gamma}&=&\sum_{\gamma_1}\frac{1}{E_{\gamma}-E_{\gamma_1}}\left|g_1^{\gamma\gamma_1}\right|^2,\\
G^{(3)}_{\gamma}&=&\sum_{\gamma_1,\gamma_2}\frac{1}{(E_{\gamma}-E_{\gamma_1})(E_{\gamma}-E_{\gamma_2})}
g_1^{\gamma\gamma_1}g_1^{\gamma_1\gamma_2}g_1^{\gamma_2\gamma}.\eeqa
It is clear that $G_\gamma^{(a)}$ has the energy dimension, and we
will see that it can be called the $a$th revised energy. Let us add
the terms (\ref{0th1}), (\ref{0th2}) and the related terms in
$A_4(t\e^{-\I E_{\gamma} t},{\rm D}),A_4(t^2\e^{-\I
E_{\gamma}t},{\rm D}),A_5(t\e^{-\I E_{\gamma} t},{\rm D})$,
$A_5(t^2\e^{-\I E_{\gamma} t},{\rm D})$, $A_6(t^2\e^{-\I E_{\gamma}
t},{\rm D})$ and $A_6(t^3 \e^{-\I E_{\gamma} t})$ given in Appendix
\ref{a1} together, that is, \beqa A_{\rm
I0}^{\gamma\gamma^\prime}(t)&=&\e^{-\I E_{\gamma}t}\left[1+(-\I
t)\left(G^{(2)}_\gamma+G^{(3)}_\gamma+G^{(4)}_\gamma+G^{(5)}_\gamma\right)\right.\nonumber\\
& &\left.+ \frac{(-\I
t)^2}{2!}\left(G^{(2)}_\gamma+G^{(3)}_\gamma\right)^2+ \frac{(-\I
t)^2}{2!}2 G^{(2)}_\gamma G^{(4)}_\gamma+\cdots
\right]\delta_{\gamma\gamma^\prime} ,\eeqa Although we have not
finished more calculations, from the mathematical symmetry and
physical concept, we can think \beqa A_{\rm
I0}^{\gamma\gamma^\prime}&=&\e^{-\I E_{\gamma}t}\left[1+(-\I
t)\left(G^{(2)}_\gamma+G^{(3)}_\gamma+G^{(4)}_\gamma+G^{(5)}_\gamma\right)\right.\nonumber\\
& &\left.+ \frac{(-\I
t)^2}{2!}\left(G^{(2)}_\gamma+G^{(3)}_\gamma+G^{(4)}_\gamma+G^{(5)}_\gamma\right)^2+\cdots
\right]\delta_{\gamma\gamma^\prime} ,\eeqa New terms added to the
above equation ought to, we think, appear at $A_7(t)$, $A_8(t)$,
$A_9(t)$ and $A_{10}(t)$, or come from the point of view introducing
the higher approximations. So we have \beq A_{\rm
I0}^{\gamma\gamma^\prime}(t)=\e^{-\I\left(E_\gamma+G^{(2)}_\gamma
+{G}^{(3)}_\gamma+G^{(4)}_\gamma+G^{(5)}_\gamma\right)
t}\delta_{\gamma\gamma^\prime}\eeq and then obtain the improved form
of the zeroth order perturbed solution of dynamics \beq \label{ips0}
\ket{\Psi_{E_T}^{(0)}(t)}_{\rm I}=\sum_{\gamma\gamma^\prime}A_{\rm
I0}a_{\gamma^\prime}(t)\ket{\Phi^\gamma}.\eeq

It is clear that $G^{(2)}_\gamma$ is real. In fact, $G^{(3)}_\gamma$
is also real. In order to prove it, we exchange the dummy indexes
$\gamma_1$ and $\gamma_2$ and take the complex conjugate of
$G^{(3)}_\gamma$, that is \beqa
{G^{(3)}_{\gamma}}^*&=&\sum_{\gamma_1,\gamma_2}\frac{1}{(E_{\gamma}-E_{\gamma_1})(E_{\gamma}-E_{\gamma_2})}
\left(g_1^{\gamma\gamma_2}\right)^*\left(g_1^{\gamma_2\gamma_1}\right)^*
\left(g_1^{\gamma_1\gamma}\right)^*\nonumber\\
&=&
\sum_{\gamma_1,\gamma_2}\frac{1}{(E_{\gamma}-E_{\gamma_1})(E_{\gamma}-E_{\gamma_2})}
g_1^{\gamma\gamma_1}g_1^{\gamma_1\gamma_2}g_1^{\gamma_2\gamma}\nonumber\\
&=& G^{(3)}_\gamma, \eeqa where we have used the relations
$\left(g_1^{\beta_1\beta_2}\right)^*=g_1^{\beta_2\beta_1}$ for any
$\beta_1$ and $\beta_2$ since $H_1$ is Hermit. Similar analyses can
be applied to $G^{(4)}_\gamma$ and $G^{(5)}_\gamma$. These mean that
$\e^{-\I
\left(G^{(2)}_\gamma+G^{(3)}_\gamma+G^{(4)}_\gamma+G^{(5)}_\gamma\right)
t}$ is still an oscillatory factor.

\subsection{Improved form of the first order perturbed solution of
dynamics}

Furthermore, in order to absorb the partial contributions from the
approximation higher than zeroth order, we need to consider the
contributions from off-diagonal elements in the higher order
approximations.

Well-known usual first order perturbing part of dynamics is \beqa
\ket{\Psi^{(1)}(t)}&=&\sum_{\gamma,\gamma^\prime}\left[\frac{\e^{-\I
E_\gamma t}}{E_{\gamma}-E_{\gamma^\prime}}-\frac{\e^{-\I
E_{\gamma^\prime}
t}}{E_{\gamma}-E_{\gamma^\prime}}\right]H_1^{\gamma\gamma^\prime}\ket{\Phi^\gamma}
= \sum_{\gamma,\gamma^\prime}\left[\left(\frac{\e^{-\I E_\gamma
t}}{E_{\gamma}-E_{\gamma^\prime}}-\frac{\e^{-\I E_{\gamma^\prime}
t}}{E_{\gamma}-E_{\gamma^\prime}}\right)g_1^{\gamma\gamma^\prime}\right]\ket{\Phi^\gamma}.\eeqa
It must be emphasized that $H_1$ is taken as only with the
off-diagonal part for simplicity. That is, we have used the
Hamiltonian redivision skill if the perturbing Hamiltonian matrix
has the diagonal elements.

Thus, from $A_3(t\e^{-\I E_{\gamma} t},{\rm N})$ and $A_4(t\e^{-\I
E_{\gamma} t},{\rm N})$, $A_4(t^2\e^{-\I E_{\gamma}t},{\rm D})$,
$A_5(t^2\e^{-\I E_{\gamma} t},{\rm N})$, $A_6(t^2\e^{-\I E_{\gamma}
t},{\rm N})$,  which are defined and calculated in the Appendix
\ref{a1}, it follows that \beqa A_{\rm
I1}^{\gamma\gamma^\prime}(t)&=&\frac{\e^{-\I
E_{\gamma}t}}{\left(E_{\gamma}-E_{\gamma^\prime}\right)}\left[1+(-\I
t)\left(G^{(2)}_\gamma+G^{(3)}_\gamma+G^{(4)}_\gamma\right)+
\frac{(-\I t)^2}{2!}\left(G^{(2)}_\gamma\right)^2+ \frac{(-\I
t)^2}{2!}2 G^{(2)}_\gamma G^{(3)}_\gamma+\cdots
\right]g_1^{\gamma\gamma^\prime}\nonumber\\
& &-\frac{\e^{-\I
E_{\gamma^\prime}t}}{\left(E_{\gamma}-E_{\gamma^\prime}\right)}\left[1+(-\I
t)\left(G^{(2)}_{\gamma^\prime}+G^{(3)}_{\gamma^\prime}+G^{(4)}_{\gamma^\prime}\right)+
\frac{(-\I t)^2}{2!}\left(G^{(2)}_{\gamma^\prime}\right)^2+
\frac{(-\I t)^2}{2!}2 G^{(2)}_{\gamma^\prime}
G^{(3)}_{\gamma^\prime}+\cdots
\right]g_1^{\gamma\gamma^\prime}.\hskip 0.5cm\eeqa  Therefore, by
rewriting \beq A_{\rm
I1}^{\gamma\gamma^\prime}(t)=\left(\frac{\e^{-\I
\left(E_\gamma+G^{(2)}_\gamma+G^{(3)}_\gamma+G^{(4)}_\gamma\right)
t}}{E_{\gamma}-E_{\gamma^\prime}}-\frac{\e^{-\I
\left(E_{\gamma^\prime}+G^{(2)}_{\gamma^\prime}+G^{(3)}_{\gamma^\prime}+G^{(4)}_{\gamma^\prime}\right)
t}}{E_{\gamma}-E_{\gamma^\prime}}\right)g_1^{\gamma\gamma^\prime},\eeq
we obtain the improved form of the first order perturbed solution of
dynamics \beqa \label{ips1} \ket{\Psi^{(1)}(t)}_{\rm
I}&=&\sum_{\gamma,\gamma^\prime}A_{\rm I1}^{\gamma\gamma^\prime}(t)
a_{\gamma^\prime}\ket{\Phi^\gamma}.\hskip 0.5cm\eeqa

\subsection{Improved second order- and third order perturbed solution}

Likewise, it is not difficult to obtain \beqa A_{\rm
I2}^{\gamma,\gamma^\prime}(t)&=&\sum_{\gamma_1}\left\{-\left[\frac{\e^{-\I
\left(E_\gamma+G^{(2)}_\gamma+G^{(3)}_\gamma\right) t}-\e^{-\I
\left(E_{\gamma_1}+G^{(2)}_{\gamma_1}+G^{(3)}_{\gamma_1}\right)
t}}{\left(E_{\gamma}-E_{\gamma_1}\right)^2}\right]
g_1^{\gamma\gamma_1}g_1^{\gamma_1\gamma}\delta_{\gamma\gamma^\prime}\right.\nonumber\\
& &+\left[\frac{\e^{-\I
\left(E_{\gamma}+G^{(2)}_{\gamma}+G^{(3)}_{\gamma}\right)
t}}{\left(E_{\gamma}-E_{\gamma_1}\right)\left(E_{\gamma}-E_{\gamma^\prime}\right)}-\frac{\e^{-\I
\left(E_{\gamma_1}+G^{(2)}_{\gamma_1}+G^{(3)}_{\gamma_1}\right)
t}}{\left(E_{\gamma}-E_{\gamma_1}\right)\left(E_{\gamma_1}-E_{\gamma^\prime}\right)}\right.\nonumber\\
& &\left.\left. +\frac{\e^{-\I
\left(E_{\gamma^\prime}+G^{(2)}_{\gamma^\prime}+G^{(3)}_{\gamma^\prime}\right)
t}}{\left(E_{\gamma}-E_{\gamma^\prime}\right)\left(E_{\gamma_1}-E_{\gamma^\prime}\right)}\right]
g_1^{\gamma\gamma_1}g_1^{\gamma_1\gamma^\prime}\eta_{\gamma\gamma^\prime}\right\},\eeqa
\beqa A_{\rm
I3}^{\gamma,\gamma^\prime}(t)&=&\sum_{\gamma_1,\gamma_2}\left[-\frac{\e^{-\I
\left(E_\gamma+G^{(2)}_\gamma\right)
t}}{\left(E_{\gamma}-E_{\gamma_1}\right)\left(E_{\gamma}-E_{\gamma_2}\right)^2}-\frac{\e^{-\I
\left(E_\gamma+G^{(2)}_\gamma\right)
t}}{\left(E_{\gamma}-E_{\gamma_1}\right)^2\left(E_{\gamma}-E_{\gamma_2}\right)}\right.\nonumber\\
& &\left.+\frac{\e^{-\I \left(E_{\gamma_1}+G^{(2)}_{\gamma_1}\right)
t}}{\left(E_{\gamma}-E_{\gamma_1}\right)^2\left(E_{\gamma_1}-E_{\gamma_2}\right)}-\frac{\e^{-\I
\left(E_{\gamma_2}+G^{(2)}_{\gamma_2}\right)
t}}{\left(E_{\gamma}-E_{\gamma_2}\right)^2\left(E_{\gamma_1}-E_{\gamma_2}\right)}\right]
g_1^{\gamma\gamma_1}g_1^{\gamma_1\gamma_2}g_1^{\gamma_2\gamma}\delta_{\gamma\gamma^\prime}
\nonumber\\
& &-\sum_{\gamma_1}\left[\frac{\e^{-\I
\left(E_\gamma+G^{(2)}_\gamma\right)
t}}{\left(E_{\gamma}-E_{\gamma_1}\right)\left(E_{\gamma}-E_{\gamma^\prime}\right)^2}+\frac{\e^{-\I
\left(E_\gamma+G^{(2)}_\gamma\right)
t}}{\left(E_{\gamma}-E_{\gamma_1}\right)^2\left(E_{\gamma}-E_{\gamma^\prime}\right)}\right]
g_1^{\gamma\gamma_1}g_1^{\gamma_1\gamma}g_1^{\gamma\gamma^\prime}\nonumber\\
& &+\sum_{\gamma_1,\gamma_2}\left[\frac{\e^{-\I
\left(E_{\gamma}+G^{(2)}_{\gamma}\right)
t}\eta_{\gamma\gamma_2}}{\left(E_{\gamma}-E_{\gamma_1}\right)\left(E_{\gamma}-E_{\gamma_2}\right)
\left(E_{\gamma}-E_{\gamma^\prime}\right)}-\frac{\e^{-\I
\left(E_{\gamma_1}+G^{(2)}_{\gamma_1}\right)
t}\eta_{\gamma_1\gamma^\prime}}{\left(E_{\gamma}-E_{\gamma_1}\right)\left(E_{\gamma_1}-E_{\gamma_2}\right)
\left(E_{\gamma_1}-E_{\gamma^\prime}\right)}\right.\nonumber\\
& &\left. +\frac{\e^{-\I
\left(E_{\gamma_2}+G^{(2)}_{\gamma_2}\right)
t}\eta_{\gamma\gamma_2}}{\left(E_{\gamma}-E_{\gamma_2}\right)\left(E_{\gamma_1}-E_{\gamma_2}\right)
\left(E_{\gamma_2}-E_{\gamma^\prime}\right)}\right]
g_1^{\gamma\gamma_1}g_1^{\gamma_1\gamma_2}g_1^{\gamma_2\gamma^\prime}\eta_{\gamma\gamma^\prime}.\eeqa
Therefore, the improved forms of the second- and third order
perturbed solutions are, respectively, \beqa \label{ips2}
\ket{\Psi^{(2)}(t)}_{\rm I}&=&\sum_{\gamma,\gamma^\prime}A_{\rm
I2}^{\gamma,\gamma^\prime}(t)
a_{\gamma^\prime}\ket{\Phi^\gamma},\\
\label{ips3} \ket{\Psi^{(3)}(t)}_{\rm
I}&=&\sum_{\gamma,\gamma^\prime}A_{\rm I3}^{\gamma,\gamma^\prime}(t)
a_{\gamma^\prime}\ket{\Phi^\gamma}.\eeqa

\subsection{Summary}

Obviously, our improved form of perturbed solution of dynamics up to
the third order approximation is \beqa
\ket{\Psi(t)}&=&\sum_{i=0}^3\ket{\Psi^{(i)}(t)}_{\rm
I}+\mathcal{O}(H_1^4).\eeqa However, this solution absorbs the
contributions from the whole $A_l^{\gamma\gamma^\prime}(t\e)$,
$A_l^{\gamma\gamma^\prime}(t^2\e)$ parts up to the fifth order
approximation and the whole $A_l^{\gamma\gamma^\prime}(t^2\e)$,
$A_l^{\gamma\gamma^\prime}(t^3\e)$ parts in the sixth order
approximation. After considering the contractions and
anti-contractions, we get the result corresponds to the replacement
\beq \e^{-\I E_{\gamma_i} t}\rightarrow \e^{-\I
\widetilde{E}_{\gamma_i} t}, \eeq in the
$A_l^{\gamma\gamma^\prime}(\e)$ part, where \beq
\widetilde{E}_{\gamma_i}={E}_{\gamma_i}+h^{\gamma_i}+\sum_{a=2}G_{\gamma_i}^{(a)},
\eeq $i=0,1,2,\cdots$, and $\gamma_0=\gamma$. Here, we have absorbed
the possible contributions from the diagonal elements of the
perturbing Hamiltonian matrix. Although the upper bound of summation
index $a$ is different from the approximation order in the finished
calculations, we can conjecture that it may be taken to at least 5
based on the consideration from the physical concept and
mathematical symmetry. For $a\geq 5$, their forms should be similar.
From our point of view, such form is so delicate that its form
happens impossibly by accident. Perhaps, there is a fundamental
formula within it. Nevertheless, we have no idea of how to prove it
strictly and generally at present.

Actually, as soon as we carry out further calculations, we can
absorb the contributions from higher order approximations. Moreover,
these calculations are not difficult and are programmable because we
only need to calculate the limitation and summation. Therefore, the
advantages of our solution have been made clear in our improved
forms of perturbed solution of dynamics. In other words, they offer
clear evidences to show our improved scheme is better than the
existing method in the precision and efficiency. In the following
several sections, we will clearly demonstrate these problems.

\section{Improved transition probability and revised Fermi's golden rule}\label{sec4}

One of the interesting applications of our perturbed solution is the
calculation of transition probability in general quantum systems
independent of time. It ameliorates the well-known conclusion
because our solution absorbs the contributions from the high order
approximations of the perturbing Hamiltonian. Moreover, in terms of
our improved forms of perturbed solution, it is easy to obtain the
high order transition probability. In addition, for the case of
sudden perturbation, our scheme is also suitable.

Let us start with the following perturbing expansion of state
evolution with time $t$, \beq
\ket{\Psi(t)}=\sum_{\gamma}c_\gamma(t)\ket{\Phi^\gamma}=\sum_{n=0}^\infty
\sum_{\gamma}c_\gamma^{(n)}(t)\ket{\Phi^\gamma}.\eeq When we take
the initial state as $\ket{\Phi^\beta}$, from our improved first
order perturbed solution, we immediately obtain \beq c_{\gamma,{\rm
I}}^{(1)}=\frac{\e^{-\I \widetilde{E}_\gamma t}-\e^{-\I
\widetilde{E}_{\beta}t}}{E_{\gamma}-E_{\beta}}
g_1^{\gamma\beta},\eeq where \beq
\widetilde{E}_{\gamma_i}=E_{\gamma_i}+h_1^{\gamma_i}+G_{\gamma_i}^{(2)}
+G_{\gamma_i}^{(3)}+G_{\gamma_i}^{(4)}.\eeq Here, we use the
subscript ``I" for distinguishing it from the usual result. Omitting
a unimportant phase factor $\e^{-\I \widetilde{E}_{\gamma}t}$, we
can rewrite it as \beq c_{\gamma,{\rm I}}^{(1)}=\frac{
g_1^{\gamma\beta}}{E_{\gamma}-E_{\beta}}
\left(1-\e^{\I\widetilde{\omega}_{\gamma\beta}t}\right),\eeq where
$\widetilde{\omega}_{\gamma\beta}=\widetilde{E}_{\gamma}-\widetilde{E}_{\beta}.$
Obviously it is different from the well known conclusion \beq
c_{\gamma}^{(1)}=\frac{ g_1^{\gamma\beta}}{E_{\gamma}-E_{\beta}}
\left(1-\e^{\I {\omega}_{\gamma\beta}t}\right),\eeq where $
\omega_{\gamma\beta}=E_{\gamma}-E_{\beta}$. Therefore, our result
contains the partial contributions from the high order
approximations.

Considering the transition probability from $\ket{\Phi^\beta}$ to
$\ket{\Phi^\gamma}$ after time $T$, we have \beq P_{\rm
I}^{\gamma\beta}(t)=\frac{\left|g_1^{\gamma\beta}\right|^2}{\omega_{\gamma\beta}^2}
\left|1-\e^{\I\widetilde{\omega}_{\gamma\beta}T}\right|^2
=\left|g_1^{\gamma\beta}\right|^2\frac{\sin^2\left(\widetilde{\omega}_{\gamma\beta}T/2\right)}
{\left({\omega}_{\gamma\beta}/2\right)^2}.\eeq

In terms of the relation \beq
\sin^2x-\sin^2y=\frac{1}{2}\left[\cos(2y)-\cos(2x)\right],\eeq we
have the revision part of transition probability \beq
\vartriangle\!\!P_{\rm I}^{\gamma\beta}(t)
=2\left|g_1^{\gamma\beta}\right|^2\frac{\cos\left({\omega}_{\gamma\beta}T\right)
-\cos\left(\widetilde{\omega}_{\gamma\beta}T\right)}
{\left({\omega}_{\gamma\beta}\right)^2}.\eeq

If plotting \beq
\frac{\sin^2\left(\widetilde{\omega}_{\gamma\beta}T/2\right)}{\left({\omega}_{\gamma\beta}/2\right)^2}=
\left(\frac{\widetilde{\omega}_{\gamma\beta}}{{\omega}_{\gamma\beta}}\right)^2
\frac{\sin^2\left(\widetilde{\omega}_{\gamma\beta}T/2\right)}
{\left(\widetilde{\omega}_{\gamma\beta}/2\right)^2},\eeq we can see
that it has a well-defined peak centered at
$\widetilde{\omega}_{\gamma\beta}=0$. Just as what has been done in
the usual case, we can extend the integral range as
$-\infty\rightarrow\infty$. Thus, the revised Fermi's golden rule
\beq w=w_{\rm F}+\vartriangle\!\!w,\eeq where the usual Fermi's
golden rule is \cite{Fermi} \beq w_{\rm F}=2\pi\rho(E_\beta)
\left|g_1^{\gamma\beta}\right|^2,\eeq in which $w$ means the
transition velocity, $\rho(E_\gamma)$ is the density of final state
and we have used the integral formula \beq
\int_{-\infty}^\infty\frac{\sin^2x}{x^2}=\pi. \eeq while the
revision part is \beq \vartriangle\!\!w=2\int_{-\infty}^\infty \d
E_\gamma \rho\left(E_{\gamma}\right)\left|g_1^{\gamma\beta}\right|^2
\frac{\cos\left({\omega}_{\gamma\beta}T\right)
-\cos\left(\widetilde{\omega}_{\gamma\beta}T\right)}
{T\left({\omega}_{\gamma\beta}\right)^2}.\eeq It is clear that $
\widetilde{\omega}_{\gamma\beta}$ is a function of $E_\gamma$, and
then a function of $\omega_{\gamma\beta}$. For simplicity, we only
consider $\widetilde{\omega}_{\gamma\beta}$ to its second order
approximation, that is \beq
\widetilde{\omega}_{\gamma\beta}=\widetilde{\omega}\left(\omega_{\gamma\beta}\right)
={\omega}_{\gamma\beta}+\sum_{\gamma_1}\left[
\frac{\left|g_1^{\gamma\gamma_1}\right|^2}{{\omega}_{\gamma\beta}-{\omega}_{\gamma_1\beta}}
-\frac{\left|g_1^{\beta\gamma_1}\right|^2}{{\omega}_{\beta\gamma_1}}\right]+\mathcal{O}(H_1^3).\eeq
Again based on $\d E_\gamma=\d{\omega}_{\gamma\beta}$, we have \beq
\vartriangle\!\!w=2\int_{-\infty}^\infty \d \omega_{\gamma\beta}
\rho\left(\omega_{\gamma\beta}+E_{\beta}\right)\left|g_1^{\gamma\beta}\right|^2
\frac{\cos\left[{\omega}_{\gamma\beta}T\right]
-\cos\left[\widetilde{\omega}_{\gamma\beta}\left(\omega\right)T\right]}
{T\left({\omega}_{\gamma\beta}\right)^2}.\eeq It seems to not to be
easy to deduce the general form of this integral. In order to
simplify it, we can use the fact that
$\widetilde{\omega}_{\gamma\beta}-{\omega}_{\gamma\beta}$ is a
smaller quantity since \beq
\vartriangle\!\!\omega_{\gamma\beta}=\widetilde{\omega}_{\gamma\beta}-{\omega}_{\gamma\beta}
=\sum_{i=2}^4 \left(G_{\gamma}^{(i)}-G_{\beta}^{(i)}\right).\eeq For
example, we can approximatively take \beq
\cos\left({\omega}_{\gamma\beta}T\right)
-\cos\left(\widetilde{\omega}_{\gamma\beta}T\right)\approx
T\left(\widetilde{\omega}_{\gamma\beta}-{\omega}_{\gamma\beta}\right)
\sin\left(\widetilde{\omega}_{\gamma\beta}T-{\omega}_{\gamma\beta}T\right),\eeq
then calculate the integral. We will study it in our other
manuscript (in preparing).

Obviously, the revision comes from the contributions of high order
approximations. The physical effect resulted from our solution,
whether is important or unimportant, should be investigated in some
concrete quantum systems. Recently, we reconsider the transition
probability and perturbed energy for a Hydrogen atom in a constant
magnetic field \cite{Ourtp1}. We find the results obtained by using
our improved scheme are indeed more satisfying in the calculation
precision and efficiency. We will discuss more examples in our
future manuscripts (in preparing).

It is clear that the relevant results can be obtained from the usual
results via replacing $\omega_{\gamma\beta}$ in the exponential
power by using $\widetilde{\omega}_{\gamma\beta}$. Thus, one thing
is true --- with the time $t$ evolving,
$\e^{\pm\I\left(\widetilde{\omega}_{\gamma\beta}t/2\right)}$ term in
the improved transition probability can be very different from
$\e^{\pm\I\left({\omega}_{\gamma\beta}t/2\right)}$ in the
traditional one, which might lead to totally different results. To
save the space, we do not intend to discuss more here.

In fact, there is no any difficulty to obtain the second- and three
order transition probability in terms of our improved forms of
perturbed solution in the previous section. More higher order
transition probability can be given effectively and accurately by
our scheme.

\section{Improved forms of perturbed energy and perturbed state}\label{sec5}

Now we study how to calculate the improved forms of perturbed energy
and perturbed state. For simplicity, we only study them concerning
the improved second order approximation. Based on the experience
from the skill one in Sec. \ref{sec6}, we can, in fact, set a new
$\widetilde{E}$ and then use the technology in the usual
perturbative theory. That is, we denote \beq
\widetilde{E}_{\gamma_i}=E_{\gamma_i}+G_{\gamma_i}^{(2)}
+G_{\gamma_i}^{(3)}.\eeq
\beqa \label{ipsto2o}
\ket{\Psi(t)}&=&\sum_{\gamma,\gamma^\prime}\left\{\e^{-\I\widetilde{E}_\gamma
t}\delta_{\gamma\gamma^\prime}+\left[\frac{\e^{-\I
\widetilde{E}_\gamma t}-\e^{-\I
\widetilde{E}_{\gamma^\prime}t}}{E_{\gamma}-E_{\gamma^\prime}}\right]
g_1^{\gamma\gamma^\prime}-\sum_{\gamma_1}\frac{\e^{-\I
\widetilde{E}_\gamma t}-\e^{-\I
\widetilde{E}_{\gamma_1}t}}{\left(E_{\gamma}-E_{\gamma_1}\right)^2}
g_1^{\gamma\gamma_1}g_1^{\gamma_1\gamma}\delta_{\gamma\gamma^\prime}
\right.\nonumber\\
& & +\sum_{\gamma_1}\left[\frac{\e^{-\I
\widetilde{E}_{\gamma}t}}{\left(E_{\gamma}-E_{\gamma_1}\right)
\left(E_{\gamma}-E_{\gamma^\prime}\right)}-\frac{\e^{-\I
\widetilde{E}_{\gamma_1}t}}{\left(E_{\gamma}-E_{\gamma_1}\right)
\left(E_{\gamma_1}-E_{\gamma^\prime}\right)}\right.\nonumber\\
& &\left.\left. +\frac{\e^{-\I
\widetilde{E}_{\gamma^\prime}t}}{\left(E_{\gamma}-E_{\gamma^\prime}\right)
\left(E_{\gamma_1}-E_{\gamma^\prime}\right)}\right]
g_1^{\gamma\gamma_1}g_1^{\gamma_1\gamma}\eta_{\gamma\gamma^\prime}\right\}
a_{\gamma^\prime}\ket{\Phi^\gamma}+\mathcal{O}(H_1^3).\eeqa Note
that $E_{\gamma_i}$ can contain the diagonal element
$h_1^{\gamma_i}$ of the original $H_1$, and we do not obviously
write $h_1^{\gamma_i}$ and take new $H_1$ matrix as off-diagonal in
the $H_0$ representation.

Because that \beq \ket{\Psi(t)}=\sum_{\gamma,\gamma^\prime}\e^{-\I
{E}_T
t}\delta_{\gamma\gamma^\prime}a_{\gamma^\prime}\ket{\Phi^\gamma},\eeq
we have \beqa\label{ipee} E_T
a_{\gamma}&=&\widetilde{E}_{\gamma}a_\gamma+\sum_{\gamma^\prime}\left\{\frac{
\widetilde{E}_\gamma-\widetilde{E}_{\gamma^\prime}}{E_{\gamma}-E_{\gamma^\prime}}
g_1^{\gamma\gamma^\prime}-\sum_{\gamma_1}\frac{\widetilde{E}_\gamma-\widetilde{E}_{\gamma_1}}
{\left(E_{\gamma}-E_{\gamma_1}\right)^2}
g_1^{\gamma\gamma_1}g_1^{\gamma_1\gamma}\delta_{\gamma\gamma^\prime}
+\sum_{\gamma_1}\left[\frac{
\widetilde{E}_{\gamma}}{\left(E_{\gamma}-E_{\gamma_1}\right)
\left(E_{\gamma}-E_{\gamma^\prime}\right)}\right.\right.\nonumber\\
& &\left.\left.-\frac{
\widetilde{E}_{\gamma_1}}{\left(E_{\gamma}-E_{\gamma_1}\right)
\left(E_{\gamma_1}-E_{\gamma^\prime}\right)}
+\frac{\widetilde{E}_{\gamma^\prime}}{\left(E_{\gamma}-E_{\gamma^\prime}\right)
\left(E_{\gamma_1}-E_{\gamma^\prime}\right)}\right]
g_1^{\gamma\gamma_1}g_1^{\gamma_1\gamma}\eta_{\gamma\gamma^\prime}\right\}
a_{\gamma^\prime}.\eeqa

In the usual perturbation theory, $H_1$ is taken as a perturbing
part with the form \beq H_1=\lambda v,\eeq where $\lambda$ is a real
number that is called the perturbing parameter. It must be
emphasized that $\widetilde{E}_{\gamma_i}$ can be taken as
explicitly independent perturbing parameter $\lambda$, because we
introduce $\lambda$ as a formal multiplier after redefinition. In
other words,  $\widetilde{E}_{\gamma_i}$ has absorbed those terms
adding to it and formed a new quantity. This way has been seen in
our Hamiltonian redivision skill. Without loss of generality, we
further take $H_1$ only with the off-diagonal form, that is \beq
H_1^{\gamma_1\gamma_2}={g}_1^{\gamma_1\gamma_2}=\lambda
{v}^{\gamma_1\gamma_2}.\eeq Then, we expand both the desired
expansion coefficients $a_\gamma$ and the energy eigenvalues $E_T$
in a power series of perturbation parameter $\lambda$: \beqa
E_T&=&\sum_{l=0}^\infty \lambda^l E_{T,{\rm I}}^{(l)},\quad
a_\gamma=\sum_{l=0}^\infty \lambda^l a_{\gamma;{\rm I}}^{(l)}.\eeqa

\subsubsection{Improved 0th approximation}

If we set $\lambda=0$, eq.(\ref{ipee}) yields \beq E_{T,{\rm
I}}^{(0)} a_{\gamma;{\rm I}}^{(0)}=\widetilde{E}_\gamma
a_{\gamma;{\rm I}}^{(0)},\eeq where $\gamma$ runs over all levels.
Actually, let us focus on the level $\gamma=\beta$, then \beq
\label{0the} E_{T,{\rm I}}^{(0)}=\widetilde{E}_\beta .\eeq When the
initial state is taken as $\ket{\Phi^\beta}$, \beq \label{0ths}
a_{\gamma;{\rm I}}^{(0)}=\delta_{\gamma\beta}.\eeq Obviously, the
improved form of perturbed energy is different from the results in
the usual perturbative theory because it absorbs the contributions
from the higher order approximations. However, the so-call improved
form of perturbed state is the same as the usual result.

\subsubsection{Improved 1st approximation}

Again from eq.(\ref{ipee}) it follows that \beq E_{T,{\rm I}}^{(0)}
a_{\gamma;{\rm I}}^{(1)} +E_{T,{\rm I}}^{(1)}a_{\gamma;{\rm
I}}^{(0)} =\widetilde{E}_\gamma a_{\gamma;{\rm I}}^{(1)}
+\sum_{\gamma^\prime}\frac{\widetilde{E}_\gamma-\widetilde{E}_{\gamma^\prime}}{E_\gamma-E_{\gamma^\prime}}
v^{\gamma\gamma^\prime}a_{\gamma^\prime;{\rm I}}^{(0)}.\eeq When
$\gamma=\beta$, it is easy to obtain \beq E_{T,{\rm I}}^{(1)}=0.
\eeq If $\gamma\neq\beta$, then \beq a_{\gamma;{\rm
I}}^{(1)}=-\frac{1}
{\left(E_\gamma-E_{\beta}\right)}{v}^{\gamma\beta}.\eeq It is clear
that the first order results are the same as the one in the usual
perturbative theory.

\subsubsection{Improved 2nd approximation}

Likewise, the following equation \beqa & & E_{T,{\rm I}}^{(2)}
a_{\gamma;{\rm I}}^{(0)}+ E_{T,{\rm
I}}^{(1)}a_{\gamma;I}^{(1)}+E_{T,{\rm I}}^{(0)}a_{\gamma;{\rm
I}}^{(2)} =\widetilde{E}_\gamma a_{\gamma;{\rm I}}^{(2)}
+\sum_{\gamma^\prime}\frac{\widetilde{E}_\gamma-\widetilde{E}_{\gamma^\prime}}{E_\gamma-E_{\gamma^\prime}}
{v}^{\gamma\gamma^\prime}a_{\gamma^\prime;{\rm I}}^{(1)}
-\sum_{\gamma_1,\gamma^\prime}\frac{\widetilde{E}_\gamma-\widetilde{E}_{\gamma_1}}
{\left(E_{\gamma}-E_{\gamma_1}\right)^2}
v^{\gamma\gamma_1}v^{\gamma_1\gamma}\delta_{\gamma\gamma^\prime}a_{\gamma^\prime;{\rm
I}}^{(0)} \nonumber\\
& &\quad +\sum_{\gamma_1,\gamma^\prime}\left[\frac{
\widetilde{E}_{\gamma}}{\left(E_{\gamma}-E_{\gamma_1}\right)
\left(E_{\gamma}-E_{\gamma^\prime}\right)} -\frac{
\widetilde{E}_{\gamma_1}}{\left(E_{\gamma}-E_{\gamma_1}\right)
\left(E_{\gamma_1}-E_{\gamma^\prime}\right)}
+\frac{\widetilde{E}_{\gamma^\prime}}{\left(E_{\gamma}-E_{\gamma^\prime}\right)
\left(E_{\gamma_1}-E_{\gamma^\prime}\right)}\right]
v^{\gamma\gamma_1}v^{\gamma_1\gamma}\eta_{\gamma\gamma^\prime}
a_{\gamma^\prime;{\rm I}}^{(0)}.\hskip 0.5cm \eeqa is obtained and
it yields \beq E_{T;{\rm I}}^{(2)}=0,\eeq if we take $\gamma=\beta$.
When $\gamma\neq\beta$, we have \beqa a_{\gamma;{\rm
I}}^{(2)}&=&\sum_{\gamma_1}\frac{1}
{\left(E_{\gamma}-E_{\beta}\right)
\left(E_{\gamma_1}-E_{\beta}\right)}
v^{\gamma\gamma_1}v^{\gamma_1\beta}\eta_{\gamma\beta}.\eeqa It is
consistent with the off-diagonal part of usual result. In fact,
since we have taken $H_1^{\gamma\gamma^\prime}$ to be off-diagonal,
it does not have a diagonal part. However, we think its form is more
appropriate. In addition, we do not consider the revision part
introduced by normalization. While $E_{T;{\rm I}}^{(2)}=0$ is a new
result.

\subsubsection{Summary}

Now we can see, up to the improved second order approximation: \beq
E_{T,\beta}\approx \widetilde{E}_\beta=E_{\beta}+G_{\beta}^{(2)}
+G_{\beta}^{(3)}.\eeq Compared with the usual one, they are
consistent at the former two orders. It is not strange since the
physical law is the same. However, our improved form of perturbed
energy contains a third order term. In other words, our solution
might be effective in order to obtain the contribution from high
order approximations. The possible physical reason is that a
redefined form of the solution is obtained.

In special, when we allow the $H_1^{\gamma\gamma^\prime}$ to have
the diagonal elements, the improved second order approximation
becomes \beq E_{T,\beta}\approx
E_{\beta}+h_1^{\beta}+G_{\beta}^{(2)} +G_{\beta}^{(3)}.\eeq

Likewise, if we redefine \beq
\widetilde{E}_{\gamma_i}=E_{\gamma_i}+h_1^{\gamma_i}+G_{\gamma_i}^{(2)}
+G_{\gamma_i}^{(3)}+G_{\gamma_i}^{(4)}.\eeq Thus, only considering
the first order approximation, we can obtain \beq E_{T,\beta}\approx
E_{\beta}+h_1^{\beta}+G_{\beta}^{(2)}
+G_{\beta}^{(3)}+G_{\beta}^{(4)}. \eeq In fact, the reason is our
conjecture in the previous section. The correct form of redefined
$\widetilde{E}_{\gamma_i}$ should be \beq E_{T,\beta}\approx
E_{\beta}+h_1^{\beta}+G_{\beta}^{(2)}
+G_{\beta}^{(3)}+G_{\beta}^{(4)} +G_{\beta}^{(5)}+\cdots. \eeq This
implies that our improved scheme absorbs the partial even whole
significant contributions from the high order approximations. In
addition, based on the fact that the improved second approximation
is actually zero, it is possible that this implies our solution will
fade down more rapidly than the solution in the usual perturbative
theory.

Actually, the main advantage of our solution is in dynamical
development. The contributions from the high order approximations
play more important roles in the relevant physical problems such as
the entanglement dynamics and decoherence process. For the improved
perturbed energy, its high order part has obvious physical meaning.
But, for the improved form of perturbed state, we find them to be
the same as the existed perturbation theory up to the second
approximation.

\section{Example and application}\label{sec6}

In order to concretely illustrate that our exact solution and the
improved scheme of perturbation theory are indeed more effective and
more accurate, let us study a simple example: two-state system,
which appears in the most of quantum mechanics textbooks. Its
Hamiltonian can be written as\beq H=\left(\begin{array}{cc}
E_1&V_{12}\\
V_{21}& E_2\end{array}\right),\eeq where we have used the the basis
formed by the unperturbed energy eigenvectors, that is \beq
\ket{\Phi^1}=\left(\begin{array}{c}1\\0\end{array}\right),\quad
\ket{\Phi^2}=\left(\begin{array}{c}0\\1\end{array}\right). \eeq In
other words: \beq H_0\ket{\Phi^\gamma}=E_\gamma\ket{\Phi^\gamma},
\quad (\gamma=1,2)\eeq \vskip -0.5cm \noindent where \vskip
-0.5cm\beq H_0=\left(\begin{array}{cc}
E_1&0\\
0& E_2\end{array}\right).\eeq Thus, this means the perturbing
Hamiltonian is taken as \beq H_1=\left(\begin{array}{cc}
0& V_{12}\\
V_{21}& 0\end{array}\right).\eeq This two state system has the
following eigen equation \beq
H\ket{\Psi^\gamma}=E^T_{\gamma}\ket{\Psi^\gamma}. \eeq It is easy to
obtain its solution: corresponding eigenvectors and eigenvalues
\beqa
\ket{\Psi^1}&=&\frac{1}{\sqrt{4|V|^2+(\omega_{21}+{\omega}_{21}^T)^2}}
\left(\begin{array}{c}\omega_{21}+{\omega}_{21}^T\\-2V_{21}\end{array}\right),\\
\ket{\Psi^2}&=&\frac{1}{\sqrt{4|V|^2+(\omega_{21}-{\omega}_{21}^T)^2}}
\left(\begin{array}{c}\omega_{21}-{\omega}_{21}^T\\-2V_{21}\end{array}\right);
\eeqa
\beqa E_1^T=\frac{1}{2}\left(E_1+E_2-{\omega}_{21}^T\right),\quad
E_2^T=\frac{1}{2}\left(E_1+E_2+{\omega}_{21}^T\right); \eeqa where
$|V|=|V_{12}|=|V_{21}|$, $\omega_{21}=E_2-E_1$,
${\omega}_{21}^T=E_2^T-E_1^T=\sqrt{4|V|^2+\omega_{21}^2}$, and we
have set $E_2> E_1$ without loss of generality.

Obviously the transition probability from state 1 to state 2 is
\beqa P^T(1\rightarrow 2)&=&\left|\bra{\Phi^2}\e^{-\I H
t}\ket{\Phi^1}\right|^2=\left|\sum_{\gamma_1,\gamma_2=1}^2\diracsp{\Phi^2}{\Psi^{\gamma_1}}
\bra{\Psi^{\gamma_1}}\e^{-\I H
t}\ket{\Psi^{\gamma_2}}\diracsp{\Psi^{\gamma_2}}{\Phi^1}\right|^2
=|V|^2\frac{\sin^2\left({\omega}_{21}^T
t/2\right)}{(\omega_{21}^T/2)^2}. \eeqa

In the usual perturbation theory, up to the second order
approximation, the well-known perturbed energies are \beqa
E_1^P=E_1-\frac{|V|^2}{\omega_{21}},\quad
E_2^P=E_1+\frac{|V|^2}{\omega_{21}}. \eeqa While, under the first
order approximation, the transition probability from state 1 to
state 2 is \beq P(1\rightarrow
2)=|V|^2\frac{\sin^2\left(\omega_{21}t/2\right)}{(\omega_{21}/2)^2}.
\eeq

Using our improved scheme, only to the first order approximation, we
get the corresponding perturbed energies \beqa
\widetilde{E}_1=E_1-\frac{|V|^2}{\omega_{21}}+\frac{|V|^4}{\omega_{21}^3},\quad
\widetilde{E}_2=E_1+\frac{|V|^2}{\omega_{21}}-\frac{|V|^4}{\omega_{21}^3},
\eeqa where we have used the facts that \beqa
G_1^{(2)}&=&-\frac{|V|^2}{\omega_{21}}=-G_2^{(2)},\quad
G_1^{(3)}=G_2^{(3)}=0,\quad
G_1^{(4)}=\frac{|V|^4}{\omega_{21}^3}=-G_2^{(4)}. \eeqa Obviously,
under the first order approximation, our scheme yields the
transition probability from state 1 to state 2 as \beq P_{\rm
I}(1\rightarrow
2)=|V|^2\frac{\sin^2\left(\widetilde{\omega}_{21}t/2\right)}{(\omega_{21}/2)^2}.
\eeq where
$\widetilde{\omega}_{21}=\widetilde{E}_2-\widetilde{E}_1$. Therefore
we can say our scheme is more effective. Moreover, we notice that
\beqa
{E}_1^T&=&E_1-\frac{|V|^2}{\omega_{21}}+\frac{|V|^4}{\omega_{21}^3}
+\mathcal{O}(|V|^6) =\widetilde{E}_1+\mathcal{O}(|V|^6)
={E}_1^P+\frac{|V|^4}{\omega_{21}^3}+\mathcal{O}(|V|^6),\\
\widetilde{E}_2^T&=&E_1+\frac{|V|^2}{\omega_{21}}
-\frac{|V|^4}{\omega_{21}^3}+\mathcal{O}(|V|^6)
=\widetilde{E}_2+\mathcal{O}(|V|^6) =
{E}_2^P-\frac{|V|^4}{\omega_{21}^3}+\mathcal{O}(|V|^6). \eeqa \vskip
-0.5cm and \beqa P^T(1\rightarrow
2)&=&|V|^2\frac{\sin^2\left(\omega_{21}t/2\right)}{(\omega_{21}/2)^2}
+|V|^2\left[\frac{\sin\left(
\omega_{21}t\right)}{2(\omega_{21}/2)^2}-\frac{\sin^2\left(
\omega_{21}t/2\right)}{(\omega_{21}/2)^3}\right]
\left(\widetilde{\omega}_{21}-{\omega}_{21}\right)
+\mathcal{O}[\left(\widetilde{\omega}_{21}-{\omega}_{21}\right)^2]\\
&=& P_{\rm I}(1\rightarrow 2)-|V|^2\frac{\sin^2\left(
\omega_{21}t/2\right)}{(\omega_{21}/2)^3}
\left(\widetilde{\omega}_{21}-{\omega}_{21}\right)
+\mathcal{O}[\left(\widetilde{\omega}_{21}-{\omega}_{21}\right)^2]\\
&=& P(1\rightarrow 2)+|V|^2\left[\frac{\sin\left(
\omega_{21}t\right)}{2(\omega_{21}/2)^2}-\frac{\sin^2\left(
\omega_{21}t/2\right)}{(\omega_{21}/2)^3}\right]
\left(\widetilde{\omega}_{21}-{\omega}_{21}\right)
+\mathcal{O}[\left(\widetilde{\omega}_{21}-{\omega}_{21}\right)^2].\eeqa
Therefore, we can say that our scheme is more accurate.

\section{Discussion and conclusion}\label{sec7}

In this paper, our improved scheme of perturbation theory is
proposed based on our exact solution in general quantum systems
\cite{My1}. Because our exact solution has a general term that is a
$c$-number function and proportional to power of the perturbing
Hamiltonian, this provides the probability considering the partial
contributions from the high order even all order approximations.
While our dynamical rearrangement and summation method helps us to
realize this probability. Just as the contributions from the high
order even all order approximations are absorbed to the lower order
approximations, our scheme becomes an improved one.

It must be emphasized that our improved scheme of perturbation
theory is proposed largely dependent on the facts that we find and
develop a series of skills and methods. From that the Hamiltonian
redivision skill overcomes the flaw in the usual perturbation
theory, improves the calculation precision, extends the applicable
range and removes the possible degeneracies to that the perturbing
Hamiltonian matrix product decomposition method separates the
contraction terms and anti-contraction terms, eliminates the
apparent divergences in the power series of the perturbing
Hamiltonian and provides the groundwork of ``dynamical rearrangement
and summation", we have seen these ideas, skills and methods to be
very useful.

Actually, the start point that delays to introduce the perturbing
parameter as possible plays an important even key role in our
improved scheme of perturbation theory. It enlightens us to seek for
the above skills and methods.

From our exact solution transferring to our improved scheme of
perturbation theory we does not directly use the cut-off
approximation, but first deals with the power series of perturbation
so that the contributions from some high order even all order
approximations can be absorbed into the lower orders than the
cut-off order as possible. Hence, our improved scheme of
perturbation theory is physically reasonable, mathematically clear
and methodologically skillful. This provides the guarantee achieving
high efficiency and high precision. Through finding the improved
forms of perturbed solutions of dynamics, we generally demonstrate
this conclusion. Furthermore, we prove the correctness of this
conclusion via calculating the improved form of transition
probability, perturbed energy and perturbed state. Specially, we
obtain the revised Fermi's golden rule. Moreover, we illustrate the
advantages of our improved scheme in an easy understanding example
of two-state system. All of this implies the physical reasons and
evidences why our improved scheme of perturbation theory is actually
calculable, operationally efficient, conclusively more accurate.

From the features of our improved scheme, we believe that it will
have interesting applications in the calculation of entanglement
dynamics and decoherence process as well as the other physical
quantities dependent on the expanding coefficients.

In fact, a given lower order approximation of improved form of the
perturbed solution absorbing the partial contributions from the
higher order even all order approximations is obtained by our
dynamical rearrangement and summation method, just like ``Fynmann
figures" summation" that has been done in the quantum field theory.
It is emphasized that these contributions have to be significant in
physics. Considering time evolution form is our physical ideas and
absorbing the high order approximations with the factors $t^a\e^{-\I
E_{\gamma_i}t}$, $(a=1,2,\cdots)$ to the improved lower order
approximations definitely can advance the precision. Therefore,
using our dynamical rearrangement and summation method is
appropriate and reasonable in our view.

For a concrete example, except for some technological and
calculational works, it needs the extensive physical background
knowledge to account for the significance of related results. That
is, since the differences of the related conclusions between our
improved scheme and the usual perturbation theory are in high order
approximation parts, we have to study the revisions (differences) to
find out whether they are important or unimportant to the studied
problems. In addition, our conjecture about the perturbed energy is
based on physical symmetry and mathematical consideration, it is
still open at the strict sense. As to the degenerate cases including
specially, vanishing the off-diagonal element of the perturbing
Hamiltonian matrix between any two degenerate levels, we have
discussed how to deal with them, except for the very complicated
cases that the degeneracy can not be completely removed by the
diagonalization of the degenerate subspaces trick and the
Hamiltonian redivision skill as well as the off-diagonal element of
the perturbing Hamiltonian matrix between any two degenerate levels
are not vanishing when the remained degeneracies are allowed.

It must be emphasized that the study on the time evolution operator
plays a central role in quantum dynamics and perturbation theory.
Because of the universal significance of our general and explicit
expression of the time evolution operator, we wish that it will have
more applications in quantum theory. Besides the above studies
through the perturbative method, it is more interesting to apply our
exact solution to the formalization study of quantum dynamics in
order to further and more powerfully show the advantages of our
exact solution.

In summary, our results can be thought of as theoretical
developments of perturbation theory, and they are helpful for
understanding the theory of quantum mechanics and providing some
powerful tools for the calculation of physical quantities in general
quantum systems. Together with our exact solution \cite{My1} and
open system dynamics \cite{My3}, they can finally form the
foundation of theoretical formulism of quantum mechanics in general
quantum systems. Further study on quantum mechanics of general
quantum systems is on progressing.

\section*{Acknowledgments}

We are grateful all the collaborators of our quantum theory group in
the Institute for Theoretical Physics of our university. This work
was funded by the National Fundamental Research Program of China
under No. 2001CB309310, and partially supported by the National
Natural Science Foundation of China under Grant No. 60573008.

\begin{appendix}

\renewcommand{\theequation}{\thesection\arabic{equation}}

\section{The calculations of the high order terms}\label{a1}

Since we have taken the $H_1$ only with the off-diagonal part, it is
enough to calculate the contributions from them. In Sec. \ref{sec6}
the contributions from the first, second and third order
approximations have been given. In this appendix, we would like to
find the contributions from the fourth to the sixth order
approximations. The calculational technologies used by us are mainly
to the limit process, dummy index changing and summation, as well as
the replacement
$g_1^{\gamma_i\gamma_j}\eta_{\gamma_i\gamma_j}=g_1^{\gamma_i\gamma_j}$
since $g_1^{\gamma_i\gamma_j}$ has been off-diagonal. These
calculations are not difficult, but are a little lengthy.

\subsection{$l=4$ case}

For the fourth order approximation, its contributions from the first
decompositions consists of eight terms: \beqa \label{A4fc}
A_4^{\gamma\gamma^\prime}&=&A_4^{\gamma\gamma^\prime}(ccc)
+A_4^{\gamma\gamma^\prime}(ccn)+A_4^{\gamma\gamma^\prime}(cnc)
+A_4^{\gamma\gamma^\prime}(ncc)\nonumber\\
& & +A_4^{\gamma\gamma^\prime}(cnn)+A_4^{\gamma\gamma^\prime}(ncn)
+A_4^{\gamma\gamma^\prime}(nnc)+{A}_4^{\gamma\gamma^\prime}(nnn).
\eeqa Its the former four terms have no the nontrivial second
contractions, and its the fifth and seven terms have one nontrivial
second contraction as follows \beqa {A}_4^{\gamma\gamma^\prime}(cnn)
&=&{A}_4^{\gamma\gamma^\prime}(cnn,kc) +
{A}_4^{\gamma\gamma^\prime}(cnn;kn),\\
{A}_4^{\gamma\gamma^\prime}(ncn)
&=&{A}_4^{\gamma\gamma^\prime}(ncn,c)
+{A}_4^{\gamma\gamma^\prime}(ncn,n), \\
A_4^{\gamma\gamma^\prime}(nnc)
&=&{A}_4^{\gamma\gamma^\prime}(nnc,ck) +
{A}_4^{\gamma\gamma^\prime}(nnc,nk). \eeqa In addition, the last
term in eq.(\ref{A4fc}) has two nontrivial second contractions, and
its fourth term has also the third contraction. Hence \beqa
{A}_4^{\gamma\gamma^\prime}(nnn)&=&{A}_4^{\gamma\gamma^\prime}(nnn,cc)
+{A}_4^{\gamma\gamma^\prime}(nnn,cn){A}_4^{\gamma\gamma^\prime}(nnn,nc)
+{A}_4^{\gamma\gamma^\prime}(nnn,nn),\\
{A}_4^{\gamma\gamma^\prime}(nnn,nn)&=&{A}_4^{\gamma\gamma^\prime}(nnn,nn,c)
+{A}_4^{\gamma\gamma^\prime}(nnn,nn,n).\eeqa All together, we have
the fifteen terms that are the contributions from whole contractions
and anti-contractions of the fourth order approximation.

First, let us calculate the former four terms only with the first
contractions and anti-contractions, that is, with more than two
$\delta$ functions (or less than two $\eta$ functions)
\beqa\label{A4ccc} {A}_4^{\gamma\gamma^\prime}(ccc)&=
&\sum_{\gamma_1,\cdots,\gamma_{5}}\left[
\sum_{i=1}^{5}(-1)^{i-1}\frac{\e^{-\I E_{\gamma_i}
t}}{d_i(E[\gamma,4])}\right]
\left[\prod_{j=1}^4g_1^{\gamma_j\gamma_{j+1}}\right]\left(
\prod_{k=1}^{3}\delta_{\gamma_k\gamma_{k+2}}\right)
\delta_{\gamma_1\gamma}\delta_{\gamma_{5}\gamma^\prime}\nonumber\\
&=&\sum_{\gamma_1}\left[\frac{3\e^{-\I
E_{\gamma}t}}{\left(E_\gamma-E_{\gamma_1}\right)^4}-\frac{3\e^{-\I
E_{\gamma_1}t}}{\left(E_\gamma-E_{\gamma_1}\right)^4}-(-\I
t)\frac{2\e^{-\I
E_{\gamma}t}}{\left(E_\gamma-E_{\gamma_1}\right)^3}\right.\nonumber\\
& &\left.-(-\I t)\frac{\e^{-\I
E_{\gamma_1}t}}{\left(E_\gamma-E_{\gamma_1}\right)^3}+\frac{(-\I
t)^2}{2}\frac{\e^{-\I
E_{\gamma}t}}{\left(E_\gamma-E_{\gamma_1}\right)^2}\right]
\left|g_1^{\gamma\gamma_1}\right|^4\delta_{\gamma\gamma^\prime}.\eeqa
\beqa\label{A4ccn} {A}_4^{\gamma\gamma^\prime}(ccn)&=
&\sum_{\gamma_1,\cdots,\gamma_{5}}\left[
\sum_{i=1}^{5}(-1)^{i-1}\frac{\e^{-\I E_{\gamma_i}
t}}{d_i(E[\gamma,4])}\right]
\left[\prod_{j=1}^4g_1^{\gamma_j\gamma_{j+1}}\right]
\delta_{\gamma_1\gamma_3}{\delta}_{\gamma_2\gamma_4}\eta_{\gamma_3\gamma_5}
\delta_{\gamma_1\gamma}\delta_{\gamma_{5}\gamma^\prime}\nonumber\\
&=&\sum_{\gamma_1}\left[-\frac{\e^{-\I
E_{\gamma}t}}{\left(E_\gamma-E_{\gamma_1}\right)^2
\left(E_{\gamma}-E_{\gamma^\prime}\right)^2}-\frac{2\e^{-\I
E_{\gamma}t}}{\left(E_\gamma-E_{\gamma_1}\right)^3
\left(E_{\gamma}-E_{\gamma^\prime}\right)}-\frac{\e^{-\I
E_{\gamma_1}t}}{\left(E_\gamma-E_{\gamma_1}\right)^2
\left(E_{\gamma_1}-E_{\gamma^\prime}\right)^2}\right.\nonumber\\
& & +\frac{2\e^{-\I
E_{\gamma_1}t}}{\left(E_\gamma-E_{\gamma_1}\right)^3\left(E_{\gamma_1}-E_{\gamma^\prime}\right)}
+\frac{\e^{-\I
E_{\gamma^\prime}t}}{\left(E_\gamma-E_{\gamma^\prime}\right)^2\left(E_{\gamma_1}-E_{\gamma^\prime}\right)^2}+(-\I
t)\frac{\e^{-\I E_{\gamma}t}}{\left(E_\gamma-E_{\gamma_1}\right)^2
\left(E_{\gamma}-E_{\gamma^\prime}\right)}
\nonumber\\
& & \left. +(-\I t)\frac{\e^{-\I
E_{\gamma_1}t}}{\left(E_\gamma-E_{\gamma_1}\right)^2
\left(E_{\gamma_1}-E_{\gamma^\prime}\right)}
\right]\left|g_1^{\gamma\gamma_1}\right|^2
g_1^{\gamma\gamma_1}g_1^{\gamma_1\gamma^\prime}\eta_{\gamma\gamma^\prime}.
\hskip 1.2cm\eeqa
\beqa\label{A4cnc} {A}_4^{\gamma\gamma^\prime}(cnc)&=
&\sum_{\gamma_1,\cdots,\gamma_{5}}\left[
\sum_{i=1}^{5}(-1)^{i-1}\frac{\e^{-\I E_{\gamma_i}
t}}{d_i(E[\gamma,4])}\right]
\left[\prod_{j=1}^4g_1^{\gamma_j\gamma_{j+1}}\right]
\delta_{\gamma_1\gamma_3}\eta_{\gamma_2\gamma_4}\delta_{\gamma_3\gamma_5}
\delta_{\gamma_1\gamma}\delta_{\gamma_{5}\gamma^\prime}\nonumber\\
&=&\sum_{\gamma_1,\gamma_2}\left[\frac{\e^{-\I
E_{\gamma}t}}{\left(E_\gamma-E_{\gamma_1}\right)
\left(E_{\gamma}-E_{\gamma_2}\right)^3} +\frac{\e^{-\I
E_{\gamma}t}}{\left(E_\gamma-E_{\gamma_1}\right)^2
\left(E_{\gamma}-E_{\gamma_2}\right)^2}+\frac{\e^{-\I
E_{\gamma}t}}{\left(E_\gamma-E_{\gamma_1}\right)^3\left(E_{\gamma}-E_{\gamma_2}\right)}
\right.\nonumber\\
& &-\frac{\e^{-\I
E_{\gamma_1}t}}{\left(E_\gamma-E_{\gamma_1}\right)^3\left(E_{\gamma_1}-E_{\gamma_2}\right)}
+\frac{\e^{-\I
E_{\gamma_2}t}}{\left(E_\gamma-E_{\gamma_2}\right)^3\left(E_{\gamma_1}-E_{\gamma_2}\right)}-(-\I
t)\frac{\e^{-\I
E_{\gamma}t}}{\left(E_\gamma-E_{\gamma_1}\right)\left(E_{\gamma}-E_{\gamma_2}\right)^2}\nonumber\\
& &\left. -(-\I t)\frac{\e^{-\I
E_{\gamma}t}}{\left(E_\gamma-E_{\gamma_1}\right)^2\left(E_{\gamma}-E_{\gamma_2}\right)}
+\frac{(-\I t)^2}{2}\frac{\e^{-\I
E_{\gamma}t}}{\left(E_\gamma-E_{\gamma_1}\right)
\left(E_{\gamma}-E_{\gamma_2}\right)}\right]\nonumber\\
& &\times\left|g_1^{\gamma\gamma_1}\right|^2
\left|g_1^{\gamma\gamma_2}\right|^2
\eta_{\gamma_1\gamma_2}\delta_{\gamma\gamma^\prime}.\hskip 1.2cm
\eeqa
\beqa\label{A4ncc} {A}_4^{\gamma\gamma^\prime}(ncc)&=
&\sum_{\gamma_1,\cdots,\gamma_{5}}\left[
\sum_{i=1}^{5}(-1)^{i-1}\frac{\e^{-\I E_{\gamma_i}
t}}{d_i(E[\gamma,4])}\right]
\left[\prod_{j=1}^4g_1^{\gamma_j\gamma_{j+1}}\right]
\eta_{\gamma_1\gamma_3}{\delta}_{\gamma_2\gamma_4}{\delta}_{\gamma_3\gamma_5}
\delta_{\gamma_1\gamma}\delta_{\gamma_{5}\gamma^\prime}\nonumber\\
&=&\sum_{\gamma_1}\left[\frac{\e^{-\I
E_{\gamma}t}}{\left(E_\gamma-E_{\gamma_1}\right)^2
\left(E_{\gamma}-E_{\gamma^\prime}\right)^2}+\frac{2\e^{-\I
E_{\gamma_1}t}}{\left(E_\gamma-E_{\gamma_1}\right)
\left(E_{\gamma_1}-E_{\gamma^\prime}\right)^3}-\frac{\e^{-\I
E_{\gamma_1}t}}{\left(E_\gamma-E_{\gamma_1}\right)^2
\left(E_{\gamma_1}-E_{\gamma^\prime}\right)^2}\right.\nonumber\\
& &-\frac{2\e^{-\I
E_{\gamma^\prime}t}}{\left(E_\gamma-E_{\gamma^\prime}\right)
\left(E_{\gamma_1}-E_{\gamma^\prime}\right)^3}-\frac{\e^{-\I
E_{\gamma^\prime}t}}{\left(E_\gamma-E_{\gamma^\prime}\right)^2
\left(E_{\gamma_1}-E_{\gamma^\prime}\right)^2} -(-\I t)\frac{\e^{-\I
E_{\gamma_1}t}}{\left(E_\gamma-E_{\gamma_1}\right)
\left(E_{\gamma_1}-E_{\gamma^\prime}\right)^2}\nonumber\\
& & \left. -(-\I t)\frac{\e^{-\I
E_{\gamma^\prime}t}}{\left(E_\gamma-E_{\gamma^\prime}\right)
\left(E_{\gamma_1}-E_{\gamma^\prime}\right)^2}\right]\left|g_1^{\gamma_1\gamma^\prime}\right|^2
g_1^{\gamma\gamma_1}g_1^{\gamma_1\gamma^\prime}\eta_{\gamma\gamma^\prime}.
\hskip 1.2cm\eeqa

Then, we calculate the three terms with the single first
contraction, that is, with one $\delta$ function.  Because one
$\delta$ function can not eliminate the whole apparent singularity,
we also need to find out the nontrivial second contraction- and/or
anti-contraction terms. \beqa\label{A4cnn-kc}
{A}_4^{\gamma\gamma^\prime}(cnn,kc)&=
&\sum_{\gamma_1,\cdots,\gamma_{5}}\left[
\sum_{i=1}^{5}(-1)^{i-1}\frac{\e^{-\I E_{\gamma_i}
t}}{d_i(E[\gamma,4])}\right]
\left[\prod_{j=1}^4g_1^{\gamma_j\gamma_{j+1}}\right]
{\delta}_{\gamma_1\gamma_3}\eta_{\gamma_2\gamma_4}
\eta_{\gamma_3\gamma_5}{\delta}_{\gamma_2\gamma^\prime}
\delta_{\gamma_1\gamma}\delta_{\gamma_{5}\gamma^\prime}\nonumber\\
&=&\sum_{\gamma_1}\left[-\frac{2\e^{-\I
E_{\gamma}t}}{\left(E_\gamma-E_{\gamma_1}\right)
\left(E_{\gamma}-E_{\gamma^\prime}\right)^3}-\frac{\e^{-\I
E_{\gamma}t}}{\left(E_\gamma-E_{\gamma_1}\right)^2
\left(E_{\gamma}-E_{\gamma^\prime}\right)^2}+\frac{\e^{-\I
E_{\gamma_1}t}}{\left(E_\gamma-E_{\gamma_1}\right)^2
\left(E_{\gamma_1}-E_{\gamma^\prime}\right)^2}\right.\nonumber\\
& &-\frac{\e^{-\I
E_{\gamma^\prime}t}}{\left(E_\gamma-E_{\gamma^\prime}\right)^2
\left(E_{\gamma_1}-E_{\gamma^\prime}\right)^2}-\frac{2\e^{-\I
E_{\gamma^\prime}t}}{\left(E_\gamma-E_{\gamma^\prime}\right)^3
\left(E_{\gamma_1}-E_{\gamma^\prime}\right)}+(-\I t)\frac{\e^{-\I
E_{\gamma}t}}{\left(E_\gamma-E_{\gamma_1}\right)
\left(E_{\gamma}-E_{\gamma^\prime}\right)^2}\nonumber\\
& &\left. -(-\I t)\frac{\e^{-\I
E_{\gamma^\prime}t}}{\left(E_\gamma-E_{\gamma^\prime}\right)^2
\left(E_{\gamma_1}-E_{\gamma^\prime}\right)}\right]
\left|g_1^{\gamma\gamma^\prime}\right|^2
g_1^{\gamma\gamma_1}g_1^{\gamma_1\gamma^\prime}.\hskip 1.2cm \eeqa
\beqa\label{A4cnn-kn} {A}_4^{\gamma\gamma^\prime}(cnn,kn)&=
&\sum_{\gamma_1,\cdots,\gamma_{5}}\left[
\sum_{i=1}^{5}(-1)^{i-1}\frac{\e^{-\I E_{\gamma_i}
t}}{d_i(E[\gamma,4])}\right]
\left[\prod_{j=1}^4g_1^{\gamma_j\gamma_{j+1}}\right]
{\delta}_{\gamma_1\gamma_3}\eta_{\gamma_2\gamma_4}
\eta_{\gamma_3\gamma_5}\eta_{\gamma_2\gamma^\prime}
\delta_{\gamma_1\gamma}\delta_{\gamma_{5}\gamma^\prime}\nonumber\\
&=&\sum_{\gamma_1,\gamma_2}\left[-\frac{\e^{-\I
E_{\gamma}t}}{\left(E_\gamma-E_{\gamma_1}\right)
\left(E_\gamma-E_{\gamma_2}\right)
\left(E_{\gamma}-E_{\gamma^\prime}\right)^2}-\frac{\e^{-\I
E_{\gamma}t}}{\left(E_\gamma-E_{\gamma_1}\right)
\left(E_\gamma-E_{\gamma_2}\right)^2
\left(E_{\gamma}-E_{\gamma^\prime}\right)}\right.\nonumber\\
& & -\frac{\e^{-\I
E_{\gamma}t}}{\left(E_\gamma-E_{\gamma_1}\right)^2
\left(E_\gamma-E_{\gamma_2}\right)
\left(E_{\gamma}-E_{\gamma^\prime}\right)}+\frac{\e^{-\I
E_{\gamma_1}t}}{\left(E_\gamma-E_{\gamma_1}\right)^2
\left(E_{\gamma_1}-E_{\gamma_2}\right)
\left(E_{\gamma_1}-E_{\gamma^\prime}\right)}\nonumber\\
& & -\frac{\e^{-\I
E_{\gamma_2}t}}{\left(E_\gamma-E_{\gamma_2}\right)^2
\left(E_{\gamma_1}-E_{\gamma_2}\right)
\left(E_{\gamma_2}-E_{\gamma^\prime}\right)}+\frac{\e^{-\I
E_{\gamma^\prime}t}}{\left(E_\gamma-E_{\gamma^\prime}\right)^2
\left(E_{\gamma_1}-E_{\gamma^\prime}\right)
\left(E_{\gamma_2}-E_{\gamma^\prime}\right)}\nonumber\\
& &\left. +(-\I t)\frac{\e^{-\I
E_{\gamma}t}}{\left(E_\gamma-E_{\gamma_1}\right)
\left(E_{\gamma}-E_{\gamma_2}\right)\left(E_{\gamma}-E_{\gamma^\prime}\right)}\right]
\left|g_1^{\gamma\gamma_1}\right|^2
g_1^{\gamma\gamma_2}g_1^{\gamma_2\gamma^\prime}
\eta_{\gamma_1\gamma_2}\eta_{\gamma_1\gamma^\prime}
\eta_{\gamma\gamma^\prime}. \hskip 1.2cm\eeqa
\beqa\label{A4ncn-c} {A}_4^{\gamma\gamma^\prime}(ncn,c)&=
&\sum_{\gamma_1,\cdots,\gamma_{5}}\left[
\sum_{i=1}^{5}(-1)^{i-1}\frac{\e^{-\I E_{\gamma_i}
t}}{d_i(E[\gamma,4])}\right]
\left[\prod_{j=1}^4g_1^{\gamma_j\gamma_{j+1}}\right]
\eta_{\gamma_1\gamma_3}{\delta}_{\gamma_2\gamma_4}\eta_{\gamma_3\gamma_5}
\delta_{\gamma_1\gamma}\delta_{\gamma_{5}\gamma^\prime}{\delta}_{\gamma\gamma^\prime}\nonumber\\
&=&\sum_{\gamma_1,\gamma_2}\left[-\frac{\e^{-\I
E_{\gamma}t}}{\left(E_{\gamma}-E_{\gamma_1}\right)^2
\left(E_{\gamma}-E_{\gamma_2}\right)^2}-\frac{2\e^{-\I
E_{\gamma}t}}{\left(E_{\gamma}-E_{\gamma_1}\right)^3
\left(E_{\gamma}-E_{\gamma_2}\right)}-\frac{\e^{-\I
E_{\gamma_1}t}}{\left(E_{\gamma}-E_{\gamma_1}\right)^2
\left(E_{\gamma_1}-E_{\gamma_2}\right)^2}\right.\nonumber\\
& &+\frac{2\e^{-\I
E_{\gamma_1}t}}{\left(E_{\gamma}-E_{\gamma_1}\right)^3
\left(E_{\gamma_1}-E_{\gamma_2}\right)} +\frac{\e^{-\I
E_{\gamma_2}t}}{\left(E_{\gamma}-E_{\gamma_2}\right)^2
\left(E_{\gamma_1}-E_{\gamma_2}\right)^2}+(-\I t)\frac{\e^{-\I
E_{\gamma}t}}{\left(E_{\gamma}-E_{\gamma_1}\right)^2
\left(E_{\gamma}-E_{\gamma_2}\right)}\nonumber\\
& &\left. +(-\I t)\frac{\e^{-\I
E_{\gamma_1}t}}{\left(E_{\gamma}-E_{\gamma_1}\right)^2
\left(E_{\gamma_1}-E_{\gamma_2}\right)}\right]\left|g_1^{\gamma\gamma_1}\right|^2
\left|g_1^{\gamma_1\gamma_2}\right|^2\eta_{\gamma\gamma_2}\delta_{\gamma\gamma^\prime}.
\hskip 1.2cm\eeqa
\beqa\label{A4ncn-n} {A}_4^{\gamma\gamma^\prime}(ncn,n)&=
&\sum_{\gamma_1,\cdots,\gamma_{5}}\left[
\sum_{i=1}^{5}(-1)^{i-1}\frac{\e^{-\I E_{\gamma_i}
t}}{d_i(E[\gamma,4])}\right]
\left[\prod_{j=1}^4g_1^{\gamma_j\gamma_{j+1}}\right]
\eta_{\gamma_1\gamma_3}{\delta}_{\gamma_2\gamma_4}\eta_{\gamma_3\gamma_5}
\delta_{\gamma_1\gamma}\delta_{\gamma_{5}\gamma^\prime}\eta_{\gamma\gamma^\prime}\nonumber\\
&=&\sum_{\gamma_1,\gamma_2}\left[\frac{\e^{-\I
E_{\gamma}t}}{\left(E_\gamma-E_{\gamma_1}\right)^2
\left(E_\gamma-E_{\gamma_2}\right)
\left(E_{\gamma}-E_{\gamma^\prime}\right)}+\frac{\e^{-\I
E_{\gamma_1}t}}{\left(E_\gamma-E_{\gamma_1}\right)
\left(E_{\gamma_1}-E_{\gamma_2}\right)
\left(E_{\gamma_1}-E_{\gamma^\prime}\right)^2}\right.\nonumber\\
& & +\frac{\e^{-\I
E_{\gamma_1}t}}{\left(E_\gamma-E_{\gamma_1}\right)
\left(E_{\gamma_1}-E_{\gamma_2}\right)^2
\left(E_{\gamma_1}-E_{\gamma^\prime}\right)}-\frac{\e^{-\I
E_{\gamma_1}t}}{\left(E_\gamma-E_{\gamma_1}\right)^2
\left(E_{\gamma_1}-E_{\gamma_2}\right)
\left(E_{\gamma_1}-E_{\gamma^\prime}\right)}\nonumber\\
& & -\frac{\e^{-\I
E_{\gamma_2}t}}{\left(E_\gamma-E_{\gamma_2}\right)
\left(E_{\gamma_1}-E_{\gamma_2}\right)^2
\left(E_{\gamma_2}-E_{\gamma^\prime}\right)}+\frac{\e^{-\I
E_{\gamma^\prime}t}}{\left(E_\gamma-E_{\gamma^\prime}\right)
\left(E_{\gamma_1}-E_{\gamma^\prime}\right)^2
\left(E_{\gamma_2}-E_{\gamma^\prime}\right)}\nonumber\\
& &\left.-(-\I t)\frac{\e^{-\I
E_{\gamma_1}t}}{\left(E_\gamma-E_{\gamma_1}\right)
\left(E_{\gamma_1}-E_{\gamma_2}\right)\left(E_{\gamma_1}-E_{\gamma^\prime}\right)}\right]
\left|g_1^{\gamma_1\gamma_2}\right|^2
g_1^{\gamma\gamma_1}g_1^{\gamma_1\gamma^\prime}
\eta_{\gamma\gamma_2}\eta_{\gamma_2\gamma^\prime}
\eta_{\gamma\gamma^\prime}. \hskip 1.2cm\eeqa
\beqa\label{A4nnc-ck} {A}_4^{\gamma\gamma^\prime}(nnc,ck)&=
&\sum_{\gamma_1,\cdots,\gamma_{5}}\left[
\sum_{i=1}^{5}(-1)^{i-1}\frac{\e^{-\I E_{\gamma_i}
t}}{d_i(E[\gamma,4])}\right]
\left[\prod_{j=1}^4g_1^{\gamma_j\gamma_{j+1}}\right]
\eta_{\gamma_1\gamma_3}
\eta_{\gamma_2\gamma_4}{\delta}_{\gamma_3\gamma_5}
{\delta}_{\gamma_1\gamma_4}
\delta_{\gamma_1\gamma}\delta_{\gamma_{5}\gamma^\prime}\nonumber\\
&=&\sum_{\gamma_1}\left[-\frac{2\e^{-\I
E_{\gamma}t}}{\left(E_\gamma-E_{\gamma_1}\right)
\left(E_{\gamma}-E_{\gamma^\prime}\right)^3}-\frac{\e^{-\I
E_{\gamma}t}}{\left(E_\gamma-E_{\gamma_1}\right)^2
\left(E_{\gamma}-E_{\gamma^\prime}\right)^2}+\frac{\e^{-\I
E_{\gamma_1}t}}{\left(E_\gamma-E_{\gamma_1}\right)^2
\left(E_{\gamma_1}-E_{\gamma^\prime}\right)^2}\right.\nonumber\\
& &-\frac{\e^{-\I
E_{\gamma^\prime}t}}{\left(E_\gamma-E_{\gamma^\prime}\right)^2
\left(E_{\gamma_1}-E_{\gamma^\prime}\right)^2}-\frac{2\e^{-\I
E_{\gamma^\prime}t}}{\left(E_\gamma-E_{\gamma^\prime}\right)^3
\left(E_{\gamma_1}-E_{\gamma^\prime}\right)}+(-\I t)\frac{\e^{-\I
E_{\gamma}t}}{\left(E_\gamma-E_{\gamma_1}\right)
\left(E_{\gamma}-E_{\gamma^\prime}\right)^2}\nonumber\\
& &\left. -(-\I t)\frac{\e^{-\I
E_{\gamma^\prime}t}}{\left(E_\gamma-E_{\gamma^\prime}\right)^2
\left(E_{\gamma_1}-E_{\gamma^\prime}\right)}\right]
\left|g_1^{\gamma\gamma^\prime}\right|^2
g_1^{\gamma\gamma_1}g_1^{\gamma_1\gamma^\prime}.\hskip 1.2cm \eeqa
\beqa\label{A4nnc-nk} {A}_4^{\gamma\gamma^\prime}(nnc,nk)&=
&\sum_{\gamma_1,\cdots,\gamma_{5}}\left[
\sum_{i=1}^{5}(-1)^{i-1}\frac{\e^{-\I E_{\gamma_i}
t}}{d_i(E[\gamma,4])}\right]
\left[\prod_{j=1}^4g_1^{\gamma_j\gamma_{j+1}}\right]
\eta_{\gamma_1\gamma_3}
\eta_{\gamma_2\gamma_4}{\delta}_{\gamma_3\gamma_5}
\eta_{\gamma_1\gamma_4}
\delta_{\gamma_1\gamma}\delta_{\gamma_{5}\gamma^\prime}\nonumber\\
&=&\sum_{\gamma_1,\gamma_2}\left[\frac{\e^{-\I
E_{\gamma}t}}{\left(E_\gamma-E_{\gamma_1}\right)
\left(E_\gamma-E_{\gamma_2}\right)
\left(E_{\gamma}-E_{\gamma^\prime}\right)^2}-\frac{\e^{-\I
E_{\gamma_1}t}}{\left(E_\gamma-E_{\gamma_1}\right)
\left(E_{\gamma_1}-E_{\gamma_2}\right)
\left(E_{\gamma_1}-E_{\gamma^\prime}\right)^2}\right.\nonumber\\
& & +\frac{\e^{-\I
E_{\gamma_2}t}}{\left(E_\gamma-E_{\gamma_2}\right)
\left(E_{\gamma_1}-E_{\gamma_2}\right)
\left(E_{\gamma_2}-E_{\gamma^\prime}\right)^2}-\frac{\e^{-\I
E_{\gamma^\prime}t}}{\left(E_\gamma-E_{\gamma^\prime}\right)
\left(E_{\gamma_1}-E_{\gamma^\prime}\right)
\left(E_{\gamma_2}-E_{\gamma^\prime}\right)^2}\nonumber\\
& & -\frac{\e^{-\I
E_{\gamma^\prime}t}}{\left(E_\gamma-E_{\gamma^\prime}\right)
\left(E_{\gamma_1}-E_{\gamma^\prime}\right)^2
\left(E_{\gamma_2}-E_{\gamma^\prime}\right)}-\frac{\e^{-\I
E_{\gamma^\prime}t}}{\left(E_\gamma-E_{\gamma^\prime}\right)^2
\left(E_{\gamma_1}-E_{\gamma^\prime}\right)
\left(E_{\gamma_2}-E_{\gamma^\prime}\right)}\nonumber\\
& &\left. -(-\I t)\frac{\e^{-\I
E_{\gamma^\prime}t}}{\left(E_\gamma-E_{\gamma^\prime}\right)
\left(E_{\gamma_1}-E_{\gamma^\prime}\right)\left(E_{\gamma_2}-E_{\gamma^\prime}\right)}\right]
\left|g_1^{\gamma_2\gamma^\prime}\right|^2
g_1^{\gamma\gamma_1}g_1^{\gamma_1\gamma^\prime}
\eta_{\gamma\gamma_2}\eta_{\gamma_1\gamma_2}
\eta_{\gamma\gamma^\prime}.\hskip 1.2cm \eeqa

Finally, we calculate the ${A}_4^{\gamma\gamma^\prime}(nnn)$ by
considering the two second decompositions, that is, its former three
terms \vskip -0.5cm \beqa\label{A4nnn-cc}
{A}_4^{\gamma\gamma^\prime}(nnn,cc)&=
&\sum_{\gamma_1,\cdots,\gamma_{5}}\left[
\sum_{i=1}^{5}(-1)^{i-1}\frac{\e^{-\I E_{\gamma_i}
t}}{d_i(E[\gamma,4])}\right]
\left[\prod_{j=1}^4g_1^{\gamma_j\gamma_{j+1}}\right]\left(
\prod_{k=1}^{3}\eta_{\gamma_k\gamma_{k+2}}\right)
\delta_{\gamma_1\gamma_4}\delta_{\gamma_2\gamma_5}
\delta_{\gamma_1\gamma}\delta_{\gamma_{5}\gamma^\prime}
\nonumber\\
&=&\sum_{\gamma_1}\left[-\frac{2\e^{-\I
E_{\gamma}t}}{\left(E_{\gamma}-E_{\gamma_1}\right)
\left(E_{\gamma}-E_{\gamma^\prime}\right)^3}-\frac{\e^{-\I
E_{\gamma}t}}{\left(E_{\gamma}-E_{\gamma_1}\right)^2
\left(E_{\gamma}-E_{\gamma^\prime}\right)^2}+\frac{\e^{-\I
E_{\gamma_1}t}}{\left(E_{\gamma}-E_{\gamma_1}\right)^2
\left(E_{\gamma_1}-E_{\gamma^\prime}\right)^2}\right.\nonumber\\
& &-\frac{\e^{-\I
E_{\gamma^\prime}t}}{\left(E_{\gamma}-E_{\gamma^\prime}\right)^2
\left(E_{\gamma_1}-E_{\gamma^\prime}\right)^2}-\frac{2\e^{-\I
E_{\gamma^\prime}t}}{\left(E_{\gamma}-E_{\gamma^\prime}\right)^3
\left(E_{\gamma_1}-E_{\gamma^\prime}\right)}+(-\I t)\frac{\e^{-\I
E_{\gamma}t}}{\left(E_{\gamma}-E_{\gamma_1}\right)
\left(E_{\gamma}-E_{\gamma^\prime}\right)^2}\nonumber\\
& &\left.-(-\I t)\frac{\e^{-\I
E_{\gamma^\prime}t}}{\left(E_{\gamma}-E_{\gamma^\prime}\right)^2
\left(E_{\gamma_1}-E_{\gamma^\prime}\right)}\right]g_1^{\gamma\gamma^\prime}
g_1^{\gamma^\prime\gamma_1}g_1^{\gamma_1\gamma}g_1^{\gamma\gamma^\prime}.
\hskip 1.2cm\eeqa
\vskip -0.5cm \beqa\label{A4nnn-cn}
{A}_4^{\gamma\gamma^\prime}(nnn,cn)&=
&\sum_{\gamma_1,\cdots,\gamma_{5}}\left[
\sum_{i=1}^{5}(-1)^{i-1}\frac{\e^{-\I E_{\gamma_i}
t}}{d_i(E[\gamma,4])}\right]
\left[\prod_{j=1}^4g_1^{\gamma_j\gamma_{j+1}}\right]\left(
\prod_{k=1}^{3}\eta_{\gamma_k\gamma_{k+2}}\right)
{\delta}_{\gamma_1\gamma_4}\eta_{\gamma_2\gamma_5}
\delta_{\gamma_1\gamma}\delta_{\gamma_{5}\gamma^\prime}
\nonumber\\
&=&\sum_{\gamma_1,\gamma_2}\left[-\frac{\e^{-\I
E_{\gamma}t}}{\left(E_{\gamma}-E_{\gamma_1}\right)\left(E_{\gamma}-E_{\gamma_2}\right)
\left(E_{\gamma}-E_{\gamma^\prime}\right)^2}-\frac{\e^{-\I
E_{\gamma}t}}{\left(E_{\gamma}-E_{\gamma_1}\right)\left(E_{\gamma}-E_{\gamma_2}\right)^2
\left(E_{\gamma}-E_{\gamma^\prime}\right)}\right.\nonumber\\
& &-\frac{\e^{-\I
E_{\gamma}t}}{\left(E_{\gamma}-E_{\gamma_1}\right)^2\left(E_{\gamma}-E_{\gamma_2}\right)
\left(E_{\gamma}-E_{\gamma^\prime}\right)}+\frac{\e^{-\I
E_{\gamma_1}t}}{\left(E_{\gamma}-E_{\gamma_1}\right)^2\left(E_{\gamma_1}-E_{\gamma_2}\right)
\left(E_{\gamma_1}-E_{\gamma^\prime}\right)}\nonumber\\
& &-\frac{\e^{-\I
E_{\gamma_2}t}}{\left(E_{\gamma}-E_{\gamma_2}\right)^2\left(E_{\gamma_1}-E_{\gamma_2}\right)
\left(E_{\gamma_2}-E_{\gamma^\prime}\right)}+\frac{\e^{-\I
E_{\gamma^\prime}t}}{\left(E_{\gamma}-E_{\gamma^\prime}\right)^2\left(E_{\gamma_1}-E_{\gamma^\prime}\right)
\left(E_{\gamma_2}-E_{\gamma^\prime}\right)} \nonumber\\
& & \left. +(-\I t)\frac{\e^{-\I
E_{\gamma}t}}{\left(E_{\gamma}-E_{\gamma_1}\right)\left(E_{\gamma}-E_{\gamma_2}\right)
\left(E_{\gamma}-E_{\gamma^\prime}\right)}\right]
g_1^{\gamma\gamma_1}g_1^{\gamma_1\gamma_2}g_1^{\gamma_2\gamma}
g_1^{\gamma\gamma^\prime}\eta_{\gamma_1\gamma^\prime}\eta_{\gamma_2\gamma^\prime}.
\hskip 0.8cm\eeqa
\vskip -0.5cm \beqa\label{A4nnn-nc}
{A}_4^{\gamma\gamma^\prime}(nnn,nc)&=
&\sum_{\gamma_1,\cdots,\gamma_{5}}\left[
\sum_{i=1}^{5}(-1)^{i-1}\frac{\e^{-\I E_{\gamma_i}
t}}{d_i(E[\gamma,4])}\right]
\left[\prod_{j=1}^4g_1^{\gamma_j\gamma_{j+1}}\right]\left(
\prod_{k=1}^{3}\eta_{\gamma_k\gamma_{k+2}}\right)
\eta_{\gamma_1\gamma_4}{\delta}_{\gamma_2\gamma_5}
\delta_{\gamma_1\gamma}\delta_{\gamma_{5}\gamma^\prime}
\nonumber\\
&=&\sum_{\gamma_1,\gamma_2}\left[\frac{\e^{-\I
E_{\gamma}t}}{\left(E_{\gamma}-E_{\gamma_1}\right)\left(E_{\gamma}-E_{\gamma_2}\right)
\left(E_{\gamma}-E_{\gamma^\prime}\right)^2}-\frac{\e^{-\I
E_{\gamma_1}t}}{\left(E_{\gamma}-E_{\gamma_1}\right)\left(E_{\gamma_1}-E_{\gamma_2}\right)
\left(E_{\gamma_1}-E_{\gamma^\prime}\right)^2}\right.\nonumber\\
& &+\frac{\e^{-\I
E_{\gamma_2}t}}{\left(E_{\gamma}-E_{\gamma_2}\right)\left(E_{\gamma_1}-E_{\gamma_2}\right)
\left(E_{\gamma_2}-E_{\gamma^\prime}\right)^2}-\frac{\e^{-\I
E_{\gamma^\prime}t}}{\left(E_{\gamma}-E_{\gamma^\prime}\right)
\left(E_{\gamma_1}-E_{\gamma^\prime}\right)
\left(E_{\gamma_2}-E_{\gamma^\prime}\right)^2}\nonumber\\
& & -\frac{\e^{-\I
E_{\gamma^\prime}t}}{\left(E_{\gamma}-E_{\gamma^\prime}\right)
\left(E_{\gamma_1}-E_{\gamma^\prime}\right)^2
\left(E_{\gamma_2}-E_{\gamma^\prime}\right)}-\frac{\e^{-\I
E_{\gamma^\prime}t}}{\left(E_{\gamma}-E_{\gamma^\prime}\right)^2
\left(E_{\gamma_1}-E_{\gamma^\prime}\right)
\left(E_{\gamma_2}-E_{\gamma^\prime}\right)}\nonumber\\
 & &\left. -(-\I t)\frac{\e^{-\I
E_{\gamma^\prime}t}}{\left(E_{\gamma}-E_{\gamma^\prime}\right)\left(E_{\gamma_1}-E_{\gamma^\prime}\right)
\left(E_{\gamma_2}-E_{\gamma^\prime}\right)}\right]
g_1^{\gamma\gamma^\prime}g_1^{\gamma^\prime\gamma_1}g_1^{\gamma_1\gamma_2}
g_1^{\gamma_2\gamma^\prime}\eta_{\gamma\gamma_1}\eta_{\gamma\gamma_2}.
\hskip 0.8cm\eeqa
while the fourth term has the third decomposition,
that is \beqa\label{A4nnn-nn-c}
{A}_4^{\gamma\gamma^\prime}(nnn,nn,c)&=
&\sum_{\gamma_1,\cdots,\gamma_{5}}\left[
\sum_{i=1}^{5}(-1)^{i-1}\frac{\e^{-\I E_{\gamma_i}
t}}{d_i(E[\gamma,4])}\right]
\left[\prod_{j=1}^4g_1^{\gamma_j\gamma_{j+1}}\right]\left(
\prod_{k=1}^{3}\eta_{\gamma_k\gamma_{k+2}}\right)
\delta_{\gamma_1\gamma}\delta_{\gamma_{5}\gamma^\prime}
\delta_{\gamma\gamma^\prime}\nonumber\\
&=&\sum_{\gamma_1,\gamma_2,\gamma_3}\left[-\frac{\e^{-\I
E_{\gamma}t}}{\left(E_{\gamma}-E_{\gamma_1}\right)\left(E_{\gamma}-E_{\gamma_2}\right)
\left(E_{\gamma}-E_{\gamma_3}\right)^2}-\frac{\e^{-\I
E_{\gamma}t}}{\left(E_{\gamma}-E_{\gamma_1}\right)\left(E_{\gamma}-E_{\gamma_2}\right)^2
\left(E_{\gamma}-E_{\gamma_3}\right)}\right.\nonumber\\
& &-\frac{\e^{-\I
E_{\gamma}t}}{\left(E_{\gamma}-E_{\gamma_1}\right)^2\left(E_{\gamma}-E_{\gamma_2}\right)
\left(E_{\gamma}-E_{\gamma_3}\right)}+\frac{\e^{-\I
E_{\gamma_1}t}}{\left(E_{\gamma}-E_{\gamma_1}\right)^2\left(E_{\gamma_1}-E_{\gamma_2}\right)
\left(E_{\gamma_1}-E_{\gamma_3}\right)}\nonumber\\
& &-\frac{\e^{-\I
E_{\gamma_2}t}}{\left(E_{\gamma}-E_{\gamma_2}\right)^2\left(E_{\gamma_1}-E_{\gamma_2}\right)
\left(E_{\gamma_2}-E_{\gamma_3}\right)}+\frac{\e^{-\I
E_{\gamma_3}t}}{\left(E_{\gamma}-E_{\gamma_3}\right)^2\left(E_{\gamma_1}-E_{\gamma_3}\right)
\left(E_{\gamma_2}-E_{\gamma_3}\right)}\nonumber\\
& &\left. +(-\I t)\frac{\e^{-\I
E_{\gamma}t}}{\left(E_{\gamma}-E_{\gamma_1}\right)\left(E_{\gamma}-E_{\gamma_2}\right)
\left(E_{\gamma}-E_{\gamma_3}\right)}\right]
g_1^{\gamma\gamma_1}g_1^{\gamma_1\gamma_2}g_1^{\gamma_2\gamma_3}
g_1^{\gamma_3\gamma}\eta_{\gamma\gamma_2} \eta_{\gamma_1\gamma_3}
{\delta}_{\gamma\gamma^\prime}.\hskip 1.2cm \eeqa
\beqa\label{A4nnn-nn-n} {A}_4^{\gamma\gamma^\prime}(nnn,nn,n)&=
&\sum_{\gamma_1,\cdots,\gamma_{5}}\left[
\sum_{i=1}^{5}(-1)^{i-1}\frac{\e^{-\I E_{\gamma_i}
t}}{d_i(E[\gamma,4])}\right]
\left[\prod_{j=1}^4g_1^{\gamma_j\gamma_{j+1}}\right]\left(
\prod_{k=1}^{3}\eta_{\gamma_k\gamma_{k+2}}\right)
\delta_{\gamma_1\gamma}\delta_{\gamma_{5}\gamma^\prime}
\eta_{\gamma\gamma^\prime}\nonumber\\
&=&\sum_{\gamma_1,\gamma_2,\gamma_3}\left[\frac{\e^{-\I
E_{\gamma}t}}{\left(E_{\gamma}-E_{\gamma_1}\right)\left(E_{\gamma}-E_{\gamma_2}\right)
\left(E_{\gamma}-E_{\gamma_3}\right)\left(E_{\gamma}-E_{\gamma^\prime}\right)}\right.\nonumber\\
& &-\frac{\e^{-\I
E_{\gamma_1}t}}{\left(E_{\gamma}-E_{\gamma_1}\right)\left(E_{\gamma_1}-E_{\gamma_2}\right)
\left(E_{\gamma_1}-E_{\gamma_3}\right)\left(E_{\gamma_1}-E_{\gamma^\prime}\right)}\nonumber\\
& &+\frac{\e^{-\I
E_{\gamma_2}t}}{\left(E_{\gamma}-E_{\gamma_2}\right)\left(E_{\gamma_1}-E_{\gamma_2}\right)
\left(E_{\gamma_2}-E_{\gamma_3}\right)\left(E_{\gamma_2}-E_{\gamma^\prime}\right)}\nonumber\\
& &-\frac{\e^{-\I
E_{\gamma_3}t}}{\left(E_{\gamma}-E_{\gamma_3}\right)\left(E_{\gamma_1}-E_{\gamma_3}\right)
\left(E_{\gamma_2}-E_{\gamma_3}\right)\left(E_{\gamma_3}-E_{\gamma^\prime}\right)}\nonumber\\
& &\left. +\frac{\e^{-\I
E_{\gamma^\prime}t}}{\left(E_{\gamma}-E_{\gamma^\prime}\right)\left(E_{\gamma_1}-E_{\gamma^\prime}\right)
\left(E_{\gamma_2}-E_{\gamma^\prime}\right)\left(E_{\gamma_3}-E_{\gamma^\prime}\right)}\right]\nonumber\\[7pt]
& & \times
g_1^{\gamma\gamma_1}g_1^{\gamma_1\gamma_2}g_1^{\gamma_2\gamma_3}
g_1^{\gamma_3\gamma^\prime}\eta_{\gamma\gamma_2}\eta_{\gamma\gamma_3}\eta_{\gamma\gamma^\prime}
\eta_{\gamma_1\gamma_3}\eta_{\gamma_1\gamma^\prime}
\eta_{\gamma_2\gamma^\prime}. \eeqa

Now, all 15 contractions and/or anti-contractions in the fourth
order approximation have been calculated out.

In order to absorb the contributions from the fourth order
approximation to the improved forms of lower order perturbed
solutions, we first decompose $A_4^{\gamma\gamma^\prime}$, which is
a summation of all above terms, into the three parts according to
their factor forms in $\e^{-\I E_{\gamma_i}t}, (-\I t) \e^{-\I
E_{\gamma_i}t} $ and $(-\I t)^2\e^{-\I E_{\gamma_i}t}/2$, that is
\beq\label{A4dto}
A_4^{\gamma\gamma^\prime}=A_4^{\gamma\gamma^\prime}(\e)+A_4^{\gamma\gamma^\prime}(t\e)
+A_4^{\gamma\gamma^\prime}(t^2\e).\eeq Secondly, we decompose its
every term into three parts according to the factor forms in
$\e^{-\I E_{\gamma}t}, \e^{-\I E_{\gamma_1}t}$
($\sum_{\gamma_1}\e^{-\I E_{\gamma_1}t}$) and $\e^{-\I
E_{\gamma^\prime}t}$, that is \beqa
A_4^{\gamma\gamma^\prime}(\e)&=&A_4^{\gamma\gamma^\prime}(\e^{-\I
E_{\gamma}t})+A_4^{\gamma\gamma^\prime}(\e^{-\I
E_{\gamma_1}t})+A_4^{\gamma\gamma^\prime}(\e^{-\I E_{\gamma^\prime}t}),\\
A_4^{\gamma\gamma^\prime}(t\e)&=&A_4^{\gamma\gamma^\prime}(t\e^{-\I
E_{\gamma}t})+A_4^{\gamma\gamma^\prime}(t\e^{-\I
E_{\gamma_1}t})+A_4^{\gamma\gamma^\prime}(t\e^{-\I E_{\gamma^\prime}t}),\\
A_4^{\gamma\gamma^\prime}(t^2\e)&=&A_4^{\gamma\gamma^\prime}(t^2\e^{-\I
E_{\gamma}t})+A_4^{\gamma\gamma^\prime}(t^2\e^{-\I
E_{\gamma_1}t})+A_4^{\gamma\gamma^\prime}(t^2\e^{-\I
E_{\gamma^\prime}t}). \eeqa Finally, we again decompose every term
in the above equations into the diagonal and off-diagonal parts
about $\gamma$ and $\gamma^\prime$, that is \beqa
A_4^{\gamma\gamma^\prime}(\e^{-\I
E_{\gamma_i}t})&=&A_4^{\gamma\gamma^\prime}(\e^{-\I
E_{\gamma_i}t};{\rm D})+A_4^{\gamma\gamma^\prime}(\e^{-\I
E_{\gamma_i}t};{\rm N}),\\
A_4^{\gamma\gamma^\prime}(t\e^{-\I
E_{\gamma_i}t})&=&A_4^{\gamma\gamma^\prime}(t\e^{-\I
E_{\gamma_i}t};{\rm D})+A_4^{\gamma\gamma^\prime}(t\e^{-\I
E_{\gamma_i}t};{\rm N}), \\
A_4^{\gamma\gamma^\prime}(t^2\e^{-\I
E_{\gamma_i}t})&=&A_4^{\gamma\gamma^\prime}(t^2\e^{-\I
E_{\gamma_i}t};{\rm D})+A_4^{\gamma\gamma^\prime}(t^2\e^{-\I
E_{\gamma_i}t};{\rm N}).  \eeqa where $E_{\gamma_i}$ takes $
E_{\gamma}, E_{\gamma_1}$ and $E_{\gamma^\prime}$.

If we do not concern the improved forms of perturbed solutions equal
to or higer than the fourth order one, we only need to write down
the second and third terms in eq.(\ref{A4dto}) and calculate their
diagonal and off-diagonal parts respectively. Based on the
calculated results above, it is easy to obtain \beqa
\label{A4gammaD} A_4^{\gamma\gamma^\prime}\left(t\e^{-\I E_\gamma
t};{\rm D}\right)&=& \left(-\I t\right)\e^{-\I E_\gamma
t}\left[\sum_{\gamma_1}\frac{-2
\left|g_1^{\gamma\gamma_1}\right|^4}{\left(E_{\gamma}-E_{\gamma_1}\right)^3}
-\sum_{\gamma_1,\gamma_2}\frac{\left|g_1^{\gamma\gamma_1}\right|^2
\left|g_1^{\gamma\gamma_2}\right|^2\eta_{\gamma_1\gamma_2}}
{\left(E_{\gamma}-E_{\gamma_1}\right)\left(E_{\gamma}-E_{\gamma_2}\right)^2}\right.
\nonumber\\
& &
-\sum_{\gamma_1,\gamma_2}\frac{\left|g_1^{\gamma\gamma_1}\right|^2
\left|g_1^{\gamma\gamma_2}\right|^2\eta_{\gamma_1\gamma_2}}
{\left(E_{\gamma}-E_{\gamma_1}\right)^2\left(E_{\gamma}-E_{\gamma_2}\right)}
+\sum_{\gamma_1,\gamma_2}\frac{\left|g_1^{\gamma\gamma_1}\right|^2
\left|g_1^{\gamma_1\gamma_2}\right|^2\eta_{\gamma\gamma_2}}
{\left(E_{\gamma}-E_{\gamma_1}\right)^2\left(E_{\gamma}-E_{\gamma_2}\right)}\nonumber\\
&
&\left.+\sum_{\gamma_1,\gamma_2,\gamma_3}\frac{g_1^{\gamma\gamma_1}
g_1^{\gamma_1\gamma_2}g_1^{\gamma_2\gamma_3}g_1^{\gamma_3\gamma}\eta_{\gamma\gamma_2}\eta_{\gamma_1\gamma_3}}
{\left(E_{\gamma}-E_{\gamma_1}\right)\left(E_{\gamma}-E_{\gamma_2}\right)
\left(E_{\gamma}-E_{\gamma_3}\right)}\right]\delta_{\gamma\gamma^\prime}.\eeqa
Substituting the relation
$\eta_{\beta_1\beta_2}=1-\delta_{\beta_1\beta_2}$, using the
technology of index exchanging and introducing the definitions of
so-called the $a$th revision energy $G_\gamma^{(a)}$: \beqa
{G}^{(2)}_\gamma&=&\sum_{\gamma_1}\frac{\left|g_1^{\gamma\gamma_1}\right|^2}
{E_{\gamma}-E_{\gamma_1}}\eeqa
\beqa G^{(4)}_\gamma &=
&\sum_{\gamma_1,\gamma_2,\gamma_3}\frac{g_1^{\gamma\gamma_1}
g_1^{\gamma_1\gamma_2}g_1^{\gamma_2\gamma_3}g_1^{\gamma_3\gamma}\eta_{\gamma\gamma_2}}
{\left(E_{\gamma}-E_{\gamma_1}\right)\left(E_{\gamma}-E_{\gamma_2}\right)
\left(E_{\gamma}-E_{\gamma_3}\right)}-\sum_{\gamma_1,\gamma_2}
\frac{g_1^{\gamma\gamma_1}g_1^{\gamma_1\gamma}g_1^{\gamma\gamma_2}g_1^{\gamma_2\gamma}}
{\left(E_{\gamma}-E_{\gamma_1}\right)^2\left(E_{\gamma}-E_{\gamma_2}\right)},
\eeqa we can simplify Eq. (\ref{A4gammaD}) to the following concise
form: \beqa A_4^{\gamma\gamma^\prime}\left(t\e^{-\I E_\gamma t};{\rm
D}\right)&=&-\left(-\I t\right)\e^{-\I E_\gamma
t}\left[\sum_{\gamma_1}\frac{
G^{(2)}_\gamma}{\left(E_{\gamma}-E_{\gamma_1}\right)^2}
\left|g_1^{\gamma\gamma_1}\right|^2-G^{(4)}_\gamma\right]\delta_{\gamma\gamma^\prime}.
\eeqa

Similar calculation and simplification lead to \beqa
A_4^{\gamma\gamma^\prime}\left(t\e^{-\I E_{\gamma} t}; {\rm
N}\right)=(-\I t)\e^{-\I E_{\gamma} t}\left[
\frac{{G}^{(3)}_{\gamma}g_1^{\gamma\gamma^\prime}}{\left(E_{\gamma}-E_{\gamma^\prime}\right)}
+\sum_{\gamma_1} \frac{{G}^{(2)}_{\gamma}}
{\left(E_{\gamma}-E_{\gamma_1}\right)\left(E_{\gamma}-E_{\gamma^\prime}\right)}
\right], \eeqa where \beqa G^{(3)}_\gamma &=
&\sum_{\gamma_1,\gamma_2}\frac{g_1^{\gamma\gamma_1}
g_1^{\gamma_1\gamma_2}g_1^{\gamma_2\gamma}}
{\left(E_{\gamma}-E_{\gamma_1}\right)\left(E_{\gamma}-E_{\gamma_2}\right)}.
\eeqa For saving the space, the corresponding detail is omitted. In
fact, it is not difficult, but it is necessary to be careful enough,
specially in the cases of higher order approximations.

In the same way, we can obtain: \beqa
A_4^{\gamma\gamma^\prime}\left(t\e^{-\I E_{\gamma_1} t}; {\rm D}
\right)&=&\left(-\I
t\right)\sum_{\gamma_1}\frac{G^{(2)}_{\gamma_1}\e^{-\I E_{\gamma_1}
t}}{\left(E_{\gamma}-E_{\gamma_1}\right)^2}
\left|g_1^{\gamma\gamma_1}\right|^2\delta_{\gamma\gamma^\prime},\\
A_4^{\gamma\gamma^\prime}\left(t\e^{-\I E_{\gamma_1} t} ; {\rm
N}\right)&=&-(-\I t)\sum_{\gamma_1}
\frac{{G}^{(2)}_{\gamma_1}\e^{-\I E_{\gamma_1} t}}
{\left(E_{\gamma}-E_{\gamma_1}\right)\left(E_{\gamma_1}-E_{\gamma^\prime}\right)}
g_1^{\gamma\gamma_1}g_1^{\gamma_1\gamma^\prime}\eta_{\gamma\gamma^\prime}.
\eeqa
\beqa A_4^{\gamma\gamma^\prime}\left(t\e^{-\I E_{\gamma^\prime}t};
{\rm D}\right)&=&0,\\
A_4^{\gamma\gamma^\prime}\left(t\e^{-\I E_{\gamma^\prime}t}; {\rm
N}\right)&=&-(-\I t)\e^{-\I E_{\gamma^\prime}
t}\left[\sum_{\gamma_1}
\frac{{G}^{(3)}_{\gamma^\prime}}{\left(E_{\gamma}-E_{\gamma^\prime}\right)}
g_1^{\gamma\gamma^\prime}\right.\nonumber\\
& &\left. +\sum_{\gamma_1} \frac{{G}^{(2)}_{\gamma^\prime}}
{\left(E_{\gamma}-E_{\gamma^\prime}\right)\left(E_{\gamma_1}-E_{\gamma^\prime}\right)}
g_1^{\gamma\gamma_1}g_1^{\gamma_1\gamma^\prime}\eta_{\gamma\gamma^\prime}\right].
\eeqa

For the terms with the factor $t^2\e$ , only one is nonzero, that is
\beqa A_4^{\gamma\gamma^\prime}\left(t^2\e\right)
&=&A_4^{\gamma\gamma^\prime}\left(t^2\e^{-\I E_{\gamma} t}; D\right)
=\frac{(-\I G^{(2)}_\gamma t)^2}{2!}\e^{-\I E_{\gamma} t}. \eeqa
since \beqa A_4^{\gamma\gamma^\prime}\left(t^2\e^{-\I E_{\gamma_1}
t};{\rm D}\right)&=&A_4^{\gamma\gamma^\prime}\left(t^2\e^{-\I
E_{\gamma^\prime} t};{\rm D}\right)=0,\\
A_4^{\gamma\gamma^\prime}\left(t^2\e^{-\I E_{\gamma} t};{\rm
N}\right)&=&A_4^{\gamma\gamma^\prime}\left(t^2\e^{-\I E_{\gamma_1}
t};{\rm D}\right)=A_4^{\gamma\gamma^\prime}\left(t^2\e^{-\I
E_{\gamma^\prime} t};{\rm D}\right)=0. \eeqa

We can see that these terms can be absorbed into (or merged with)
the lower order approximations to obtain the improved forms of
perturbed solutions.

\subsection{l=5 case}

Now let we consider the case of the fifth order approximation
($l=5$). From eq.(\ref{gpd}) it follows that the first
decompositions of $g$-product have $2^4=16$ terms. They can be
divided into 5 groups \beq A_5^{\gamma\gamma^\prime}= \sum_{i=0}^4
\mathcal{A}_5^{\gamma\gamma^\prime}(i;\eta),\eeq where $i$ indicates
the number of $\eta$ functions. Obviously \beq
\mathcal{A}_5^{\gamma\gamma^\prime}(0;\eta)={A}_5^{\gamma\gamma^\prime}(cccc),\eeq
\beqa
\mathcal{A}_5^{\gamma\gamma^\prime}(1;\eta)={A}_5^{\gamma\gamma^\prime}(cccn)
+{A}_5^{\gamma\gamma^\prime}(ccnc)+{A}_5^{\gamma\gamma^\prime}(cncc)
+{A}_5^{\gamma\gamma^\prime}(nccc),\eeqa \beqa
\mathcal{A}_5^{\gamma\gamma^\prime}(2;\eta)&=&{A}_5^{\gamma\gamma^\prime}(ccnn)
+{A}_5^{\gamma\gamma^\prime}(cncn)+{A}_5^{\gamma\gamma^\prime}(cnnc)\nonumber\\
& & +{A}_5^{\gamma\gamma^\prime}(nccn)
+{A}_5^{\gamma\gamma^\prime}(ncnc)
+{A}_5^{\gamma\gamma^\prime}(nncc),\eeqa \beqa
\mathcal{A}_5^{\gamma\gamma^\prime}(3;\eta)&=&{A}_5^{\gamma\gamma^\prime}(cnnn)
+{A}_5^{\gamma\gamma^\prime}(ncnn)
+{A}_5^{\gamma\gamma^\prime}(nncn)
+{A}_5^{\gamma\gamma^\prime}(nnnc), \eeqa \beq
\mathcal{A}_5^{\gamma\gamma^\prime}(4;\eta)={A}_5^{\gamma\gamma^\prime}(nnnn).\eeq
Here, we have used the notations stated in Sec. \ref{sec6}.

By calculation, we obtain the
$\mathcal{A}^{\gamma\gamma^\prime}_5(0,\eta)$ and every term of
$\mathcal{A}^{\gamma\gamma^\prime}_5(1,\eta)$ have only nontrivial
first contractions and/or anti-contractions. But, we can find that
every term of $\mathcal{A}^{\gamma\gamma^\prime}_5(2,\eta)$ can have
one nontrivial second or third or fourth contraction or
anti-contraction, that is \beqa
A^{\gamma\gamma^\prime}_5(ccnn)&=&A^{\gamma\gamma^\prime}_5(ccnn,kkc)+A^{\gamma\gamma^\prime}_5(ccnn,kkn),\\
A^{\gamma\gamma^\prime}_5(cncn)&=&A^{\gamma\gamma^\prime}_5(cncn,kc)+A^{\gamma\gamma^\prime}_5(cncn,kc),\\
A^{\gamma\gamma^\prime}_5(cnnc)&=&A^{\gamma\gamma^\prime}_5(cnnc,kck)+A^{\gamma\gamma^\prime}_5(cnnc,kck),\\
A^{\gamma\gamma^\prime}_5(nccn)&=&A^{\gamma\gamma^\prime}_5(nccn,c)+A^{\gamma\gamma^\prime}_5(nccn,n),\\
A^{\gamma\gamma^\prime}_5(ncnc)&=&A^{\gamma\gamma^\prime}_5(ncnc,ck)+A^{\gamma\gamma^\prime}_5(ncnc,nk),\\
A^{\gamma\gamma^\prime}_5(nncc)&=&A^{\gamma\gamma^\prime}_5(nncc,ckk)+A^{\gamma\gamma^\prime}_5(nncc,nkk).
\eeqa Similarly, every term of
$\mathcal{A}^{\gamma\gamma^\prime}_5(3,\eta)$ can have two higher
order contractions and/or anti-contractions: \beqa
A^{\gamma\gamma^\prime}_5(cnnn)&=&A^{\gamma\gamma^\prime}_5(cnnn,kcc)
+A^{\gamma\gamma^\prime}_5(cnnn,kcn)\nonumber\\
& &+A^{\gamma\gamma^\prime}_5(cnnn,knc)
+A^{\gamma\gamma^\prime}_5(cnnn,knn),\\
A^{\gamma\gamma^\prime}_5(ncnn)&=&A^{\gamma\gamma^\prime}_5(ncnn,kkc,ck)
+A^{\gamma\gamma^\prime}_5(ncnn,kkn,ck)\nonumber\\
& & +A^{\gamma\gamma^\prime}_5(ncnn,kkc,nk)+A^{\gamma\gamma^\prime}_5(ncnn,kkn,nk),\\
A^{\gamma\gamma^\prime}_5(nncn)&=&A^{\gamma\gamma^\prime}_5(nncn,ckk,kc)
+A^{\gamma\gamma^\prime}_5(nncn,ckk,kn)\nonumber\\
& &+A^{\gamma\gamma^\prime}_5(nncn,nkk,kc)+A^{\gamma\gamma^\prime}_5(nncn,nkk,kn),\\
A^{\gamma\gamma^\prime}_5(nnnc)&=&A^{\gamma\gamma^\prime}_5(nnnc,cck)
+A^{\gamma\gamma^\prime}_5(nnnc,cnk)\nonumber\\
& &+A^{\gamma\gamma^\prime}_5(nnnc,nck)
+A^{\gamma\gamma^\prime}_5(nnnc,nnk).\eeqa Moreover, their last
terms, with two higher order anti-contractions, can have one
nontrivial more higher contraction or anti-contraction: \beqa
A^{\gamma\gamma^\prime}_5(cnnn,knn)&=&A^{\gamma\gamma^\prime}_5(cnnn,knn,kc)
+A^{\gamma\gamma^\prime}_5(cnnn,knn.kn),\\
A^{\gamma\gamma^\prime}_5(ncnn,kkn,nk)&=&A^{\gamma\gamma^\prime}_5(ncnn,kkn,nk,c)
+A^{\gamma\gamma^\prime}_5(ncnn,kkn,nk,n),\\
A^{\gamma\gamma^\prime}_5(nncn,nkk,kn)&=&A^{\gamma\gamma^\prime}_5(nncn,nkk,kn,c)
+A^{\gamma\gamma^\prime}_5(nncn,nkk,kn,n),\\
A^{\gamma\gamma^\prime}_5(nnnc,nnk)&=&A^{\gamma\gamma^\prime}_5(nnnc,nnk,ck)
+A^{\gamma\gamma^\prime}_5(nnnc,nnk,nk). \eeqa In the case of
$A^{\gamma\gamma^\prime}_5(nnnn)$, there are three terms
corresponding to the second decompositions that result in \beqa
\hskip -0.3in
A^{\gamma\gamma^\prime}_5(nnnn)&=&A^{\gamma\gamma^\prime}_5(nnnn,ccc)
+A^{\gamma\gamma^\prime}_5(nnnn,ccn)+A^{\gamma\gamma^\prime}_5(nnnn,cnc)
+A^{\gamma\gamma^\prime}_5(nnnn,ncc)\nonumber\\
&
&+A^{\gamma\gamma^\prime}_5(nnnn,cnn)+A^{\gamma\gamma^\prime}_5(nnnn,ncn)
+A^{\gamma\gamma^\prime}_5(nnnn,nnc)+A^{\gamma\gamma^\prime}_5(nnnn,nnn).
\eeqa In the above expression, from the fifth term to the seventh
term have the third- or fourth- contraction and anti-contraction,
the eighth term has two third contractions and anti-contractions:
\beqa
A^{\gamma\gamma^\prime}_5(nnnn,cnn)&=&A^{\gamma\gamma^\prime}_5(nnnn,cnn,kc)
+A^{\gamma\gamma^\prime}_5(nnnn,cnn,kn),\\
A^{\gamma\gamma^\prime}_5(nnnn,ncn)&=&A^{\gamma\gamma^\prime}_5(nnnn,ncn,c)
+A^{\gamma\gamma^\prime}_5(nnnn,ncn,n),\\
A^{\gamma\gamma^\prime}_5(nnnn,nnc)&=&A^{\gamma\gamma^\prime}_5(nnnn,nnc,ck)
+A^{\gamma\gamma^\prime}_5(nnnn,nnc,nk),\\
A^{\gamma\gamma^\prime}_5(nnnn,nnn)&=&A^{\gamma\gamma^\prime}_5(nnnn,nnn,cc)
+A^{\gamma\gamma^\prime}_5(nnnn,nnn,cn)\nonumber\\
& &
+A^{\gamma\gamma^\prime}_5(nnnn,nnn,nc)+A^{\gamma\gamma^\prime}_5(nnnn,nnn,nn).
\eeqa In addition, $A^{\gamma\gamma^\prime}_5(nnnn,nnn,nn)$ consists
of the fourth contraction and anti-contraction \beq
A^{\gamma\gamma^\prime}_5(nnnn,nnn,nn)=A^{\gamma\gamma^\prime}_5(nnnn,nnn,nn,c)
+A^{\gamma\gamma^\prime}_5(nnnn,nnn,nn,n).\eeq According to the
above analysis, we obtain that the contribution from the five order
approximation is made of $52$ terms after finding out all of
contractions and anti-contractions.

Just like we have done in the $l=4$ case, we decompose
\beq\label{A5dto}
A_5^{\gamma\gamma^\prime}=A_5^{\gamma\gamma^\prime}(\e)+A_5^{\gamma\gamma^\prime}(t\e)
+A_5^{\gamma\gamma^\prime}(t^2\e),\eeq where \beqa
A_5^{\gamma\gamma^\prime}(\e)&=&A_5^{\gamma\gamma^\prime}(\e^{-\I
E_{\gamma}t})+A_5^{\gamma\gamma^\prime}(\e^{-\I
E_{\gamma_1}t})+A_5^{\gamma\gamma^\prime}(\e^{-\I
E_{\gamma_2}t})+A_5^{\gamma\gamma^\prime}(\e^{-\I E_{\gamma^\prime}t}),\\
A_4^{\gamma\gamma^\prime}(t\e)&=&A_5^{\gamma\gamma^\prime}(t\e^{-\I
E_{\gamma}t})+A_5^{\gamma\gamma^\prime}(t\e^{-\I
E_{\gamma_1}t})+A_5^{\gamma\gamma^\prime}(t\e^{-\I
E_{\gamma_2}t})+A_5^{\gamma\gamma^\prime}(t\e^{-\I E_{\gamma^\prime}t}),\\
A_5^{\gamma\gamma^\prime}(t^2\e)&=&A_5^{\gamma\gamma^\prime}(t^2\e^{-\I
E_{\gamma}t})+A_5^{\gamma\gamma^\prime}(t^2\e^{-\I
E_{\gamma_1}t})+A_5^{\gamma\gamma^\prime}(t^2\e^{-\I
E_{\gamma_2}t})+A_5^{\gamma\gamma^\prime}(t^2\e^{-\I
E_{\gamma^\prime}t}). \eeqa While, every term in the above equations
has its diagonal and off-diagonal parts about $\gamma$ and
$\gamma^\prime$, that is \beqa A_5^{\gamma\gamma^\prime}(\e^{-\I
E_{\gamma_i}t})&=&A_5^{\gamma\gamma^\prime}(\e^{-\I
E_{\gamma_i}t};{\rm D})+A_5^{\gamma\gamma^\prime}(\e^{-\I
E_{\gamma_i}t};{\rm N}),\\
A_5^{\gamma\gamma^\prime}(t\e^{-\I
E_{\gamma_i}t})&=&A_5^{\gamma\gamma^\prime}(t\e^{-\I
E_{\gamma_i}t};{\rm D})+A_5^{\gamma\gamma^\prime}(t\e^{-\I
E_{\gamma_i}t};{\rm N}), \\
A_5^{\gamma\gamma^\prime}(t^2\e^{-\I
E_{\gamma_i}t})&=&A_5^{\gamma\gamma^\prime}(t^2\e^{-\I
E_{\gamma_i}t};{\rm D})+A_5^{\gamma\gamma^\prime}(t^2\e^{-\I
E_{\gamma_i}t};{\rm N}).  \eeqa where $E_{\gamma_i}$ takes $
E_{\gamma}, E_{\gamma_1}, E_{\gamma_2}$ and $E_{\gamma^\prime}$.

If we do not concern the improved forms of perturbed solution higher
than the fourth order one, we only need to write down the second and
third terms in eq.(\ref{A5dto}). We can calculate them and the
results are put in the supplementary of Ref. \cite{My2}.

Based on these contraction- and anti contraction- expressions, we
can, in terms of the rearrangement and summation, obtain \beqa
A_5(t\e^{-\I E_{\gamma} t},{\rm D})&=& -(-\I
G_\gamma^{(3)}t)\sum_{\gamma_1}\frac{\e^{-\I E_{\gamma}
t}}{\left(E_{\gamma}-E_{\gamma_1}\right)^2}g_1^{\gamma\gamma_1}g_1^{\gamma_1\gamma}
\delta_{\gamma\gamma^\prime}\nonumber\\
& &  -(-\I
G_\gamma^{(2)}t)\sum_{\gamma_1,\gamma_2}\left[\frac{\e^{-\I
E_{\gamma}
t}}{\left(E_{\gamma}-E_{\gamma_1}\right)^2\left(E_{\gamma}-E_{\gamma_2}\right)}\right.
\nonumber\\ & & \left. +\frac{\e^{-\I E_{\gamma}
t}}{\left(E_{\gamma}-E_{\gamma_1}\right)\left(E_{\gamma}-E_{\gamma_2}\right)^2}\right]
g_1^{\gamma\gamma_1}g_1^{\gamma_1\gamma_2} g_1^{\gamma_2\gamma}
\delta_{\gamma\gamma^\prime}+ (-\I
G_\gamma^{(5)}t)\delta_{\gamma\gamma^\prime}\eeqa where \beqa
G_\gamma^{(5)}&=&\sum_{\gamma_1,\gamma_2,\gamma_3,\gamma_4}
\frac{g_1^{\gamma\gamma_1}g_1^{\gamma_1\gamma_2}
g_1^{\gamma_2\gamma_3}g_1^{\gamma_3\gamma_4}g_1^{\gamma_4\gamma}
\eta_{\gamma\gamma_2}\eta_{\gamma\gamma_3}}
{\left(E_{\gamma}-E_{\gamma_1}\right)\left(E_{\gamma}-E_{\gamma_2}\right)
\left(E_{\gamma}-E_{\gamma_3}\right)\left(E_{\gamma}-E_{\gamma_4}\right)}\nonumber\\
& & -\sum_{\gamma_1,\gamma_2,\gamma_3}\left[
\frac{g_1^{\gamma\gamma_1}g_1^{\gamma\gamma_2}
g_1^{\gamma_1\gamma}g_1^{\gamma_2\gamma_3}g_1^{\gamma_3\gamma}}
{\left(E_{\gamma}-E_{\gamma_1}\right)^2\left(E_{\gamma}-E_{\gamma_2}\right)
\left(E_{\gamma}-E_{\gamma_3}\right)}
+\frac{g_1^{\gamma\gamma_1}g_1^{\gamma\gamma_2}
g_1^{\gamma_1\gamma}g_1^{\gamma_2\gamma_3}g_1^{\gamma_3\gamma}}
{\left(E_{\gamma}-E_{\gamma_1}\right)\left(E_{\gamma}-E_{\gamma_2}\right)^2
\left(E_{\gamma}-E_{\gamma_3}\right)}\right.\nonumber\\
& &\left.+\frac{g_1^{\gamma\gamma_1}g_1^{\gamma\gamma_2}
g_1^{\gamma_1\gamma}g_1^{\gamma_2\gamma_3}g_1^{\gamma_3\gamma}}
{\left(E_{\gamma}-E_{\gamma_1}\right)\left(E_{\gamma}-E_{\gamma_2}\right)
\left(E_{\gamma}-E_{\gamma_3}\right)^2}\right].\hskip 1.2cm\eeqa
\beqa A_5(t\e^{-\I E_{\gamma_1} t},{\rm D})&=& (-\I
G_\gamma^{(3)}t)\sum_{\gamma_1}\frac{\e^{-\I E_{\gamma_1}
t}}{\left(E_{\gamma}-E_{\gamma_1}\right)^2}g_1^{\gamma\gamma_1}g_1^{\gamma_1\gamma}
\delta_{\gamma\gamma^\prime}\nonumber\\
& &  +(-\I G_\gamma^{(2)}t)\sum_{\gamma_1,\gamma_2}\frac{\e^{-\I
E_{\gamma_1}
t}}{\left(E_{\gamma}-E_{\gamma_1}\right)^2\left(E_{\gamma_1}-E_{\gamma_2}\right)}
g_1^{\gamma\gamma_1}g_1^{\gamma_1\gamma_2} g_1^{\gamma_2\gamma}
\delta_{\gamma\gamma^\prime}\eeqa
\beqa A_5(t\e^{-\I E_{\gamma_2} t},{\rm D})&=& -(-\I
G_\gamma^{(2)}t)\sum_{\gamma_1,\gamma_2}\frac{\e^{-\I E_{\gamma_2}
t}}{\left(E_{\gamma}-E_{\gamma_2}\right)^2\left(E_{\gamma_1}-E_{\gamma_2}\right)}
g_1^{\gamma\gamma_1}g_1^{\gamma_1\gamma_2} g_1^{\gamma_2\gamma}
\delta_{\gamma\gamma^\prime}\eeqa
\beqa A_5(t\e^{-\I E_{\gamma^\prime} t},{\rm D})&=& 0\eeqa
\beqa A_5(t\e^{-\I E_{\gamma} t},{\rm N})&=&(-\I G_\gamma^{(4)}
t)\frac{\e^{-\I E_{\gamma}
t}g_1^{\gamma\gamma^\prime}}{\left(E_{\gamma}-E_{\gamma^\prime}\right)}+
(-\I G_\gamma^{(3)} t)\sum_{\gamma_1}\frac{\e^{-\I E_{\gamma}
t}g_1^{\gamma\gamma_1}g_1^{\gamma_1\gamma^\prime}
\eta^{\gamma\gamma^\prime}}{\left(E_{\gamma}-E_{\gamma_1}\right)\left(E_{\gamma}-E_{\gamma^\prime}\right)}
\nonumber\\
& &-(-\I G_\gamma^{(2)} t)\sum_{\gamma_1}\left[\frac{\e^{-\I
E_{\gamma} t} g_1^{\gamma\gamma_1}g_1^{\gamma_1\gamma}
g_1^{\gamma\gamma^\prime}}{\left(E_{\gamma}-E_{\gamma_1}\right)^2\left(E_{\gamma}-E_{\gamma^\prime}\right)}
+\frac{\e^{-\I E_{\gamma} t}
g_1^{\gamma\gamma_1}g_1^{\gamma_1\gamma}
g_1^{\gamma\gamma^\prime}}{\left(E_{\gamma}-E_{\gamma_1}\right)
\left(E_{\gamma}-E_{\gamma^\prime}\right)^2}\right]\nonumber\\
& & +(-\I G_\gamma^{(2)} t)\sum_{\gamma_1,\gamma_2}\frac{\e^{-\I
E_{\gamma} t}g_1^{\gamma\gamma_1}g_1^{\gamma_1\gamma_2}
g_1^{\gamma_2\gamma^\prime}\eta_{\gamma\gamma_2}\eta_{\gamma\gamma^\prime}}
{\left(E_{\gamma}-E_{\gamma_1}\right)\left(E_{\gamma}-E_{\gamma_2}\right)
\left(E_{\gamma}-E_{\gamma^\prime}\right)} \eeqa
\beqa A_5(t\e^{-\I E_{\gamma_1} t},{\rm N})&=&-(-\I
t)\sum_{\gamma_1}\frac{G_{\gamma_1}^{(3)}\e^{-\I
E_{\gamma_1}t}g_1^{\gamma\gamma_1}g_1^{\gamma_1\gamma^\prime}
\eta_{\gamma\gamma^\prime}}{\left(E_{\gamma}-E_{\gamma_1}\right)
\left(E_{\gamma_1}-E_{\gamma^\prime}\right)}
\nonumber\\
& &-(-\I t)\sum_{\gamma_1,\gamma_2}\frac{G_{\gamma_1}^{(2)}\e^{-\I
E_{\gamma_1}t}g_1^{\gamma\gamma_1}g_1^{\gamma_1\gamma_2}
g_1^{\gamma_2\gamma^\prime}\eta_{\gamma_1\gamma^\prime}\eta_{\gamma\gamma^\prime}}
{\left(E_{\gamma}-E_{\gamma_1}\right)\left(E_{\gamma_1}-E_{\gamma_2}\right)
\left(E_{\gamma_1}-E_{\gamma^\prime}\right)} \eeqa
\beqa A_5(t\e^{-\I E_{\gamma_2} t},{\rm N})&=&(-\I
t)\sum_{\gamma_1,\gamma_2}\frac{G_{\gamma_2}^{(2)}\e^{-\I
E_{\gamma_2}t}g_1^{\gamma\gamma_1}g_1^{\gamma_1\gamma_2}
g_1^{\gamma_2\gamma^\prime}\eta_{\gamma\gamma_2}\eta_{\gamma\gamma^\prime}}
{\left(E_{\gamma}-E_{\gamma_2}\right)\left(E_{\gamma_1}-E_{\gamma_2}\right)
\left(E_{\gamma_2}-E_{\gamma^\prime}\right)} \eeqa
\beqa A_5(t\e^{-\I E_{\gamma^\prime} t},{\rm N})&=&-(-\I
G_{\gamma^\prime}^{(4)} t)\frac{\e^{-\I E_{\gamma^\prime}
t}g_1^{\gamma\gamma^\prime}}{\left(E_{\gamma}-E_{\gamma^\prime}\right)}+
(-\I G_{\gamma^\prime}^{(3)} t)\sum_{\gamma_1}\frac{\e^{-\I
E_{\gamma^\prime} t}g_1^{\gamma\gamma_1}g_1^{\gamma_1\gamma^\prime}
\eta^{\gamma\gamma^\prime}}{\left(E_{\gamma}-E_{\gamma^\prime}\right)
\left(E_{\gamma_1}-E_{\gamma^\prime}\right)}
\nonumber\\
& &+(-\I G_{\gamma^\prime}^{(2)}
t)\sum_{\gamma_1}\left[\frac{\e^{-\I E_{\gamma^\prime} t}
g_1^{\gamma^\prime\gamma_1}g_1^{\gamma_1\gamma^\prime}
g_1^{\gamma\gamma^\prime}}{\left(E_{\gamma}-E_{\gamma^\prime}\right)^2
\left(E_{\gamma_1}-E_{\gamma^\prime}\right)} +\frac{\e^{-\I
E_{\gamma^\prime} t}
g_1^{\gamma^\prime\gamma_1}g_1^{\gamma_1\gamma^\prime}
g_1^{\gamma\gamma^\prime}}{\left(E_{\gamma}-E_{\gamma^\prime}\right)
\left(E_{\gamma_1}-E_{\gamma^\prime}\right)^2}\right]\nonumber\\
& & -(-\I G_{\gamma^\prime}^{(2)}
t)\sum_{\gamma_1,\gamma_2}\frac{\e^{-\I E_{\gamma^\prime}
t}g_1^{\gamma\gamma_1}g_1^{\gamma_1\gamma_2}
g_1^{\gamma_2\gamma^\prime}\eta_{\gamma_1\gamma^\prime}\eta_{\gamma\gamma^\prime}}
{\left(E_{\gamma}-E_{\gamma^\prime}\right)\left(E_{\gamma_1}-E_{\gamma^\prime}\right)
\left(E_{\gamma_2}-E_{\gamma^\prime}\right)} \eeqa

For the parts with $t^2\e$, we have \beqa A_5(t^2\e^{-\I E_{\gamma}
t},{\rm D})&=&\frac{(-\I t)^2}{2!} 2
G_{\gamma}^{(2)}G_{\gamma}^{(3)}\delta_{\gamma\gamma^\prime}\e^{-\I
E_{\gamma} t},\\
A_5(t^2\e^{-\I E_{\gamma_1} t},{\rm D})&=&A_5(t^2\e^{-\I
E_{\gamma_2} t},{\rm D})=A_5(t^2\e^{-\I E_{\gamma^\prime} t},{\rm
D})=0.\eeqa
\beqa A_5(t^2\e^{-\I E_{\gamma} t},{\rm N})&=&\frac{(-\I t)^2}{2!}
\left(G_{\gamma}^{(2)}\right)^2\frac{\e^{-\I
E_{\gamma} t}}{E_{\gamma}-E_{\gamma^\prime}}g_1^{\gamma\gamma^\prime},\\
A_5(t^2\e^{-\I E_{\gamma_2} t},{\rm N})&=&A_5(t^2\e^{-\I
E_{\gamma^\prime} t},{\rm N})=0,\\
A_5(t^2\e^{-\I E_{\gamma^\prime} t},{\rm N})&=&-\frac{(-\I t)^2}{2!}
\left(G_{\gamma^\prime}^{(2)}\right)^2\frac{\e^{-\I
E_{\gamma^\prime}
t}}{E_{\gamma}-E_{\gamma^\prime}}g_1^{\gamma\gamma^\prime}.\eeqa

It is clear that the above diagonal and off-diagonal part about
$A_5^{\gamma\gamma^\prime}(t\e)$ and
$A_5^{\gamma\gamma^\prime}(t\e)$ indeed has the expected forms and
can be absorbed reasonably into the lower order approximations in
order to obtain the improved forms of perturbed solutions.

\subsection{$l=6$ case}

Now let we consider the case of the sixth order approximation
($l=6$). From eq.(\ref{gpd}) it follows that the first
decompositions of $g$-product have $2^5=32$ terms. Like the $l=5$
case, they can be divided into 6 groups \beq
A_6^{\gamma\gamma^\prime}= \sum_{i=0}^4
\mathcal{A}_6^{\gamma\gamma^\prime}(i;\eta),\eeq where $i$ indicates
the number of $\eta$ functions. Obviously \beq
\mathcal{A}_6^{\gamma\gamma^\prime}(0;\eta)={A}_6^{\gamma\gamma^\prime}(ccccc),\eeq
\beqa
\mathcal{A}_6^{\gamma\gamma^\prime}(1;\eta)&=&{A}_6^{\gamma\gamma^\prime}(ccccn)
+{A}_6^{\gamma\gamma^\prime}(cccnc)+{A}_6^{\gamma\gamma^\prime}(ccncc)\nonumber\\
& &
+{A}_6^{\gamma\gamma^\prime}(cnccc)+{A}_6^{\gamma\gamma^\prime}(ncccc),\eeqa
\beqa \mathcal{A}_6^{\gamma\gamma^\prime}(2;\eta)&=&
{A}_6^{\gamma\gamma^\prime}(cccn)+
{A}_6^{\gamma\gamma^\prime}(ccncn)+
{A}_6^{\gamma\gamma^\prime}(cnccn)+
{A}_6^{\gamma\gamma^\prime}(ncccn)\nonumber\\ & & +
{A}_6^{\gamma\gamma^\prime}(ccnnc) +
{A}_6^{\gamma\gamma^\prime}(cncnc) +
 {A}_6^{\gamma\gamma^\prime}(nccnc)+
 {A}_6^{\gamma\gamma^\prime}(cnncc)\nonumber\\ & &+
 {A}_6^{\gamma\gamma^\prime}(ncncc) +
 {A}_6^{\gamma\gamma^\prime}(nnccc),\eeqa \beqa
\mathcal{A}_6^{\gamma\gamma^\prime}(3;\eta)&=&
{A}_6^{\gamma\gamma^\prime}(ccnnn)
+{A}_6^{\gamma\gamma^\prime}(cncnn) +
{A}_6^{\gamma\gamma^\prime}(cnncn) +
{A}_6^{\gamma\gamma^\prime}(cnnnc)\nonumber\\ & &  +
{A}_6^{\gamma\gamma^\prime}(nccnn) +
{A}_6^{\gamma\gamma^\prime}(ncncn) +
{A}_6^{\gamma\gamma^\prime}(ncnnc) +
{A}_6^{\gamma\gamma^\prime}(nnccn)\nonumber\\ & &  +
{A}_6^{\gamma\gamma^\prime}(nncnc) +
{A}_6^{\gamma\gamma^\prime}(nnncc), \eeqa \beqa
\mathcal{A}_6^{\gamma\gamma^\prime}(4;\eta)&=&{A}_6^{\gamma\gamma^\prime}(cnnnn)
+ {A}_6^{\gamma\gamma^\prime}(ncnnn)
+{A}_6^{\gamma\gamma^\prime}(nncnn)\nonumber\\ & &  +
{A}_6^{\gamma\gamma^\prime}(nnncn)+
{A}_6^{\gamma\gamma^\prime}(nnnnc) \eeqa
 \beq
\mathcal{A}_6^{\gamma\gamma^\prime}(5;\eta)={A}_6^{\gamma\gamma^\prime}(nnnnn).\eeq

Furthermore considering the high order contraction or
anti-contraction, we have \beqa {A}_6^{\gamma\gamma^\prime}(cccnn)
&=& {A}_6^{\gamma\gamma^\prime}(cccnn, kkkc) +
{A}_6^{\gamma\gamma^\prime}(cccnn, kkkn),\\
{A}_6^{\gamma\gamma^\prime}(ccncn) &=&
{A}_6^{\gamma\gamma^\prime}(ccncn, kkc) +
{A}_6^{\gamma\gamma^\prime}(ccncn, kkn),\\
{A}_6^{\gamma\gamma^\prime}(ccnnc) &=&
{A}_6^{\gamma\gamma^\prime}(ccnnc, kkck) +
{A}_6^{\gamma\gamma^\prime}(ccnnc, kknk),\\
{A}_6^{\gamma\gamma^\prime}(cnccn) &=&
{A}_6^{\gamma\gamma^\prime}(cnccn, kc) +
{A}_6^{\gamma\gamma^\prime}(cnccn, kn),\\
{A}_6^{\gamma\gamma^\prime}(cncnc) &=&
{A}_6^{\gamma\gamma^\prime}(cncnc, kck) +
{A}_6^{\gamma\gamma^\prime}(cncnc, knk),\\
{A}_6^{\gamma\gamma^\prime}(cnncc) &=&
{A}_6^{\gamma\gamma^\prime}(cnncc, kckk) +
{A}_6^{\gamma\gamma^\prime}(cnncc, knkk),\\
{A}_6^{\gamma\gamma^\prime}(ncccn) &=&
{A}_6^{\gamma\gamma^\prime}(ncccn, c) +
{A}_6^{\gamma\gamma^\prime}(ncccn, n),\\
{A}_6^{\gamma\gamma^\prime}(nccnc) &=&
{A}_6^{\gamma\gamma^\prime}(nccnc, ck) +
{A}_6^{\gamma\gamma^\prime}(nccnc, nk),\\
{A}_6^{\gamma\gamma^\prime}(ncncc) &=&
{A}_6^{\gamma\gamma^\prime}(ncncc, ckk) +
{A}_6^{\gamma\gamma^\prime}(ncncc, nkk),\\
{A}_6^{\gamma\gamma^\prime}(nnccc) &=&
{A}_6^{\gamma\gamma^\prime}(nnccc, ckkk) +
{A}_6^{\gamma\gamma^\prime}(nnccc, nkkk) \eeqa
\beqa {A}_6^{\gamma\gamma^\prime}(ccnnn) &=&
 {A}_6^{\gamma\gamma^\prime}(ccnnn, kkcc)
 + {A}_6^{\gamma\gamma^\prime}(ccnnn, kkcn)
 + {A}_6^{\gamma\gamma^\prime}(ccnnn, kknc)\nonumber\\
 & & +
    {A}_6^{\gamma\gamma^\prime}(ccnnn, kknn, kkc)
    + {A}_6^{\gamma\gamma^\prime}(ccnnn, kknn,
    kkn),\eeqa
\beqa {A}_6^{\gamma\gamma^\prime}(cncnn) &=&
  {A}_6^{\gamma\gamma^\prime}(cncnn, kkkc, kck)
  + {A}_6^{\gamma\gamma^\prime}(cncnn, kkkc, knk)
  + {A}_6^{\gamma\gamma^\prime}(cncnn, kkkn, kck)\nonumber\\ & &
  + {A}_6^{\gamma\gamma^\prime}(cncnn, kkkn, knk, kc)
    + {A}_6^{\gamma\gamma^\prime}(cncnn, kkkn, knk, kn),\\
{A}_6^{\gamma\gamma^\prime}(cnncn) &=&
  {A}_6^{\gamma\gamma^\prime}(cnncn, kckk, kkc)
  + {A}_6^{\gamma\gamma^\prime}(cnncn, kckk, kkn)
  + {A}_6^{\gamma\gamma^\prime}(cnncn, knkk, kkc)\nonumber\\
  & &+
    {A}_6^{\gamma\gamma^\prime}(cnncn, knkk, kkn, kc)
    + {A}_6^{\gamma\gamma^\prime}(cnncn, knkk, kkn, kn),\\
{A}_6^{\gamma\gamma^\prime}(cnnnc) &=&
  {A}_6^{\gamma\gamma^\prime}(cnnnc, kcck)
  + {A}_6^{\gamma\gamma^\prime}(cnnnc, kcnk)
  + {A}_6^{\gamma\gamma^\prime}(cnnnc, knck)\nonumber\\
    & &  +
    {A}_6^{\gamma\gamma^\prime}(cnnnc, knnk, kck)+ {A}_6^{\gamma\gamma^\prime}(cnnnc, knnk,
    knk),\eeqa
\beqa {A}_6^{\gamma\gamma^\prime}(nccnn) &=&
  {A}_6^{\gamma\gamma^\prime}(nccnn, kkkc, ck)
  + {A}_6^{\gamma\gamma^\prime}(nccnn, kkkc, nk)
  + {A}_6^{\gamma\gamma^\prime}(nccnn, kkkn, ck)\nonumber\\
  & & +
    {A}_6^{\gamma\gamma^\prime}(nccnn, kkkn, nk, c) + {A}_6^{\gamma\gamma^\prime}(nccnn, kkkn, nk,
    n),\\
{A}_6^{\gamma\gamma^\prime}(ncncn) &=&
  {A}_6^{\gamma\gamma^\prime}(ncncn, ckc)
  + {A}_6^{\gamma\gamma^\prime}(ncncn, ckn)
  + {A}_6^{\gamma\gamma^\prime}(ncncn, nkc)\nonumber\\
  & & + {A}_6^{\gamma\gamma^\prime}(ncncn, nkn, c) +
    {A}_6^{\gamma\gamma^\prime}(ncncn, nkn, n),\\
{A}_6^{\gamma\gamma^\prime}(ncnnc) &=&
  {A}_6^{\gamma\gamma^\prime}(ncnnc, kkck, ckk)
  + {A}_6^{\gamma\gamma^\prime}(ncnnc, kkck, nkk)
  + {A}_6^{\gamma\gamma^\prime}(ncnnc, kknk, ckk)\nonumber\\
  & & +
    {A}_6^{\gamma\gamma^\prime}(ncnnc, kknk, nkk, ck)
    + {A}_6^{\gamma\gamma^\prime}(ncnnc, kknk, nkk, nk),\\
{A}_6^{\gamma\gamma^\prime}(nnccn) &=&
  {A}_6^{\gamma\gamma^\prime}(nnccn, cc)
  + {A}_6^{\gamma\gamma^\prime}(nnccn, cn)
  + {A}_6^{\gamma\gamma^\prime}(nnccn, nc)\nonumber\\
& &  + {A}_6^{\gamma\gamma^\prime}(nnccn, nn, c) +
    {A}_6^{\gamma\gamma^\prime}(nnccn, nn, n),\\
{A}_6^{\gamma\gamma^\prime}(nncnc) &=&
  {A}_6^{\gamma\gamma^\prime}(nncnc, ckkk, kck)
  + {A}_6^{\gamma\gamma^\prime}(nncnc, ckkk, knk)
  + {A}_6^{\gamma\gamma^\prime}(nncnc, nkkk, kck)\nonumber\\
  & & +
    {A}_6^{\gamma\gamma^\prime}(nncnc, nkkk, knk, ck)
    + {A}_6^{\gamma\gamma^\prime}(nncnc, nkkk, knk, nk),\\
{A}_6^{\gamma\gamma^\prime}(nnncc) &=&
  {A}_6^{\gamma\gamma^\prime}(nnncc, cckk)
  + {A}_6^{\gamma\gamma^\prime}(nnncc, cnkk)
  + {A}_6^{\gamma\gamma^\prime}(nnncc, nckk) \nonumber\\
  & & +
    {A}_6^{\gamma\gamma^\prime}(nnncc, nnkk, c) + {A}_6^{\gamma\gamma^\prime}(nnncc, nnkk, n)
\eeqa
\beqa {A}_6^{\gamma\gamma^\prime}(cnnnn) &=&
  {A}_6^{\gamma\gamma^\prime}(cnnnn, kccc)
  + {A}_6^{\gamma\gamma^\prime}(cnnnn, kccn)
  + {A}_6^{\gamma\gamma^\prime}(cnnnn, kcnc)\nonumber\\
  & &
  + {A}_6^{\gamma\gamma^\prime}(cnnnn, kncc)
  + {A}_6^{\gamma\gamma^\prime}(cnnnn, kcnn)
  + {A}_6^{\gamma\gamma^\prime}(cnnnn, kncn)\nonumber\\
  & & + {A}_6^{\gamma\gamma^\prime}(cnnnn, knnc)
  + {A}_6^{\gamma\gamma^\prime}(cnnnn, knnn),
\eeqa \beqa {A}_6^{\gamma\gamma^\prime}(cnnnn, kcnn) &=&
{A}_6^{\gamma\gamma^\prime}(cnnnn, kcnn, kkc) +
{A}_6^{\gamma\gamma^\prime}(cnnnn, kcnn, kkn),\\
{A}_6^{\gamma\gamma^\prime}(cnnnn, kncn) &=&
{A}_6^{\gamma\gamma^\prime}(cnnnn, kncn, kc) +
{A}_6^{\gamma\gamma^\prime}(cnnnn, kncn, kn),\\
{A}_6^{\gamma\gamma^\prime}(cnnnn, knnc) &=&
{A}_6^{\gamma\gamma^\prime}(cnnnn, knnc, kck) +
{A}_6^{\gamma\gamma^\prime}(cnnnn, knnc, knk), \eeqa \beqa
{A}_6^{\gamma\gamma^\prime}(cnnnn, knnn) &=&
  {A}_6^{\gamma\gamma^\prime}(cnnnn, knnn, kcc)
  + {A}_6^{\gamma\gamma^\prime}(cnnnn, knnn, kcn)\nonumber\\ & &
  + {A}_6^{\gamma\gamma^\prime}(cnnnn, knnn, knc)+
    {A}_6^{\gamma\gamma^\prime}(cnnnn, knnn, knn, kc)\nonumber\\ & &
    + {A}_6^{\gamma\gamma^\prime}(cnnnn, knnn, knn, kn).
\eeqa
\beqa {A}_6^{\gamma\gamma^\prime}(ncnnn) &=&
  {A}_6^{\gamma\gamma^\prime}(ncnnn, kkcc, ckk)
  + {A}_6^{\gamma\gamma^\prime}(ncnnn, kkcc, nkk)\nonumber\\ & &
  + {A}_6^{\gamma\gamma^\prime}(ncnnn, kkcn, ckk) +
    {A}_6^{\gamma\gamma^\prime}(ncnnn, kknc, ckk)\nonumber\\
  & &
    + {A}_6^{\gamma\gamma^\prime}(ncnnn, kkcn, nkk)
    + {A}_6^{\gamma\gamma^\prime}(ncnnn, kknc, nkk)\nonumber\\
& & +
    {A}_6^{\gamma\gamma^\prime}(ncnnn, kknn, ckk)
    + {A}_6^{\gamma\gamma^\prime}(ncnnn, kknn, nkk),
    \eeqa
\beqa {A}_6^{\gamma\gamma^\prime}(ncnnn, kkcn, nkk) &=&
{A}_6^{\gamma\gamma^\prime}(ncnnn, kkcn, nkk, c) +
{A}_6^{\gamma\gamma^\prime}(ncnnn, kkcn, nkk, n),\\
{A}_6^{\gamma\gamma^\prime}(ncnnn, kknc, nkk) &=&
  {A}_6^{\gamma\gamma^\prime}(ncnnn, kknc, nkk, ck)
  + {A}_6^{\gamma\gamma^\prime}(ncnnn, kknc, nkk, nk),\ \ \ \ \ \ \ \\
{A}_6^{\gamma\gamma^\prime}(ncnnn, kknn, ckk) &=&
{A}_6^{\gamma\gamma^\prime}(ncnnn, kknn, ckc) +
{A}_6^{\gamma\gamma^\prime}(ncnnn, kknn, ckn), \eeqa
 \beqa
{A}_6^{\gamma\gamma^\prime}(ncnnn, kknn, nkk) &=&
  {A}_6^{\gamma\gamma^\prime}(ncnnn, kknn, nkc, ck)
  + {A}_6^{\gamma\gamma^\prime}(ncnnn, kknn, nkc, nk)\nonumber\\
 & &
  +
    {A}_6^{\gamma\gamma^\prime}(ncnnn, kknn, nkn, ck)
    + {A}_6^{\gamma\gamma^\prime}(ncnnn, kknn, nkn, nk, c)\nonumber\\
 & & +
    {A}_6^{\gamma\gamma^\prime}(ncnnn, kknn, nkn, nk, n)
\eeqa
\beqa {A}_6^{\gamma\gamma^\prime}(nncnn) &=&
  {A}_6^{\gamma\gamma^\prime}(nncnn, ckkc, kck)
  + {A}_6^{\gamma\gamma^\prime}(nncnn, ckkc, knk)\nonumber\\
  & &
  + {A}_6^{\gamma\gamma^\prime}(nncnn, ckkn, kck)+
    {A}_6^{\gamma\gamma^\prime}(nncnn, nkkc, kck)\nonumber\\
  & &
    + {A}_6^{\gamma\gamma^\prime}(nncnn, ckkn, knk)
    + {A}_6^{\gamma\gamma^\prime}(nncnn, nkkc, knk)\nonumber\\
    & & +
    {A}_6^{\gamma\gamma^\prime}(nncnn, nkkn, kck)
    + {A}_6^{\gamma\gamma^\prime}(nncnn, nkkn, knk),
\eeqa \beqa {A}_6^{\gamma\gamma^\prime}(nncnn, ckkn, knk) &=&
  {A}_6^{\gamma\gamma^\prime}(nncnn, ckkn, knk, kc)
  + {A}_6^{\gamma\gamma^\prime}(nncnn, ckkn, knk, kn),\ \ \ \ \ \ \ \\
{A}_6^{\gamma\gamma^\prime}(nncnn, nkkc, knk) &=&
  {A}_6^{\gamma\gamma^\prime}(nncnn, nkkc, knk, ck)
  + {A}_6^{\gamma\gamma^\prime}(nncnn, nkkc, knk, nk),\\
{A}_6^{\gamma\gamma^\prime}(nncnn, nkkn, kck) &=&
{A}_6^{\gamma\gamma^\prime}(nncnn, nkkn, kck, c) +
{A}_6^{\gamma\gamma^\prime}(nncnn, nkkn, kck, n), \eeqa \beqa
{A}_6^{\gamma\gamma^\prime}(nncnn, nkkn, knk) &=&
  {A}_6^{\gamma\gamma^\prime}(nncnn, nkkn, knk, cc)
  + {A}_6^{\gamma\gamma^\prime}(nncnn, nkkn, knk, cn)\nonumber\\
  & &
  +
    {A}_6^{\gamma\gamma^\prime}(nncnn, nkkn, knk, nc)
    + {A}_6^{\gamma\gamma^\prime}(nncnn, nkkn, knk, nn, c)\nonumber\\
  & &  +
    {A}_6^{\gamma\gamma^\prime}(nncnn, nkkn, knk, nn, n)
\eeqa
\beqa {A}_6^{\gamma\gamma^\prime}(nnncn) &=&
  {A}_6^{\gamma\gamma^\prime}(nnncn, cckk, kkc)
  + {A}_6^{\gamma\gamma^\prime}(nnncn, cckk, kkn)\nonumber\\
  & &
  + {A}_6^{\gamma\gamma^\prime}(nnncn, cnkk, kkc)
  +
    {A}_6^{\gamma\gamma^\prime}(nnncn, nckk, kkc)\nonumber\\
  & &
    + {A}_6^{\gamma\gamma^\prime}(nnncn, cnkk, kkn)
    + {A}_6^{\gamma\gamma^\prime}(nnncn, nckk, kkn)\nonumber\\
    & & +
    {A}_6^{\gamma\gamma^\prime}(nnncn, nnkk, kkc)
    + {A}_6^{\gamma\gamma^\prime}(nnncn, nnkk, kkn),
\eeqa \beqa {A}_6^{\gamma\gamma^\prime}(nnncn, cnkk, kkn) &=&
  {A}_6^{\gamma\gamma^\prime}(nnncn, cnkk, kkn, kc)
  + {A}_6^{\gamma\gamma^\prime}(nnncn, cnkk, kkn, kn),\ \ \ \ \ \ \\
{A}_6^{\gamma\gamma^\prime}(nnncn, nckk, kkn) &=&
{A}_6^{\gamma\gamma^\prime}(nnncn, nckk, kkn, c) +
{A}_6^{\gamma\gamma^\prime}(nnncn, nckk, kkn, n),\\
{A}_6^{\gamma\gamma^\prime}(nnncn, nnkk, kkc) &=&
{A}_6^{\gamma\gamma^\prime}(nnncn, nnkk, ckc) +
{A}_6^{\gamma\gamma^\prime}(nnncn, nnkk, nkc), \eeqa \beqa
{A}_6^{\gamma\gamma^\prime}(nnncn, nnkk, kkn) &=&
  {A}_6^{\gamma\gamma^\prime}(nnncn, nnkk, ckn, kc)
  + {A}_6^{\gamma\gamma^\prime}(nnncn, nnkk, ckn, kn)\nonumber\\
  & &
  +
    {A}_6^{\gamma\gamma^\prime}(nnncn, nnkk, nkn, kc)
    + {A}_6^{\gamma\gamma^\prime}(nnncn, nnkk, nkn, kn, c)\nonumber\\
  & &  +
    {A}_6^{\gamma\gamma^\prime}(nnncn, nnkk, nkn, kn, n).
\eeqa
\beqa {A}_6^{\gamma\gamma^\prime}(nnnnc) &=&
  {A}_6^{\gamma\gamma^\prime}(nnnnc, ccck)
  + {A}_6^{\gamma\gamma^\prime}(nnnnc, ccnk)
  + {A}_6^{\gamma\gamma^\prime}(nnnnc, cnck)\nonumber\\
  & &
  + {A}_6^{\gamma\gamma^\prime}(nnnnc, ncck)+
    {A}_6^{\gamma\gamma^\prime}(nnnnc, cnnk)
    + {A}_6^{\gamma\gamma^\prime}(nnnnc, ncnk)\nonumber\\
    & &
    + {A}_6^{\gamma\gamma^\prime}(nnnnc, nnck)
    + {A}_6^{\gamma\gamma^\prime}(nnnnc, nnnk),
\eeqa \beqa {A}_6^{\gamma\gamma^\prime}(nnnnc, cnnk) &=&
{A}_6^{\gamma\gamma^\prime}(nnnnc, cnnk, kck) +
{A}_6^{\gamma\gamma^\prime}(nnnnc, cnnk, knk),\\
{A}_6^{\gamma\gamma^\prime}(nnnnc, ncnk) &=&
{A}_6^{\gamma\gamma^\prime}(nnnnc, ncnk, ck) +
{A}_6^{\gamma\gamma^\prime}(nnnnc, ncnk, nk),\\
{A}_6^{\gamma\gamma^\prime}(nnnnc, nnck) &=&
{A}_6^{\gamma\gamma^\prime}(nnnnc, nnck, c) +
{A}_6^{\gamma\gamma^\prime}(nnnnc, nnck, n),\eeqa \beqa
{A}_6^{\gamma\gamma^\prime}(nnnnc, nnnk) &=&
  {A}_6^{\gamma\gamma^\prime}(nnnnc, nnnk, cck)
  + {A}_6^{\gamma\gamma^\prime}(nnnnc, nnnk, cnk)\nonumber\\
  & &
  + {A}_6^{\gamma\gamma^\prime}(nnnnc, nnnk, nck) +
    {A}_6^{\gamma\gamma^\prime}(nnnnc, nnnk, nnk, ck)\nonumber\\
  & &
    + {A}_6^{\gamma\gamma^\prime}(nnnnc, nnnk, nnk, nk).
\eeqa
\beqa {A}_6^{\gamma\gamma^\prime}(nnnnn) &=&
  {A}_6^{\gamma\gamma^\prime}(nnnnn, cccc)
  + {A}_6^{\gamma\gamma^\prime}(nnnnn, cccn)
  + {A}_6^{\gamma\gamma^\prime}(nnnnn, ccnc)\nonumber\\ & &
  + {A}_6^{\gamma\gamma^\prime}(nnnnn, cncc)
  +
    {A}_6^{\gamma\gamma^\prime}(nnnnn, nccc)
    + {A}_6^{\gamma\gamma^\prime}(nnnnn, ccnn)\nonumber\\ & &
    + {A}_6^{\gamma\gamma^\prime}(nnnnn, cncn)
    +
    {A}_6^{\gamma\gamma^\prime}(nnnnn, cnnc)
    + {A}_6^{\gamma\gamma^\prime}(nnnnn, nccn)\nonumber\\ & &
    + {A}_6^{\gamma\gamma^\prime}(nnnnn, ncnc)
    +
    {A}_6^{\gamma\gamma^\prime}(nnnnn, nncc)
    + {A}_6^{\gamma\gamma^\prime}(nnnnn, cnnn)\nonumber\\
    & & + {A}_6^{\gamma\gamma^\prime}(nnnnn, ncnn)
    +
    {A}_6^{\gamma\gamma^\prime}(nnnnn, nncn)
    + {A}_6^{\gamma\gamma^\prime}(nnnnn, nnnc) \nonumber\\
  & &+ {A}_6^{\gamma\gamma^\prime}(nnnnn,
    nnnn),\eeqa
\beqa {A}_6^{\gamma\gamma^\prime}(nnnnn, ccnn) &=&
{A}_6^{\gamma\gamma^\prime}(nnnnn, ccnn, kkc) +
{A}_6^{\gamma\gamma^\prime}(nnnnn, ccnn, kkn),\\
{A}_6^{\gamma\gamma^\prime}(nnnnn, cncn) &=&
{A}_6^{\gamma\gamma^\prime}(nnnnn, cncn, kc) +
{A}_6^{\gamma\gamma^\prime}(nnnnn, cncn, kn),\\
{A}_6^{\gamma\gamma^\prime}(nnnnn, cnnc) &=&
{A}_6^{\gamma\gamma^\prime}(nnnnn, cnnc, kck) +
{A}_6^{\gamma\gamma^\prime}(nnnnn, cnnc, knk),\\
{A}_6^{\gamma\gamma^\prime}(nnnnn, nccn) &=&
{A}_6^{\gamma\gamma^\prime}(nnnnn, nccn, c) +
{A}_6^{\gamma\gamma^\prime}(nnnnn, nccn, n),\\
{A}_6^{\gamma\gamma^\prime}(nnnnn, ncnc) &=&
{A}_6^{\gamma\gamma^\prime}(nnnnn, ncnc, ck) +
{A}_6^{\gamma\gamma^\prime}(nnnnn, ncnc, nk),\\
{A}_6^{\gamma\gamma^\prime}(nnnnn, nncc) &=&
{A}_6^{\gamma\gamma^\prime}(nnnnn, nncc, ckk) +
{A}_6^{\gamma\gamma^\prime}(nnnnn, nncc, nkk), \eeqa \beqa
{A}_6^{\gamma\gamma^\prime}(nnnnn, cnnn) &=&
  {A}_6^{\gamma\gamma^\prime}(nnnnn, cnnn, kcc)
  + {A}_6^{\gamma\gamma^\prime}(nnnnn, cnnn, kcn)\nonumber\\& &
  + {A}_6^{\gamma\gamma^\prime}(nnnnn, cnnn, knc)
   +
    {A}_6^{\gamma\gamma^\prime}(nnnnn, cnnn, knn, kc)\nonumber\\& &
    + {A}_6^{\gamma\gamma^\prime}(nnnnn, cnnn, knn, kn),\\
{A}_6^{\gamma\gamma^\prime}(nnnnn, ncnn) &=&
  {A}_6^{\gamma\gamma^\prime}(nnnnn, ncnn, kkc, ck)
  + {A}_6^{\gamma\gamma^\prime}(nnnnn, ncnn, kkc, nk)\nonumber\\& &
  +
    {A}_6^{\gamma\gamma^\prime}(nnnnn, ncnn, kkn, ck)
    + {A}_6^{\gamma\gamma^\prime}(nnnnn, ncnn, kkn, nk, c) \nonumber\\& &+
    {A}_6^{\gamma\gamma^\prime}(nnnnn, ncnn, kkn, nk, n),\\
{A}_6^{\gamma\gamma^\prime}(nnnnn, nncn) &=&
  {A}_6^{\gamma\gamma^\prime}(nnnnn, nncn, ckk, kc)
  + {A}_6^{\gamma\gamma^\prime}(nnnnn, nncn, ckk, kn)\nonumber\\& &
  +
    {A}_6^{\gamma\gamma^\prime}(nnnnn, nncn, nkk, kc)
    + {A}_6^{\gamma\gamma^\prime}(nnnnn, nncn, nkk, kn, c)\nonumber\\& & +
    {A}_6^{\gamma\gamma^\prime}(nnnnn, nncn, nkk, kn, n),\\
{A}_6^{\gamma\gamma^\prime}(nnnnn, nnnc) &=&
  {A}_6^{\gamma\gamma^\prime}(nnnnn, nnnc, cck)
  + {A}_6^{\gamma\gamma^\prime}(nnnnn, nnnc, cnk)\nonumber\\& &
  + {A}_6^{\gamma\gamma^\prime}(nnnnn, nnnc, nck)+
    {A}_6^{\gamma\gamma^\prime}(nnnnn, nnnc, nnk, ck)\nonumber\\& &
    + {A}_6^{\gamma\gamma^\prime}(nnnnn, nnnc, nnk, nk),\\
{A}_6^{\gamma\gamma^\prime}(nnnnn, nnnn) &=&
  {A}_6^{\gamma\gamma^\prime}(nnnnn, nnnn, ccc)
  + {A}_6^{\gamma\gamma^\prime}(nnnnn, nnnn, ccn)\nonumber\\& &
  + {A}_6^{\gamma\gamma^\prime}(nnnnn, nnnn, cnc) +
    {A}_6^{\gamma\gamma^\prime}(nnnnn, nnnn, ncc)\nonumber\\& &
    + {A}_6^{\gamma\gamma^\prime}(nnnnn, nnnn, cnn)
    + {A}_6^{\gamma\gamma^\prime}(nnnnn, nnnn, ncn)\nonumber\\
    & & +
    {A}_6^{\gamma\gamma^\prime}(nnnnn, nnnn, nnc)
    + {A}_6^{\gamma\gamma^\prime}(nnnnn, nnnn, nnn),\eeqa
\beqa \!\!\!\!{A}_6^{\gamma\gamma^\prime}(nnnnn, nnnn, cnn)\!\!\!
&=&\!\!\!
  {A}_6^{\gamma\gamma^\prime}(nnnnn, nnnn, cnn, kc)
  + {A}_6^{\gamma\gamma^\prime}(nnnnn, nnnn, cnn, kn),\ \ \ \ \\
\!\!\!\!{A}_6^{\gamma\gamma^\prime}(nnnnn, nnnn, ncn)\!\!\!
&=&\!\!\! {A}_6^{\gamma\gamma^\prime}(nnnnn, nnnn, ncn, c) +
{A}_6^{\gamma\gamma^\prime}(nnnnn, nnnn, ncn, n),\ \ \ \ \ \\
\!\!\!\!{A}_6^{\gamma\gamma^\prime}(nnnnn, nnnn, nnc)\!\!\!
&=&\!\!\!
  {A}_6^{\gamma\gamma^\prime}(nnnnn, nnnn, nnc, ck)
  + {A}_6^{\gamma\gamma^\prime}(nnnnn, nnnn, nnc, nk),\eeqa
\beqa {A}_6^{\gamma\gamma^\prime}(nnnnn, nnnn, nnn) &=&
  {A}_6^{\gamma\gamma^\prime}(nnnnn, nnnn, nnn, cc)
  + {A}_6^{\gamma\gamma^\prime}(nnnnn, nnnn, nnn, cn)\nonumber\\& &
  +
    {A}_6^{\gamma\gamma^\prime}(nnnnn, nnnn, nnn, nc)
    + {A}_6^{\gamma\gamma^\prime}(nnnnn, nnnn, nnn, nn, c) \nonumber\\& &+
    {A}_6^{\gamma\gamma^\prime}(nnnnn, nnnn, nnn, nn, n).
\eeqa

Thus, we obtain that the contribution from the six order
approximation is made of $203$ terms after finding out all of
contractions and anti-contractions.

Just like we have done in the $l=4$ or 5 cases, we decompose
\beqa\label{A6dto1}
A_6^{\gamma\gamma^\prime}&=&A_6^{\gamma\gamma^\prime}(\e)+A_6^{\gamma\gamma^\prime}(t\e)
+A_6^{\gamma\gamma^\prime}(t^2\e)+A_6^{\gamma\gamma^\prime}(t^3\e)\\
\label{A6dto2}&=&A_6^{\gamma\gamma^\prime}(\e,t\e)+A_6^{\gamma\gamma^\prime}(t^2\e,t^3\e)
,\eeqa To our purpose, we only calculate the second term
$A_6^{\gamma\gamma^\prime}(t^2\e,t^3\e)$ in eq.(\ref{A6dto2}).
Without loss of generality, we decompose it into \beqa
A_6^{\gamma\gamma^\prime}(t^2\e,t^3\e)&=&A_6^{\gamma\gamma^\prime}(t^2\e^{-\I
E_{\gamma}t},t^3\e^{-\I
E_{\gamma}t})+A_6^{\gamma\gamma^\prime}(t^2\e^{-\I
E_{\gamma_1}t},t^3\e^{-\I
E_{\gamma_1}t})\nonumber\\
& &+A_6^{\gamma\gamma^\prime}(t^2\e^{-\I
E_{\gamma^\prime}t},t^3\e^{-\I E_{\gamma^\prime}t}).\eeqa While,
every term in the above equations has its diagonal and off-diagonal
parts about $\gamma$ and $\gamma^\prime$, that is \beqa
A_6^{\gamma\gamma^\prime}(t^2\e^{-\I
E_{\gamma_i}t})&=&A_6^{\gamma\gamma^\prime}(t^2\e^{-\I
E_{\gamma_i}t};{\rm D})+A_6^{\gamma\gamma^\prime}(t^2\e^{-\I
E_{\gamma_i}t};{\rm N}),\\
A_6^{\gamma\gamma^\prime}(t^3\e^{-\I
E_{\gamma_i}t})&=&A_6^{\gamma\gamma^\prime}(t^3\e^{-\I
E_{\gamma_i}t};{\rm D})+A_6^{\gamma\gamma^\prime}(t^3\e^{-\I
E_{\gamma_i}t};{\rm N}).  \eeqa where $E_{\gamma_i}$ takes $
E_{\gamma}, E_{\gamma_1}$ and $E_{\gamma^\prime}$.

Based on our calculations, we find that there are nonvanishing 91
terms and vanishing 112 terms with $t^2\e, t^3\e$ factor parts in
all of 203 contraction- and anti contraction- expressions (see in
the supplementary Ref. \cite{My2}). Therefore we can, in terms of
rearrangement and summation, obtain the following concise forms:
\beqa A_6(t^2\e^{-\I E_{\gamma}t},{\rm D})&=&\frac{(-\I
t)^2}{2!}\left(G_\gamma^{(3)}\right)^2\e^{-\I
E_{\gamma}t}\delta_{\gamma\gamma^\prime}+\frac{(-\I t)^2}{2!}2
G_\gamma^{(2)}G_\gamma^{(4)}\e^{-\I
E_{\gamma}t}\delta_{\gamma\gamma^\prime}\nonumber\\
& &-\frac{(-\I
t)^2}{2!}\sum_{\gamma_1}\frac{\left(G_\gamma^{(2)}\right)^2\e^{-\I
E_{\gamma}t}}{\left(E_\gamma-E_{\gamma_1}\right)^2}
g_1^{\gamma\gamma_1}g_1^{\gamma_1\gamma}\delta_{\gamma\gamma^\prime}.
\eeqa
\beqa A_6(t^2\e^{-\I E_{\gamma_1}t},{\rm D})&=&\frac{(-\I
t)^2}{2!}\sum_{\gamma_1}\frac{\left(G_\gamma^{(2)}\right)^2\e^{-\I
E_{\gamma_1}t}}{\left(E_\gamma-E_{\gamma_1}\right)^2}
g_1^{\gamma\gamma_1}g_1^{\gamma_1\gamma}\delta_{\gamma\gamma^\prime}.
\eeqa \beqa A_6(t^2\e^{-\I E_{\gamma^\prime}t},{\rm D})&=&0.\eeqa
\beqa A_6(t^2\e^{-\I E_{\gamma}t},{\rm N})&=&\frac{(-\I t)^2}{2!}2
G_\gamma^{(2)}G_\gamma^{(3)}\frac{\e^{-\I
E_{\gamma}t}}{\left(E_\gamma-E_{\gamma^\prime}\right)}g_1^{\gamma\gamma^\prime}\nonumber\\
& &+\frac{(-\I
t)^2}{2!}\sum_{\gamma_1}\frac{\left(G_\gamma^{(2)}\right)^2\e^{-\I
E_{\gamma}t}}{\left(E_\gamma-E_{\gamma_1}\right)\left(E_\gamma-E_{\gamma^\prime}\right)}
g_1^{\gamma\gamma_1}g_1^{\gamma_1\gamma^\prime}\eta_{\gamma\gamma^\prime}.
\eeqa
\beqa A_6(t^2\e^{-\I E_{\gamma_1}t},{\rm N})&=&-\frac{(-\I
t)^2}{2!}\sum_{\gamma_1}\frac{\left(G_{\gamma_1}^{(2)}\right)^2\e^{-\I
E_{\gamma_1}t}}{\left(E_\gamma-E_{\gamma_1}\right)\left(E_{\gamma_1}-E_{\gamma^\prime}\right)}
g_1^{\gamma\gamma_1}g_1^{\gamma_1\gamma^\prime}\eta_{\gamma\gamma^\prime}.
\eeqa
\beqa A_6(t^2\e^{-\I E_{\gamma^\prime}t},{\rm N})&=&-\frac{(-\I
t)^2}{2!}2
G_{\gamma^\prime}^{(2)}G_{\gamma^\prime}^{(3)}\frac{\e^{-\I
E_{\gamma^\prime}t}}{\left(E_\gamma-E_{\gamma^\prime}\right)}g_1^{\gamma\gamma^\prime}\nonumber\\
& &+\frac{(-\I
t)^2}{2!}\sum_{\gamma_1}\frac{\left(G_{\gamma^\prime}^{(2)}\right)^2\e^{-\I
E_{\gamma^\prime}t}}{\left(E_\gamma-E_{\gamma^\prime}\right)\left(E_{\gamma_1}-E_{\gamma^\prime}\right)}
g_1^{\gamma\gamma_1}g_1^{\gamma_1\gamma^\prime}\eta_{\gamma\gamma^\prime}.
\eeqa Their forms are indeed the same as expected and can be
absorbed  reasonably into the lower order approximations in order to
obtain the improved forms of perturbed solutions.

\end{appendix}



\section*{\Large\bf\sc Supplement}


\subsection*{1 $l=5$ case}

In the following, we respectively calculate the 52 component
expressions of $A_5^{\gamma\gamma^\prime}$, and put the second and
third terms in eq.(\ref{A5dto}) together as
$A_5^{\gamma\gamma^\prime}(t\e,t^2\e)$. \beqa
{A}_5^{\gamma\gamma^\prime}(cccc;t\e,t^2\e)&=&\left[-(-\I
t)\frac{3\e^{-\I E_{\gamma}t}+3\e^{-\I
E_{\gamma^\prime}t}}{\left(E_\gamma-E_{\gamma^\prime}\right)^4}
+\frac{(-\I t)^2}{2}\frac{\e^{-\I
E_{\gamma}t}}{\left(E_\gamma-E_{\gamma^\prime}\right)^3}\right.\nonumber\\
& &\left.-\frac{(-\I t)^2}{2}\frac{\e^{-\I
E_{\gamma^\prime}t}}{\left(E_\gamma-E_{\gamma^\prime}\right)^3}\right]
\left|g_1^{\gamma\gamma^\prime}\right|^4g_1^{\gamma\gamma^\prime}.\eeqa
\beqa {A}_5^{\gamma\gamma^\prime}(cccn;t\e,t^2\e)&=&\sum_{\gamma_1}
\left[-(-\I t)\frac{2\e^{-\I
E_{\gamma}t}}{\left(E_\gamma-E_{\gamma_1}\right)^3
\left(E_\gamma-E_{\gamma^\prime}\right)}-(-\I t)\frac{\e^{-\I
E_{\gamma}t}}{\left(E_\gamma-E_{\gamma_1}\right)^2
\left(E_\gamma-E_{\gamma^\prime}\right)^2}\right.\nonumber\\
& &\left.-(-\I t)\frac{\e^{-\I
E_{\gamma_1}t}}{\left(E_\gamma-E_{\gamma_1}\right)^3
\left(E_{\gamma_1}-E_{\gamma^\prime}\right)}+\frac{(-\I
t)^2}{2!}\frac{\e^{-\I
E_{\gamma}t}}{\left(E_\gamma-E_{\gamma_1}\right)^2\left(E_\gamma-E_{\gamma^\prime}\right)}\right]\nonumber\\
& & \times
\left|g_1^{\gamma\gamma_1}\right|^4g_1^{\gamma\gamma^\prime}\eta_{\gamma_1\gamma^\prime}.\eeqa
\beqa {A}_5^{\gamma\gamma^\prime}(ccnc;t\e,t^2\e)&=&\sum_{\gamma_1}
\left[(-\I t)\frac{\e^{-\I
E_{\gamma}t}}{\left(E_\gamma-E_{\gamma_1}\right)
\left(E_\gamma-E_{\gamma^\prime}\right)^3}-(-\I t)\frac{\e^{-\I
E_{\gamma^\prime}t}}{\left(E_\gamma-E_{\gamma^\prime}\right)^2
\left(E_{\gamma_1}-E_{\gamma^\prime}\right)^2}\right.\nonumber\\
& &\left.-(-\I t)\frac{2\e^{-\I
E_{\gamma^\prime}t}}{\left(E_\gamma-E_{\gamma^\prime}\right)^3
\left(E_{\gamma_1}-E_{\gamma^\prime}\right)}-\frac{(-\I
t)^2}{2!}\frac{\e^{-\I
E_{\gamma^\prime}t}}{\left(E_\gamma-E_{\gamma^\prime}\right)^2
\left(E_{\gamma_1}-E_{\gamma^\prime}\right)}\right]\nonumber\\
& & \times \left|g_1^{\gamma_1\gamma^\prime}\right|^2
\left|g_1^{\gamma\gamma^\prime}\right|^2g_1^{\gamma\gamma^\prime}\eta_{\gamma\gamma_1}.\eeqa
\beqa {A}_5^{\gamma\gamma^\prime}(cncc;t\e,t^2\e)&=&\sum_{\gamma_1}
\left[-(-\I t)\frac{2\e^{-\I
E_{\gamma}t}}{\left(E_\gamma-E_{\gamma_1}\right)
\left(E_\gamma-E_{\gamma^\prime}\right)^3}-(-\I t)\frac{\e^{-\I
E_{\gamma}t}}{\left(E_\gamma-E_{\gamma_1}\right)^2
\left(E_{\gamma}-E_{\gamma^\prime}\right)^2}\right.\nonumber\\
& &\left.+(-\I t)\frac{\e^{-\I
E_{\gamma^\prime}t}}{\left(E_\gamma-E_{\gamma^\prime}\right)^3
\left(E_{\gamma_1}-E_{\gamma^\prime}\right)}+\frac{(-\I
t)^2}{2!}\frac{\e^{-\I
E_{\gamma}t}}{\left(E_\gamma-E_{\gamma_1}\right)
\left(E_{\gamma}-E_{\gamma^\prime}\right)^2}\right]\nonumber\\
& & \times \left|g_1^{\gamma\gamma_1}\right|^2
\left|g_1^{\gamma\gamma^\prime}\right|^2g_1^{\gamma\gamma^\prime}\eta_{\gamma_1\gamma^\prime}.\eeqa
\beqa {A}_5^{\gamma\gamma^\prime}(nccc;t\e,t^2\e)&=&\sum_{\gamma_1}
\left[-(-\I t)\frac{\e^{-\I
E_{\gamma_1}t}}{\left(E_\gamma-E_{\gamma_1}\right)
\left(E_{\gamma_1}-E_{\gamma^\prime}\right)^3}-(-\I t)\frac{2\e^{-\I
E_{\gamma^\prime}t}}{\left(E_\gamma-E_{\gamma^\prime}\right)
\left(E_{\gamma_1}-E_{\gamma^\prime}\right)^3}\right.\nonumber\\
& &\left.-(-\I t)\frac{\e^{-\I
E_{\gamma^\prime}t}}{\left(E_\gamma-E_{\gamma^\prime}\right)^2
\left(E_{\gamma_1}-E_{\gamma^\prime}\right)^2}-\frac{(-\I
t)^2}{2!}\frac{\e^{-\I
E_{\gamma^\prime}t}}{\left(E_\gamma-E_{\gamma^\prime}\right)
\left(E_{\gamma_1}-E_{\gamma^\prime}\right)^2}\right]\nonumber\\
& & \times
\left|g_1^{\gamma_1\gamma^\prime}\right|^4g_1^{\gamma\gamma^\prime}\eta_{\gamma\gamma_1}.\eeqa
\beqa & &
{A}_5^{\gamma\gamma^\prime}(ccnn,kkc;t\e,t^2\e)\nonumber\\
& & \quad =\sum_{\gamma_1,\gamma_2} \left[-(-\I t)\frac{\e^{-\I
E_{\gamma}t}}{\left(E_\gamma-E_{\gamma_1}\right)^2
\left(E_{\gamma}-E_{\gamma_2}\right)^2}-(-\I t)\frac{2\e^{-\I
E_{\gamma}t}}{\left(E_\gamma-E_{\gamma_1}\right)^3
\left(E_{\gamma}-E_{\gamma_2}\right)}\right.\nonumber\\
& &\qquad \left.-(-\I t)\frac{\e^{-\I
E_{\gamma_1}t}}{\left(E_\gamma-E_{\gamma_1}\right)^3
\left(E_{\gamma_1}-E_{\gamma_2}\right)}+\frac{(-\I
t)^2}{2!}\frac{\e^{-\I
E_{\gamma}t}}{\left(E_\gamma-E_{\gamma_1}\right)^2
\left(E_{\gamma}-E_{\gamma_2}\right)}\right]\nonumber\\
& &\qquad \times
\left|g_1^{\gamma\gamma_1}\right|^2g_1^{\gamma\gamma_1}g_1^{\gamma_1\gamma_2}
g_1^{\gamma_2\gamma}\delta_{\gamma\gamma^\prime}.\eeqa
\beqa {A}_5^{\gamma\gamma^\prime}(ccnn,kkn;t\e,t^2\e)
&=&\sum_{\gamma_1,\gamma_2} \left[(-\I t)\frac{\e^{-\I
E_{\gamma}t}}{\left(E_\gamma-E_{\gamma_1}\right)^2
\left(E_{\gamma}-E_{\gamma_2}\right)\left(E_{\gamma}-E_{\gamma^\prime}\right)}\right.\nonumber\\
& &\left.+(-\I t)\frac{\e^{-\I
E_{\gamma_1}t}}{\left(E_\gamma-E_{\gamma_1}\right)^2
\left(E_{\gamma_1}-E_{\gamma_2}\right)\left(E_{\gamma_1}-E_{\gamma^\prime}\right)}\right]\nonumber\\
& & \times
\left|g_1^{\gamma\gamma_1}\right|^2g_1^{\gamma\gamma_1}g_1^{\gamma_1\gamma_2}
g_1^{\gamma_2\gamma^\prime}\eta_{\gamma_1\gamma^\prime}\eta_{\gamma\gamma_2}\eta_{\gamma\gamma^\prime}.\eeqa
\beqa & &
{A}_5^{\gamma\gamma^\prime}(cncn,kc;t\e,t^2\e)\nonumber\\
& & \quad =\sum_{\gamma_1} \left[-(-\I t)\frac{2\e^{-\I
E_{\gamma}t}}{\left(E_\gamma-E_{\gamma_1}\right)
\left(E_{\gamma}-E_{\gamma^\prime}\right)^3}-(-\I t)\frac{\e^{-\I
E_{\gamma}t}}{\left(E_\gamma-E_{\gamma_1}\right)^2
\left(E_{\gamma}-E_{\gamma^\prime}\right)^2}\right.\nonumber\\
& &\qquad \left.+(-\I t)\frac{\e^{-\I
E_{\gamma^\prime}t}}{\left(E_\gamma-E_{\gamma^\prime}\right)^3
\left(E_{\gamma_1}-E_{\gamma^\prime}\right)}+\frac{(-\I
t)^2}{2!}\frac{\e^{-\I
E_{\gamma}t}}{\left(E_\gamma-E_{\gamma_1}\right)
\left(E_{\gamma}-E_{\gamma^\prime}\right)^2}\right]\nonumber\\
& &\qquad \times \left|g_1^{\gamma\gamma_1}\right|^2
\left|g_1^{\gamma\gamma^\prime}\right|^2
g_1^{\gamma\gamma^\prime}\eta_{\gamma_1\gamma^\prime}.\eeqa
\beqa {A}_5^{\gamma\gamma^\prime}(cncn,kn;t\e,t^2\e)
&=&\sum_{\gamma_1,\gamma_2} \left[-(-\I t)\frac{\e^{-\I
E_{\gamma}t}}{\left(E_\gamma-E_{\gamma_1}\right)\left(E_\gamma-E_{\gamma_2}\right)
\left(E_{\gamma}-E_{\gamma^\prime}\right)^2}\right.\nonumber\\
& & -(-\I t)\frac{\e^{-\I
E_{\gamma}t}}{\left(E_\gamma-E_{\gamma_1}\right)\left(E_\gamma-E_{\gamma_2}\right)^2
\left(E_{\gamma}-E_{\gamma^\prime}\right)}\nonumber\\
& &-(-\I t)\frac{\e^{-\I
E_{\gamma}t}}{\left(E_\gamma-E_{\gamma_1}\right)^2\left(E_\gamma-E_{\gamma_2}\right)
\left(E_{\gamma}-E_{\gamma^\prime}\right)}\nonumber\\
& &\left. +\frac{(-\I t)^2}{2!}\frac{\e^{-\I
E_{\gamma}t}}{\left(E_\gamma-E_{\gamma_1}\right)\left(E_\gamma-E_{\gamma_2}\right)
\left(E_{\gamma}-E_{\gamma^\prime}\right)}\right]\nonumber\\
& & \times \left|g_1^{\gamma\gamma_1}\right|^2
\left|g_1^{\gamma\gamma_2}\right|^2
g_1^{\gamma\gamma^\prime}\eta_{\gamma_1\gamma_2}
\eta_{\gamma_1\gamma^\prime}\eta_{\gamma_2\gamma^\prime}.\eeqa
\beqa & &
{A}_5^{\gamma\gamma^\prime}(cnnc,kck;t\e,t^2\e)\nonumber\\
& & \quad =\sum_{\gamma_1} \left[(-\I t)\frac{\e^{-\I
E_{\gamma}t}}{\left(E_\gamma-E_{\gamma_1}\right)^2
\left(E_{\gamma}-E_{\gamma^\prime}\right)^2}+(-\I t)\frac{\e^{-\I
E_{\gamma_1}t}}{\left(E_\gamma-E_{\gamma_1}\right)^2
\left(E_{\gamma_1}-E_{\gamma^\prime}\right)^2}\right.\nonumber\\
& &\qquad \left.+(-\I t)\frac{\e^{-\I
E_{\gamma^\prime}t}}{\left(E_\gamma-E_{\gamma^\prime}\right)^2
\left(E_{\gamma_1}-E_{\gamma^\prime}\right)^2}\right]\left|g_1^{\gamma\gamma_1}\right|^2
\left|g_1^{\gamma_1\gamma^\prime}\right|^2
g_1^{\gamma\gamma^\prime}.\eeqa \beqa
{A}_5^{\gamma\gamma^\prime}(cnnc,knk;t\e,t^2\e)
&=&\sum_{\gamma_1,\gamma_2} \left[(-\I t)\frac{\e^{-\I
E_{\gamma}t}}{\left(E_\gamma-E_{\gamma_1}\right)
\left(E_{\gamma}-E_{\gamma_2}\right)\left(E_{\gamma}-E_{\gamma^\prime}\right)^2}\right.\nonumber\\
& &\left.+(-\I t)\frac{\e^{-\I
E_{\gamma^\prime}t}}{\left(E_\gamma-E_{\gamma^\prime}\right)^2
\left(E_{\gamma_1}-E_{\gamma^\prime}\right)\left(E_{\gamma_2}-E_{\gamma^\prime}\right)}\right]\nonumber\\
& & \times \left|g_1^{\gamma\gamma_1}\right|^2
\left|g_1^{\gamma_2\gamma^\prime}\right|^2
g_1^{\gamma\gamma^\prime}\eta_{\gamma_1\gamma^\prime}\eta_{\gamma_1\gamma_2}\eta_{\gamma\gamma_2}.\eeqa
\beqa & &
{A}_5^{\gamma\gamma^\prime}(nccn,c;t\e,t^2\e)\nonumber\\
& & \quad =\sum_{\gamma_1,\gamma_2} \left[(-\I t)\frac{\e^{-\I
E_{\gamma}t}}{\left(E_\gamma-E_{\gamma_1}\right)^2
\left(E_{\gamma}-E_{\gamma_2}\right)^2}+(-\I t)\frac{\e^{-\I
E_{\gamma_1}t}}{\left(E_\gamma-E_{\gamma_1}\right)^2
\left(E_{\gamma_1}-E_{\gamma_2}\right)^2}\right.\nonumber\\
& &\qquad \left.+(-\I t)\frac{\e^{-\I
E_{\gamma_2}t}}{\left(E_\gamma-E_{\gamma_2}\right)^2
\left(E_{\gamma_1}-E_{\gamma_2}\right)^2}\right]\left|g_1^{\gamma_1\gamma_2}\right|^2
g_1^{\gamma\gamma_1}g_1^{\gamma_1\gamma_2}g_1^{\gamma_2\gamma}
\delta_{\gamma\gamma^\prime}.\eeqa \beqa
{A}_5^{\gamma\gamma^\prime}(nccn,n;t\e,t^2\e)
&=&\sum_{\gamma_1,\gamma_2} \left[-(-\I t)\frac{\e^{-\I
E_{\gamma_1}t}}{\left(E_\gamma-E_{\gamma_1}\right)
\left(E_{\gamma_1}-E_{\gamma_2}\right)^2\left(E_{\gamma_1}-E_{\gamma^\prime}\right)}\right.\nonumber\\
& &\left.-(-\I t)\frac{\e^{-\I
E_{\gamma_2}t}}{\left(E_\gamma-E_{\gamma_2}\right)
\left(E_{\gamma_1}-E_{\gamma_2}\right)^2\left(E_{\gamma_2}-E_{\gamma^\prime}\right)}\right]\nonumber\\
& & \times
\left|g_1^{\gamma_1\gamma_2}\right|^2g_1^{\gamma\gamma_1}g_1^{\gamma_1\gamma_2}
g_1^{\gamma_2\gamma^\prime}\eta_{\gamma_1\gamma^\prime}\eta_{\gamma\gamma_2}\eta_{\gamma\gamma^\prime}.\eeqa
\beqa & &
{A}_5^{\gamma\gamma^\prime}(ncnc,ck;t\e,t^2\e)\nonumber\\
& & \quad =\sum_{\gamma_1} \left[(-\I t)\frac{\e^{-\I
E_{\gamma}t}}{\left(E_\gamma-E_{\gamma_1}\right)
\left(E_{\gamma}-E_{\gamma^\prime}\right)^3}-(-\I t)\frac{\e^{-\I
E_{\gamma^\prime}t}}{\left(E_\gamma-E_{\gamma^\prime}\right)^2
\left(E_{\gamma_1}-E_{\gamma^\prime}\right)^2}\right.\nonumber\\
& &\qquad \left.-(-\I t)\frac{2\e^{-\I
E_{\gamma^\prime}t}}{\left(E_\gamma-E_{\gamma^\prime}\right)^3
\left(E_{\gamma_1}-E_{\gamma^\prime}\right)}-\frac{(-\I
t)^2}{2!}\frac{\e^{-\I
E_{\gamma^\prime}t}}{\left(E_\gamma-E_{\gamma^\prime}\right)^2
\left(E_{\gamma_1}-E_{\gamma^\prime}\right)}\right]\nonumber\\
& &\qquad\times \left|g_1^{\gamma_1\gamma^\prime}\right|^2
\left|g_1^{\gamma\gamma^\prime}\right|^2
g_1^{\gamma\gamma^\prime}\eta_{\gamma\gamma_1}.\eeqa \beqa
{A}_5^{\gamma\gamma^\prime}(ncnc,nk;t\e,t^2\e)
&=&\sum_{\gamma_1,\gamma_2} \left[-(-\I t)\frac{\e^{-\I
E_{\gamma^\prime}t}}{\left(E_\gamma-E_{\gamma^\prime}\right)
\left(E_{\gamma_1}-E_{\gamma^\prime}\right)
\left(E_{\gamma_2}-E_{\gamma^\prime}\right)^2}\right.\nonumber\\
& & -(-\I t)\frac{\e^{-\I
E_{\gamma^\prime}t}}{\left(E_\gamma-E_{\gamma^\prime}\right)
\left(E_{\gamma_1}-E_{\gamma^\prime}\right)^2
\left(E_{\gamma_2}-E_{\gamma^\prime}\right)}\nonumber\\
& &-(-\I t)\frac{\e^{-\I
E_{\gamma^\prime}t}}{\left(E_\gamma-E_{\gamma^\prime}\right)^2
\left(E_{\gamma_1}-E_{\gamma^\prime}\right)
\left(E_{\gamma_2}-E_{\gamma^\prime}\right)}\nonumber\\
& &\left. -\frac{(-\I t)^2}{2!}\frac{\e^{-\I
E_{\gamma^\prime}t}}{\left(E_\gamma-E_{\gamma^\prime}\right)
\left(E_{\gamma_1}-E_{\gamma^\prime}\right)
\left(E_{\gamma_2}-E_{\gamma^\prime}\right)}\right]\nonumber\\
& & \times \left|g_1^{\gamma_1\gamma^\prime}\right|^2
\left|g_1^{\gamma_2\gamma^\prime}\right|^2
g_1^{\gamma\gamma^\prime}\eta_{\gamma\gamma_1}
\eta_{\gamma\gamma_2}\eta_{\gamma_1\gamma_2}.\eeqa \beqa & &
{A}_5^{\gamma\gamma^\prime}(nncc,ckk;t\e,t^2\e)\nonumber\\
& & \quad =\sum_{\gamma_1,\gamma_2} \left[-(-\I t)\frac{2\e^{-\I
E_{\gamma}t}}{\left(E_\gamma-E_{\gamma_1}\right)
\left(E_{\gamma}-E_{\gamma_2}\right)^3}-(-\I t)\frac{\e^{-\I
E_{\gamma}t}}{\left(E_\gamma-E_{\gamma_1}\right)^2
\left(E_{\gamma}-E_{\gamma_2}\right)^2}\right.\nonumber\\
& &\qquad \left.+(-\I t)\frac{\e^{-\I
E_{\gamma_2}t}}{\left(E_\gamma-E_{\gamma_2}\right)^3
\left(E_{\gamma_1}-E_{\gamma_2}\right)}+\frac{(-\I
t)^2}{2!}\frac{\e^{-\I
E_{\gamma}t}}{\left(E_\gamma-E_{\gamma_1}\right)
\left(E_{\gamma}-E_{\gamma_2}\right)^2}\right]\nonumber\\
& &\qquad \times \left|g_1^{\gamma\gamma_2}\right|^2
g_1^{\gamma\gamma_1}g_1^{\gamma_1\gamma_2}g_1^{\gamma_2\gamma}\delta_{\gamma\gamma^\prime}.\eeqa
\beqa {A}_5^{\gamma\gamma^\prime}(nncc,nkk;t\e,t^2\e)
&=&\sum_{\gamma_1,\gamma_2} \left[(-\I t)\frac{\e^{-\I
E_{\gamma_2}t}}{\left(E_\gamma-E_{\gamma_2}\right)
\left(E_{\gamma_1}-E_{\gamma_2}\right)\left(E_{\gamma_2}-E_{\gamma^\prime}\right)^2}\right.\nonumber\\
& &\left.+(-\I t)\frac{\e^{-\I
E_{\gamma^\prime}t}}{\left(E_\gamma-E_{\gamma^\prime}\right)
\left(E_{\gamma_1}-E_{\gamma^\prime}\right)\left(E_{\gamma_2}-E_{\gamma^\prime}\right)^2}\right]\nonumber\\
& & \times
\left|g_1^{\gamma_2\gamma^\prime}\right|^2g_1^{\gamma\gamma_1}g_1^{\gamma_1\gamma_2}
g_1^{\gamma_2\gamma^\prime}\eta_{\gamma_1\gamma^\prime}\eta_{\gamma\gamma_2}\eta_{\gamma\gamma^\prime}.\eeqa
\beqa & &
{A}_5^{\gamma\gamma^\prime}(cnnn,kcc;t\e,t^2\e)\nonumber\\
& & \quad =\sum_{\gamma_1,\gamma_2} \left[-(-\I t)\frac{2\e^{-\I
E_{\gamma}t}}{\left(E_\gamma-E_{\gamma_1}\right)^3
\left(E_{\gamma}-E_{\gamma_2}\right)}-(-\I t)\frac{\e^{-\I
E_{\gamma}t}}{\left(E_\gamma-E_{\gamma_1}\right)^2
\left(E_{\gamma}-E_{\gamma_2}\right)^2}\right.\nonumber\\
& &\qquad \left.-(-\I t)\frac{\e^{-\I
E_{\gamma_1}t}}{\left(E_\gamma-E_{\gamma_1}\right)^3
\left(E_{\gamma_1}-E_{\gamma_2}\right)}+\frac{(-\I
t)^2}{2!}\frac{\e^{-\I
E_{\gamma}t}}{\left(E_\gamma-E_{\gamma_1}\right)^2
\left(E_{\gamma}-E_{\gamma_2}\right)}\right]\nonumber\\
& &\qquad \times \left|g_1^{\gamma\gamma_1}\right|^2
g_1^{\gamma\gamma_2}g_1^{\gamma_2\gamma_1}g_1^{\gamma_1\gamma}\delta_{\gamma\gamma^\prime}.\eeqa
\beqa {A}_5^{\gamma\gamma^\prime}(cnnn,kcn;t\e,t^2\e)
&=&\sum_{\gamma_1,\gamma_2} \left[(-\I t)\frac{\e^{-\I
E_{\gamma}t}}{\left(E_\gamma-E_{\gamma_1}\right)^2
\left(E_{\gamma}-E_{\gamma_2}\right)\left(E_{\gamma}-E_{\gamma^\prime}\right)}\right.\nonumber\\
& &\left.+(-\I t)\frac{\e^{-\I
E_{\gamma_1}t}}{\left(E_\gamma-E_{\gamma_1}\right)^2
\left(E_{\gamma_1}-E_{\gamma_2}\right)\left(E_{\gamma_1}-E_{\gamma^\prime}\right)}\right]\nonumber\\
& & \times
\left|g_1^{\gamma\gamma_1}\right|^2g_1^{\gamma\gamma_2}g_1^{\gamma_2\gamma_1}
g_1^{\gamma_1\gamma^\prime}\eta_{\gamma_2\gamma^\prime}
\eta_{\gamma\gamma^\prime}.\eeqa \beqa
{A}_5^{\gamma\gamma^\prime}(cnnn,knc;t\e,t^2\e)
&=&\sum_{\gamma_1,\gamma_2,\gamma_3} \left[-(-\I t)\frac{\e^{-\I
E_{\gamma}t}}{\left(E_\gamma-E_{\gamma_1}\right)^2\left(E_\gamma-E_{\gamma_2}\right)
\left(E_{\gamma}-E_{\gamma_3}\right)}\right.\nonumber\\
& & -(-\I t)\frac{\e^{-\I
E_{\gamma}t}}{\left(E_\gamma-E_{\gamma_1}\right)\left(E_\gamma-E_{\gamma_2}\right)^2
\left(E_{\gamma}-E_{\gamma_3}\right)}\nonumber\\
& &-(-\I t)\frac{\e^{-\I
E_{\gamma}t}}{\left(E_\gamma-E_{\gamma_1}\right)\left(E_\gamma-E_{\gamma_2}\right)
\left(E_{\gamma}-E_{\gamma_3}\right)^2}\nonumber\\
& &\left. +\frac{(-\I t)^2}{2!}\frac{\e^{-\I
E_{\gamma}t}}{\left(E_\gamma-E_{\gamma_1}\right)\left(E_\gamma-E_{\gamma_2}\right)
\left(E_{\gamma}-E_{\gamma_3}\right)}\right]\nonumber\\
& & \times \left|g_1^{\gamma\gamma_1}\right|^2
g_1^{\gamma\gamma_2}g_1^{\gamma_2\gamma_3}g_1^{\gamma_3\gamma}\eta_{\gamma_1\gamma_2}
\eta_{\gamma_1\gamma_3}\delta_{\gamma\gamma^\prime}.\eeqa
\beqa {A}_5^{\gamma\gamma^\prime}(cnnn,knn,kc;t\e,t^2\e)
&=&\sum_{\gamma_1,\gamma_2} \left[(-\I t)\frac{\e^{-\I
E_{\gamma}t}}{\left(E_\gamma-E_{\gamma_1}\right)
\left(E_{\gamma}-E_{\gamma_2}\right)\left(E_{\gamma}-E_{\gamma^\prime}\right)^2}\right.\nonumber\\
& &\left.+(-\I t)\frac{\e^{-\I
E_{\gamma^\prime}t}}{\left(E_\gamma-E_{\gamma^\prime}\right)^2
\left(E_{\gamma_1}-E_{\gamma^\prime}\right)\left(E_{\gamma_2}-E_{\gamma^\prime}\right)}\right]\nonumber\\
& &
\times\left|g_1^{\gamma\gamma^\prime}\right|^2g_1^{\gamma\gamma_1}g_1^{\gamma_1\gamma_2}
g_1^{\gamma_2\gamma^\prime}\eta_{\gamma\gamma_2}
\eta_{\gamma_1\gamma^\prime}.\eeqa \beqa &
&{A}_5^{\gamma\gamma^\prime}(cnnn,knn,kn;t\e,t^2\e) \nonumber\\ & &
\quad =\sum_{\gamma_1,\gamma_2,\gamma_3}(-\I t)\frac{\e^{-\I
E_{\gamma}t}\left|g_1^{\gamma\gamma_1}\right|^2
g_1^{\gamma\gamma_2}g_1^{\gamma_2\gamma_3}g_1^{\gamma_3\gamma^\prime}\eta_{\gamma\gamma_3}
\eta_{\gamma\gamma^\prime}\eta_{\gamma_1\gamma_2}
\eta_{\gamma_1\gamma_3}\eta_{\gamma_1\gamma^\prime}\eta_{\gamma_2\gamma^\prime}}{\left(E_\gamma-E_{\gamma_1}\right)
\left(E_{\gamma}-E_{\gamma_2}\right)\left(E_{\gamma}-E_{\gamma_3}\right)
\left(E_{\gamma}-E_{\gamma^\prime}\right)}. \eeqa
\beqa & &
{A}_5^{\gamma\gamma^\prime}(ncnn,kkc,ck;t\e,t^2\e)\nonumber\\
& & \quad =\sum_{\gamma_1} \left[(-\I t)\frac{\e^{-\I
E_{\gamma}t}}{\left(E_\gamma-E_{\gamma_1}\right)^2
\left(E_{\gamma}-E_{\gamma^\prime}\right)^2}+(-\I t)\frac{\e^{-\I
E_{\gamma_1}t}}{\left(E_\gamma-E_{\gamma_1}\right)^2
\left(E_{\gamma_1}-E_{\gamma^\prime}\right)^2}\right.\nonumber\\
& &\qquad \left.+(-\I t)\frac{\e^{-\I
E_{\gamma^\prime}t}}{\left(E_\gamma-E_{\gamma^\prime}\right)^2
\left(E_{\gamma_1}-E_{\gamma^\prime}\right)^2}\right]\left|g_1^{\gamma\gamma_1}\right|^2
\left|g_1^{\gamma_1\gamma^\prime}\right|^2
g_1^{\gamma\gamma^\prime}.\eeqa
\beqa {A}_5^{\gamma\gamma^\prime}(ncnn,kkc,nk;t\e,t^2\e)
&=&\sum_{\gamma_1,\gamma_2} \left[-(-\I t)\frac{\e^{-\I
E_{\gamma_1}t}}{\left(E_\gamma-E_{\gamma_1}\right)
\left(E_{\gamma_1}-E_{\gamma_2}\right)\left(E_{\gamma_1}-E_{\gamma^\prime}\right)^2}\right.\nonumber\\
& &\left.+(-\I t)\frac{\e^{-\I
E_{\gamma^\prime}t}}{\left(E_\gamma-E_{\gamma^\prime}\right)
\left(E_{\gamma_1}-E_{\gamma^\prime}\right)^2\left(E_{\gamma_2}-E_{\gamma^\prime}\right)}\right]\nonumber\\
& &
\times\left|g_1^{\gamma_1\gamma^\prime}\right|^2g_1^{\gamma\gamma_1}g_1^{\gamma_1\gamma_2}
g_1^{\gamma_2\gamma^\prime}\eta_{\gamma\gamma_2}\eta_{\gamma\gamma^\prime}.\eeqa
\beqa {A}_5^{\gamma\gamma^\prime}(ncnn,kkn,ck;t\e,t^2\e)
&=&\sum_{\gamma_1,\gamma_2} \left[(-\I t)\frac{\e^{-\I
E_{\gamma}t}}{\left(E_\gamma-E_{\gamma_1}\right)^2
\left(E_{\gamma}-E_{\gamma_2}\right)\left(E_{\gamma}-E_{\gamma^\prime}\right)}\right.\nonumber\\
& &\left.+(-\I t)\frac{\e^{-\I
E_{\gamma_1}t}}{\left(E_\gamma-E_{\gamma_1}\right)^2
\left(E_{\gamma_1}-E_{\gamma_2}\right)\left(E_{\gamma_1}-E_{\gamma^\prime}\right)}\right]\nonumber\\
& & \times\left|g_1^{\gamma\gamma_1}\right|^2
\left|g_1^{\gamma_1\gamma_2}\right|^2
g_1^{\gamma\gamma^\prime}\eta_{\gamma\gamma_2}\eta_{\gamma_1\gamma^\prime}\eta_{\gamma_2\gamma^\prime}.\eeqa
\beqa {A}_5^{\gamma\gamma^\prime}(ncnn,kkn,nk,c;t\e,t^2\e)
&=&\sum_{\gamma_1,\gamma_2,\gamma_3} \left[(-\I t)\frac{\e^{-\I
E_{\gamma}t}}{\left(E_\gamma-E_{\gamma_1}\right)^2
\left(E_{\gamma}-E_{\gamma_2}\right)\left(E_{\gamma}-E_{\gamma_3}\right)}\right.\nonumber\\
& &\left.+(-\I t)\frac{\e^{-\I
E_{\gamma_1}t}}{\left(E_\gamma-E_{\gamma_1}\right)^2
\left(E_{\gamma_1}-E_{\gamma_2}\right)\left(E_{\gamma_1}-E_{\gamma_3}\right)}\right]\nonumber\\
& & \times\left|g_1^{\gamma_1\gamma_2}\right|^2
g_1^{\gamma\gamma_1}g_1^{\gamma_1\gamma_3}
g_1^{\gamma_3\gamma}\eta_{\gamma\gamma_2}\eta_{\gamma_2\gamma_3}\delta_{\gamma\gamma^\prime}.\eeqa
\beqa & &{A}_5^{\gamma\gamma^\prime}(ncnn,kkn,nk,n;t\e,t^2\e)
\nonumber\\ & & \quad =-\sum_{\gamma_1,\gamma_2,\gamma_3}(-\I
t)\frac{\e^{-\I E_{\gamma_1}t}\left|g_1^{\gamma_1\gamma_2}\right|^2
g_1^{\gamma\gamma_1}g_1^{\gamma_1\gamma_3}g_1^{\gamma_3\gamma^\prime}\eta_{\gamma\gamma_2}\eta_{\gamma\gamma_3}
\eta_{\gamma\gamma^\prime}\eta_{\gamma_1\gamma^\prime}
\eta_{\gamma_2\gamma_3}\eta_{\gamma_2\gamma^\prime}}{\left(E_\gamma-E_{\gamma_1}\right)
\left(E_{\gamma_1}-E_{\gamma_2}\right)\left(E_{\gamma_1}-E_{\gamma_3}\right)
\left(E_{\gamma_1}-E_{\gamma^\prime}\right)}. \eeqa
\beqa & &
{A}_5^{\gamma\gamma^\prime}(nncn,ckk;kc,t\e,t^2\e)\nonumber\\
& & \quad =\sum_{\gamma_1} \left[(-\I t)\frac{\e^{-\I
E_{\gamma}t}}{\left(E_\gamma-E_{\gamma_1}\right)^2
\left(E_{\gamma}-E_{\gamma^\prime}\right)^2}+(-\I t)\frac{\e^{-\I
E_{\gamma_1}t}}{\left(E_\gamma-E_{\gamma_1}\right)^2
\left(E_{\gamma_1}-E_{\gamma^\prime}\right)^2}\right.\nonumber\\
& &\qquad \left.+(-\I t)\frac{\e^{-\I
E_{\gamma^\prime}t}}{\left(E_\gamma-E_{\gamma^\prime}\right)^2
\left(E_{\gamma_1}-E_{\gamma^\prime}\right)^2}\right]\left|g_1^{\gamma\gamma_1}\right|^2
\left|g_1^{\gamma_1\gamma^\prime}\right|^2
g_1^{\gamma\gamma^\prime}.\eeqa \beqa
{A}_5^{\gamma\gamma^\prime}(nncn,ckk,kn;t\e,t^2\e)
&=&\sum_{\gamma_1,\gamma_2} \left[(-\I t)\frac{\e^{-\I
E_{\gamma}t}}{\left(E_\gamma-E_{\gamma_1}\right)
\left(E_{\gamma}-E_{\gamma_2}\right)^2\left(E_{\gamma}-E_{\gamma^\prime}\right)}\right.\nonumber\\
& &\left.-(-\I t)\frac{\e^{-\I
E_{\gamma_2}t}}{\left(E_\gamma-E_{\gamma_2}\right)^2
\left(E_{\gamma_1}-E_{\gamma_2}\right)\left(E_{\gamma_2}-E_{\gamma^\prime}\right)}\right]\nonumber\\
& & \times\left|g_1^{\gamma\gamma_2}\right|^2
g_1^{\gamma\gamma_1}g_1^{\gamma_1\gamma_2}
g_1^{\gamma_2\gamma^\prime}\eta_{\gamma\gamma^\prime}\eta_{\gamma_1\gamma^\prime}.\eeqa
\beqa {A}_5^{\gamma\gamma^\prime}(nncn,nkk,kc;t\e,t^2\e)
&=&\sum_{\gamma_1,\gamma_2} \left[-(-\I t)\frac{\e^{-\I
E_{\gamma_1}t}}{\left(E_\gamma-E_{\gamma_1}\right)
\left(E_{\gamma_1}-E_{\gamma_2}\right)\left(E_{\gamma_1}-E_{\gamma^\prime}\right)^2}\right.\nonumber\\
& &\left.+(-\I t)\frac{\e^{-\I
E_{\gamma^\prime}t}}{\left(E_{\gamma}-E_{\gamma^\prime}\right)
\left(E_{\gamma_1}-E_{\gamma^\prime}\right)^2\left(E_{\gamma_2}-E_{\gamma^\prime}\right)}\right]\nonumber\\
& &
\times\left|g_1^{\gamma_1\gamma_2}\right|^2\left|g_1^{\gamma_1\gamma^\prime}\right|^2
g_1^{\gamma\gamma^\prime}\eta_{\gamma\gamma_1}\eta_{\gamma\gamma_2}\eta_{\gamma_2\gamma^\prime}.\eeqa
\beqa {A}_5^{\gamma\gamma^\prime}(nncn,nkk,kn,c;t\e,t^2\e)
&=&\sum_{\gamma_1,\gamma_2\gamma_3} \left[(-\I t)\frac{\e^{-\I
E_{\gamma}t}}{\left(E_\gamma-E_{\gamma_1}\right)
\left(E_{\gamma}-E_{\gamma_2}\right)^2
\left(E_{\gamma}-E_{\gamma_3}\right)}\right.\nonumber\\
& &\left.-(-\I t)\frac{\e^{-\I
E_{\gamma_2}t}}{\left(E_{\gamma}-E_{\gamma_2}\right)^2
\left(E_{\gamma_1}-E_{\gamma_2}\right)\left(E_{\gamma_2}-E_{\gamma_3}\right)}\right]\nonumber\\
& &
\times\left|g_1^{\gamma_2\gamma_3}\right|^2g_1^{\gamma\gamma_1}g_1^{\gamma_1\gamma_2}
g_1^{\gamma_2\gamma}\eta_{\gamma\gamma_3}\eta_{\gamma_1\gamma_3}\delta_{\gamma\gamma^\prime}.\eeqa
\beqa & &{A}_5^{\gamma\gamma^\prime}(nncn,nkk,kn,n;t\e,t^2\e)
\nonumber\\ & & \quad =\sum_{\gamma_1,\gamma_2,\gamma_3}(-\I
t)\frac{\e^{-\I E_{\gamma_2}t}\left|g_1^{\gamma_2\gamma_3}\right|^2
g_1^{\gamma\gamma_1}g_1^{\gamma_1\gamma_2}g_1^{\gamma_2\gamma^\prime}\eta_{\gamma\gamma_2}\eta_{\gamma\gamma_3}
\eta_{\gamma\gamma^\prime}
\eta_{\gamma_1\gamma_3}\eta_{\gamma_1\gamma^\prime}
\eta_{\gamma_3\gamma^\prime}}{\left(E_\gamma-E_{\gamma_2}\right)
\left(E_{\gamma_1}-E_{\gamma_2}\right)\left(E_{\gamma_2}-E_{\gamma_3}\right)
\left(E_{\gamma_2}-E_{\gamma^\prime}\right)}. \eeqa
\beqa & &
{A}_5^{\gamma\gamma^\prime}(nnnc,cck;t\e,t^2\e)\nonumber\\
& & \quad =\sum_{\gamma_1,\gamma_2} \left[-(-\I t)\frac{\e^{-\I
E_{\gamma}t}}{\left(E_\gamma-E_{\gamma_1}\right)^2
\left(E_{\gamma}-E_{\gamma_2}\right)^2}-(-\I t)\frac{2\e^{-\I
E_{\gamma}t}}{\left(E_\gamma-E_{\gamma_1}\right)^3
\left(E_{\gamma}-E_{\gamma_2}\right)}\right.\nonumber\\
& &\qquad \left.-(-\I t)\frac{\e^{-\I
E_{\gamma_1}t}}{\left(E_\gamma-E_{\gamma_1}\right)^3
\left(E_{\gamma_1}-E_{\gamma_2}\right)}+\frac{(-\I
t)^2}{2!}\frac{\e^{-\I
E_{\gamma}t}}{\left(E_\gamma-E_{\gamma_1}\right)^2
\left(E_{\gamma}-E_{\gamma_2}\right)}\right]\nonumber\\
& &\qquad \times \left|g_1^{\gamma\gamma_1}\right|^2
g_1^{\gamma\gamma_1}g_1^{\gamma_1\gamma_2}g_1^{\gamma_2\gamma}\delta_{\gamma\gamma^\prime}.\eeqa
\beqa {A}_5^{\gamma\gamma^\prime}(nnnc,cnk;t\e,t^2\e)
&=&\sum_{\gamma_1,\gamma_2,\gamma_3} \left[-(-\I t)\frac{\e^{-\I
E_{\gamma}t}}{\left(E_\gamma-E_{\gamma_1}\right)\left(E_\gamma-E_{\gamma_2}\right)
\left(E_{\gamma}-E_{\gamma_3}\right)^2}\right.\nonumber\\
& & -(-\I t)\frac{\e^{-\I
E_{\gamma}t}}{\left(E_\gamma-E_{\gamma_1}\right)\left(E_\gamma-E_{\gamma_2}\right)^2
\left(E_{\gamma}-E_{\gamma_3}\right)}\nonumber\\
& &-(-\I t)\frac{\e^{-\I
E_{\gamma}t}}{\left(E_\gamma-E_{\gamma_1}\right)^2\left(E_\gamma-E_{\gamma_2}\right)
\left(E_{\gamma}-E_{\gamma_3}\right)}\nonumber\\
& &\left. +\frac{(-\I t)^2}{2!}\frac{\e^{-\I
E_{\gamma}t}}{\left(E_\gamma-E_{\gamma_1}\right)\left(E_\gamma-E_{\gamma_2}\right)
\left(E_{\gamma}-E_{\gamma_3}\right)}\right]\nonumber\\
& & \times \left|g_1^{\gamma\gamma_3}\right|^2
g_1^{\gamma\gamma_1}g_1^{\gamma_1\gamma_2}g_1^{\gamma_2\gamma}\eta_{\gamma_1\gamma_3}
\eta_{\gamma_2\gamma_3}\delta_{\gamma\gamma^\prime}.\eeqa
\beqa {A}_5^{\gamma\gamma^\prime}(nnnc,nck;t\e,t^2\e)
&=&\sum_{\gamma_1,\gamma_2} \left[-(-\I t)\frac{\e^{-\I
E_{\gamma_1}t}}{\left(E_\gamma-E_{\gamma_1}\right)
\left(E_{\gamma_1}-E_{\gamma_2}\right)\left(E_{\gamma_1}-E_{\gamma^\prime}\right)^2}\right.\nonumber\\
& &\left.+(-\I t)\frac{\e^{-\I
E_{\gamma^\prime}t}}{\left(E_{\gamma}-E_{\gamma^\prime}\right)\left(E_{\gamma_1}-E_{\gamma^\prime}\right)^2
\left(E_{\gamma_2}-E_{\gamma^\prime}\right)}\right]\nonumber\\
& & \times
\left|g_1^{\gamma_1\gamma^\prime}\right|^2g_1^{\gamma\gamma_1}g_1^{\gamma_1\gamma_2}
g_1^{\gamma_2\gamma^\prime}\eta_{\gamma\gamma_2}
\eta_{\gamma\gamma^\prime}.\eeqa
\beqa {A}_5^{\gamma\gamma^\prime}(nnnc,nnk,ck;t\e,t^2\e)
&=&\sum_{\gamma_1,\gamma_2} \left[(-\I t)\frac{\e^{-\I
E_{\gamma}t}}{\left(E_\gamma-E_{\gamma_1}\right)
\left(E_{\gamma}-E_{\gamma_2}\right)\left(E_{\gamma}-E_{\gamma^\prime}\right)^2}\right.\nonumber\\
& &\left.+(-\I t)\frac{\e^{-\I
E_{\gamma^\prime}t}}{\left(E_{\gamma}-E_{\gamma^\prime}\right)^2\left(E_{\gamma_1}-E_{\gamma^\prime}\right)
\left(E_{\gamma_2}-E_{\gamma^\prime}\right)}\right]\nonumber\\
& & \times
\left|g_1^{\gamma\gamma^\prime}\right|^2g_1^{\gamma\gamma_1}g_1^{\gamma_1\gamma_2}
g_1^{\gamma_2\gamma^\prime}\eta_{\gamma\gamma_2}
\eta_{\gamma_1\gamma^\prime}.\eeqa
\beqa & &{A}_5^{\gamma\gamma^\prime}(nnnc,nnk,nk;t\e,t^2\e)
\nonumber\\ & & \quad =\sum_{\gamma_1,\gamma_2,\gamma_3}(-\I
t)\frac{\e^{-\I
E_{\gamma^\prime}t}\left|g_1^{\gamma_3\gamma^\prime}\right|^2
g_1^{\gamma\gamma_1}g_1^{\gamma_1\gamma_2}g_1^{\gamma_2\gamma^\prime}\eta_{\gamma\gamma_2}\eta_{\gamma\gamma_3}
\eta_{\gamma\gamma^\prime}
\eta_{\gamma_1\gamma_3}\eta_{\gamma_1\gamma^\prime}
\eta_{\gamma_2\gamma_3}}{\left(E_\gamma-E_{\gamma^\prime}\right)
\left(E_{\gamma_1}-E_{\gamma^\prime}\right)\left(E_{\gamma_2}-E_{\gamma^\prime}\right)
\left(E_{\gamma_3}-E_{\gamma^\prime}\right)}. \eeqa
\beqa & &
{A}_5^{\gamma\gamma^\prime}(nnnn,ccc;t\e,t^2\e)\nonumber\\
& & \quad =\sum_{\gamma_1} \left[(-\I t)\frac{\e^{-\I
E_{\gamma}t}}{\left(E_\gamma-E_{\gamma_1}\right)^2
\left(E_{\gamma}-E_{\gamma^\prime}\right)^2}+(-\I t)\frac{\e^{-\I
E_{\gamma_1}t}}{\left(E_\gamma-E_{\gamma_1}\right)^2
\left(E_{\gamma_1}-E_{\gamma^\prime}\right)^2}\right.\nonumber\\
& &\qquad \left.+(-\I t)\frac{\e^{-\I
E_{\gamma^\prime}t}}{\left(E_\gamma-E_{\gamma^\prime}\right)^2
\left(E_{\gamma_1}-E_{\gamma^\prime}\right)^2}\right]\left|g_1^{\gamma\gamma_1}\right|^2
\left(g_1^{\gamma\gamma_1}g_1^{\gamma_1\gamma^\prime}\right)^2
g_1^{\gamma\gamma^\prime}.\eeqa
\beqa {A}_5^{\gamma\gamma^\prime}(nnnn,ccn;t\e,t^2\e)
&=&\sum_{\gamma_1,\gamma_2} \left[(-\I t)\frac{\e^{-\I
E_{\gamma}t}}{\left(E_\gamma-E_{\gamma_1}\right)^2
\left(E_{\gamma}-E_{\gamma_2}\right)\left(E_{\gamma}-E_{\gamma^\prime}\right)}\right.\nonumber\\
& &\left.+(-\I t)\frac{\e^{-\I
E_{\gamma_1}t}}{\left(E_{\gamma}-E_{\gamma_1}\right)^2\left(E_{\gamma_1}-E_{\gamma_2}\right)
\left(E_{\gamma_1}-E_{\gamma^\prime}\right)}\right]\nonumber\\
& & \times \left(g_1^{\gamma\gamma_1}g_1^{\gamma_1\gamma_2}
g_1^{\gamma_2\gamma}\right)g_1^{\gamma\gamma_1}g_1^{\gamma_1\gamma^\prime}\eta_{\gamma\gamma^\prime}
\eta_{\gamma_2\gamma^\prime}.\eeqa
\beqa {A}_5^{\gamma\gamma^\prime}(nnnn,cnc;t\e,t^2\e)
&=&\sum_{\gamma_1,\gamma_2} \left[(-\I t)\frac{\e^{-\I
E_{\gamma}t}}{\left(E_\gamma-E_{\gamma_1}\right)
\left(E_{\gamma}-E_{\gamma_2}\right)\left(E_{\gamma}-E_{\gamma^\prime}\right)^2}\right.\nonumber\\
& &\left.+(-\I t)\frac{\e^{-\I
E_{\gamma^\prime}t}}{\left(E_{\gamma}-E_{\gamma^\prime}\right)^2\left(E_{\gamma_1}-E_{\gamma^\prime}\right)
\left(E_{\gamma_2}-E_{\gamma^\prime}\right)}\right]\nonumber\\
& & \times
\left(g_1^{\gamma\gamma_1}g_1^{\gamma_1\gamma^\prime}\right)\left(
g_1^{\gamma\gamma_2}g_1^{\gamma_2\gamma^\prime}\right)g_1^{\gamma\gamma^\prime}\eta_{\gamma_1\gamma_2}
.\eeqa
\beqa {A}_5^{\gamma\gamma^\prime}(nnnn,ncc;t\e,t^2\e)
&=&\sum_{\gamma_1,\gamma_2} \left[-(-\I t)\frac{\e^{-\I
E_{\gamma_1}t}}{\left(E_\gamma-E_{\gamma_1}\right)
\left(E_{\gamma_1}-E_{\gamma_2}\right)\left(E_{\gamma_1}-E_{\gamma^\prime}\right)^2}\right.\nonumber\\
& &\left.+(-\I t)\frac{\e^{-\I
E_{\gamma^\prime}t}}{\left(E_{\gamma}-E_{\gamma^\prime}\right)\left(E_{\gamma_1}-E_{\gamma^\prime}\right)^2
\left(E_{\gamma_2}-E_{\gamma^\prime}\right)}\right]\nonumber\\
& & \times \left(g_1^{\gamma^\prime\gamma_2}g_1^{\gamma_2\gamma_1}
g_1^{\gamma_1\gamma^\prime}\right)g_1^{\gamma\gamma_1}g_1^{\gamma_1\gamma^\prime}
\eta_{\gamma\gamma_2}\eta_{\gamma\gamma^\prime} .\eeqa
\beqa {A}_5^{\gamma\gamma^\prime}(nnnn,cnn,kc;t\e,t^2\e)
&=&\sum_{\gamma_1,\gamma_2} \left[(-\I t)\frac{\e^{-\I
E_{\gamma}t}}{\left(E_\gamma-E_{\gamma_1}\right)
\left(E_{\gamma}-E_{\gamma_2}\right)\left(E_{\gamma}-E_{\gamma^\prime}\right)^2}\right.\nonumber\\
& &\left.+(-\I t)\frac{\e^{-\I
E_{\gamma^\prime}t}}{\left(E_{\gamma}-E_{\gamma^\prime}\right)^2\left(E_{\gamma_1}-E_{\gamma^\prime}\right)
\left(E_{\gamma_2}-E_{\gamma^\prime}\right)}\right]\nonumber\\
& & \times
\left(g_1^{\gamma^\prime\gamma_1}g_1^{\gamma_1\gamma}\right)\left(
g_1^{\gamma\gamma_2}g_1^{\gamma_2\gamma^\prime}\right)g_1^{\gamma\gamma^\prime}
\eta_{\gamma_1\gamma_2}.\eeqa
\beqa & &{A}_5^{\gamma\gamma^\prime}(nnnn,cnn,kn;t\e,t^2\e)
\nonumber\\ & & \quad =\sum_{\gamma_1,\gamma_2,\gamma_3}(-\I
t)\frac{\e^{-\I E_{\gamma}t}
g_1^{\gamma\gamma_1}g_1^{\gamma_1\gamma_2}g_1^{\gamma_2\gamma}
g_1^{\gamma\gamma_3}g_1^{\gamma_3\gamma^\prime}\eta_{\gamma\gamma^\prime}\eta_{\gamma_1\gamma_3}
\eta_{\gamma_2\gamma_3}\eta_{\gamma_1\gamma^\prime}
\eta_{\gamma_2\gamma^\prime}}{\left(E_\gamma-E_{\gamma_1}\right)
\left(E_{\gamma}-E_{\gamma_2}\right)\left(E_{\gamma}-E_{\gamma_3}\right)
\left(E_{\gamma}-E_{\gamma^\prime}\right)}. \eeqa
\beqa {A}_5^{\gamma\gamma^\prime}(nnnn,ncn,c;t\e,t^2\e)
&=&\sum_{\gamma_1,\gamma_2\gamma_3} \left[(-\I t)\frac{\e^{-\I
E_{\gamma}t}}{\left(E_\gamma-E_{\gamma_1}\right)^2
\left(E_{\gamma}-E_{\gamma_2}\right)
\left(E_{\gamma}-E_{\gamma_3}\right)}\right.\nonumber\\
& &\left.+(-\I t)\frac{\e^{-\I
E_{\gamma_1}t}}{\left(E_{\gamma}-E_{\gamma_1}\right)^2
\left(E_{\gamma_1}-E_{\gamma_2}\right)\left(E_{\gamma_1}-E_{\gamma_3}\right)}\right]\nonumber\\
& &
\times\left|g_1^{\gamma\gamma_1}\right|^2g_1^{\gamma_1\gamma_2}g_1^{\gamma_2\gamma_3}
g_1^{\gamma_3\gamma_1}\eta_{\gamma\gamma_2}\eta_{\gamma\gamma_3}\delta_{\gamma\gamma^\prime}.\eeqa
\beqa & &{A}_5^{\gamma\gamma^\prime}(nnnn,ncn,n;t\e,t^2\e)
\nonumber\\ & & \quad =-\sum_{\gamma_1,\gamma_2,\gamma_3}(-\I
t)\frac{\e^{-\I E_{\gamma_1}t}
g_1^{\gamma\gamma_1}g_1^{\gamma_1\gamma_2}
g_1^{\gamma_2\gamma_3}g_1^{\gamma_3\gamma_1}g_1^{\gamma_1\gamma^\prime}
\eta_{\gamma\gamma^\prime}\eta_{\gamma\gamma_2}
\eta_{\gamma\gamma_3}\eta_{\gamma_2\gamma^\prime}
\eta_{\gamma_3\gamma^\prime}}{\left(E_\gamma-E_{\gamma_1}\right)
\left(E_{\gamma_1}-E_{\gamma_2}\right)\left(E_{\gamma_1}-E_{\gamma_3}\right)
\left(E_{\gamma_1}-E_{\gamma^\prime}\right)}. \eeqa
\beqa {A}_5^{\gamma\gamma^\prime}(nnnn,nnc,ck;t\e,t^2\e)
&=&\sum_{\gamma_1,\gamma_2} \left[(-\I t)\frac{\e^{-\I
E_{\gamma}t}}{\left(E_\gamma-E_{\gamma_1}\right)
\left(E_{\gamma}-E_{\gamma_2}\right)\left(E_{\gamma}-E_{\gamma^\prime}\right)^2}\right.\nonumber\\
& &\left.+(-\I t)\frac{\e^{-\I
E_{\gamma^\prime}t}}{\left(E_{\gamma}-E_{\gamma^\prime}\right)^2\left(E_{\gamma_1}-E_{\gamma^\prime}\right)
\left(E_{\gamma_2}-E_{\gamma^\prime}\right)}\right]\nonumber\\
& & \times
\left(g_1^{\gamma\gamma_1}g_1^{\gamma_1\gamma^\prime}\right)\left(
g_1^{\gamma^\prime\gamma_2}g_1^{\gamma_2\gamma}\right)g_1^{\gamma\gamma^\prime}\eta_{\gamma_1\gamma_2}
.\eeqa
\beqa & &{A}_5^{\gamma\gamma^\prime}(nnnn,nnc,nk;t\e,t^2\e)
\nonumber\\ & & \quad =\sum_{\gamma_1,\gamma_2,\gamma_3}(-\I
t)\frac{\e^{-\I E_{\gamma^\prime}t} g_1^{\gamma^\prime\gamma_2}
g_1^{\gamma_2\gamma_3}g_1^{\gamma_3\gamma^\prime}g_1^{\gamma\gamma_1}g_1^{\gamma_1\gamma^\prime}
\eta_{\gamma\gamma^\prime}\eta_{\gamma\gamma_2}
\eta_{\gamma\gamma_3}\eta_{\gamma_1\gamma_2}
\eta_{\gamma_1\gamma_3}}{\left(E_\gamma-E_{\gamma^\prime}\right)
\left(E_{\gamma_1}-E_{\gamma^\prime}\right)\left(E_{\gamma_2}-E_{\gamma^\prime}\right)
\left(E_{\gamma_3}-E_{\gamma^\prime}\right)}. \eeqa
\beqa {A}_5^{\gamma\gamma^\prime}(nnnn,nnn,cc;t\e,t^2\e)
&=&\sum_{\gamma_1,\gamma_2} \left[(-\I t)\frac{\e^{-\I
E_{\gamma}t}}{\left(E_\gamma-E_{\gamma_1}\right)
\left(E_{\gamma}-E_{\gamma_2}\right)\left(E_{\gamma}-E_{\gamma^\prime}\right)^2}\right.\nonumber\\
& &\left.+(-\I t)\frac{\e^{-\I
E_{\gamma^\prime}t}}{\left(E_{\gamma}-E_{\gamma^\prime}\right)^2\left(E_{\gamma_1}-E_{\gamma^\prime}\right)
\left(E_{\gamma_2}-E_{\gamma^\prime}\right)}\right]\nonumber\\
& & \times
g_1^{\gamma\gamma^\prime}g_1^{\gamma^\prime\gamma_1}g_1^{\gamma_1\gamma_2}
g_1^{\gamma_2\gamma}g_1^{\gamma\gamma^\prime}\eta_{\gamma\gamma_1}\eta_{\gamma_2\gamma^\prime}
.\eeqa
\beqa & &{A}_5^{\gamma\gamma^\prime}(nnnn,nnn,cn;t\e,t^2\e)
\nonumber\\ & & \quad =\sum_{\gamma_1,\gamma_2,\gamma_3}(-\I
t)\frac{\e^{-\I E_{\gamma}t}
g_1^{\gamma\gamma_1}g_1^{\gamma_1\gamma_2}
g_1^{\gamma_2\gamma_3}g_1^{\gamma_3\gamma}g_1^{\gamma\gamma^\prime}
\eta_{\gamma\gamma_2}
\eta_{\gamma_1\gamma_3}\eta_{\gamma_1\gamma^\prime}
\eta_{\gamma_2\gamma^\prime}\eta_{\gamma_3\gamma^\prime}}{\left(E_\gamma-E_{\gamma_1}\right)
\left(E_{\gamma}-E_{\gamma_2}\right)\left(E_{\gamma}-E_{\gamma_3}\right)
\left(E_{\gamma}-E_{\gamma^\prime}\right)}. \eeqa
\beqa & &{A}_5^{\gamma\gamma^\prime}(nnnn,nnn,nc;t\e,t^2\e)
\nonumber\\ & & \quad =\sum_{\gamma_1,\gamma_2,\gamma_3}(-\I
t)\frac{\e^{-\I E_{\gamma^\prime}t} g_1^{\gamma^\prime\gamma_1}
g_1^{\gamma_1\gamma_2}g_1^{\gamma_2\gamma_3}g_1^{\gamma_3\gamma^\prime}g_1^{\gamma\gamma^\prime}
\eta_{\gamma\gamma_1}\eta_{\gamma\gamma_2}
\eta_{\gamma\gamma_3}\eta_{\gamma_1\gamma_3}
\eta_{\gamma_2\gamma^\prime}}{\left(E_\gamma-E_{\gamma^\prime}\right)
\left(E_{\gamma_1}-E_{\gamma^\prime}\right)\left(E_{\gamma_2}-E_{\gamma^\prime}\right)
\left(E_{\gamma_3}-E_{\gamma^\prime}\right)}. \eeqa
\beqa & &{A}_5^{\gamma\gamma^\prime}(nnnn,nnn,nn,c;t\e,t^2\e)
\nonumber\\ & & \quad
=\sum_{\gamma_1,\gamma_2,\gamma_3,\gamma_4}(-\I t)\frac{\e^{-\I
E_{\gamma}t} g_1^{\gamma\gamma_1}g_1^{\gamma_1\gamma_2}
g_1^{\gamma_2\gamma_3}g_1^{\gamma_3\gamma_4}g_1^{\gamma_4\gamma}
\eta_{\gamma\gamma_2}
\eta_{\gamma\gamma_3}\eta_{\gamma_1\gamma_3}\eta_{\gamma_1\gamma_4}
\eta_{\gamma_2\gamma_4}\delta_{\gamma\gamma^\prime}}{\left(E_\gamma-E_{\gamma_1}\right)
\left(E_{\gamma}-E_{\gamma_2}\right)\left(E_{\gamma}-E_{\gamma_3}\right)
\left(E_{\gamma}-E_{\gamma_4}\right)}. \eeqa
\beqa {A}_5^{\gamma\gamma^\prime}(nnnn,nnn,nn,n;t\e,t^2\e)&=& 0
\eeqa

\subsection{$l=6$ case}

Based on our calculations, we find that there are nonvanishing 91
terms and vanishing 112 terms with $t^2\e, t^3\e$ factor parts in
all of 203 contraction- and anti contraction- expressions. In the
following, we respectively calculate them term by term, and we only
write down the non-zero expressions for saving space.

\beqa A_6(ccccc;t^2\e,t^3\e)&=&\sum_{\gamma_1}\left[\frac{(-\I
t)^3}{3!}\frac{\e^{-\I
E_{\gamma}t}}{\left(E_\gamma-E_{\gamma_1}\right)^3}-\frac{(-\I
t)^2}{2}\frac{3\e^{-\I E_{\gamma}t}}{\left(E_\gamma-E_{\gamma_1}\right)^4}\right.\nonumber\\
& &\left.+\frac{(-\I t)^2}{2}\frac{\e^{-\I
E_{\gamma_1}t}}{\left(E_\gamma-E_{\gamma_1}\right)^4}\right]
\left|g_1^{\gamma\gamma_1}\right|^6\delta_{\gamma\gamma^\prime}.\eeqa
\beqa A_6(ccccn;t^2\e,t^3\e)&=&\sum_{\gamma_1}\left[\frac{(-\I
t)^2}{2}\frac{\e^{-\I
E_{\gamma}t}}{\left(E_\gamma-E_{\gamma_1}\right)^3
\left(E_\gamma-E_{\gamma^\prime}\right)}\right.\nonumber\\
& &\left.-\frac{(-\I t)^2}{2}\frac{\e^{-\I
E_{\gamma_1}t}}{\left(E_\gamma-E_{\gamma_1}\right)^3
\left(E_{\gamma_1}-E_{\gamma^\prime}\right)}\right]
\left|g_1^{\gamma\gamma_1}\right|^4g_1^{\gamma\gamma_1}
g_1^{\gamma_1\gamma^\prime}\eta_{\gamma\gamma^\prime}.\eeqa
\beqa
A_6(cccnc;t^2\e,t^3\e)&=&\sum_{\gamma_1,\gamma_2}\left[\frac{(-\I
t)^3}{3!}\frac{\e^{-\I
E_{\gamma}t}}{\left(E_\gamma-E_{\gamma_1}\right)^2
\left(E_\gamma-E_{\gamma^\prime}\right)}-\frac{(-\I
t)^2}{2}\frac{\e^{-\I
E_{\gamma}t}}{\left(E_\gamma-E_{\gamma_1}\right)^2
\left(E_{\gamma}-E_{\gamma_2}\right)^2}\right.\nonumber\\
& &\left.-\frac{(-\I t)^2}{2}\frac{\e^{-\I
E_{\gamma}t}}{\left(E_\gamma-E_{\gamma_1}\right)^3
\left(E_{\gamma}-E_{\gamma_2}\right)}\right]
\left|g_1^{\gamma\gamma_1}\right|^4\left|g_1^{\gamma\gamma_2}\right|^2
\eta_{\gamma_1\gamma_2}\delta_{\gamma\gamma^\prime}.\eeqa
\beqa A_6(ccncc;t^2\e,t^3\e)&=&\sum_{\gamma_1}\frac{(-\I
t)^2}{2}\frac{\e^{-\I
E_{\gamma_1}t}}{\left(E_\gamma-E_{\gamma_1}\right)^2
\left(E_{\gamma_1}-E_{\gamma^\prime}\right)^2}
\left|g_1^{\gamma\gamma_1}\right|^2
\left|g_1^{\gamma_1\gamma^\prime}\right|^2g_1^{\gamma\gamma_1}
g_1^{\gamma_1\gamma^\prime}\eta_{\gamma\gamma^\prime}.\hskip
1.0cm\eeqa
\beqa
A_6(cnccc;t^2\e,t^3\e)&=&\sum_{\gamma_1,\gamma_2}\left[\frac{(-\I
t)^3}{3!}\frac{\e^{-\I
E_{\gamma}t}}{\left(E_\gamma-E_{\gamma_1}\right)
\left(E_\gamma-E_{\gamma_2}\right)^2}-\frac{(-\I
t)^2}{2}\frac{\e^{-\I
E_{\gamma}t}}{\left(E_\gamma-E_{\gamma_1}\right)
\left(E_{\gamma}-E_{\gamma_2}\right)^3}\right.\nonumber\\
& &\left.-\frac{(-\I t)^2}{2}\frac{\e^{-\I
E_{\gamma}t}}{\left(E_\gamma-E_{\gamma_1}\right)^2
\left(E_{\gamma}-E_{\gamma_2}\right)^2}\right]
\left|g_1^{\gamma\gamma_1}\right|^2\left|g_1^{\gamma\gamma_2}\right|^4
\eta_{\gamma_1\gamma_2}\delta_{\gamma\gamma^\prime}.\eeqa
\beqa A_6(ncccc;t^2\e,t^3\e)&=&\sum_{\gamma_1}\left[-\frac{(-\I
t)^2}{2}\frac{\e^{-\I
E_{\gamma_1}t}}{\left(E_\gamma-E_{\gamma_1}\right)
\left(E_{\gamma_1}-E_{\gamma^\prime}\right)^3}\right.\nonumber\\
& &\left.+\frac{(-\I t)^2}{2}\frac{\e^{-\I
E_{\gamma^\prime}t}}{\left(E_\gamma-E_{\gamma^\prime}\right)
\left(E_{\gamma_1}-E_{\gamma^\prime}\right)^3}\right]
\left|g_1^{\gamma_1\gamma^\prime}\right|^4g_1^{\gamma\gamma_1}
g_1^{\gamma_1\gamma^\prime}\eta_{\gamma\gamma^\prime}.\eeqa
\beqa A_6(cccnn,kkkc;t^2\e,t^3\e)&=&\sum_{\gamma_1}\left[\frac{(-\I
t)^2}{2}\frac{\e^{-\I
E_{\gamma}t}}{\left(E_\gamma-E_{\gamma_1}\right)
\left(E_{\gamma}-E_{\gamma^\prime}\right)^3}\right.\nonumber\\
& &\left.+\frac{(-\I t)^2}{2}\frac{\e^{-\I
E_{\gamma^\prime}t}}{\left(E_\gamma-E_{\gamma^\prime}\right)^3
\left(E_{\gamma_1}-E_{\gamma^\prime}\right)}\right]
\left|g_1^{\gamma\gamma^\prime}\right|^4g_1^{\gamma\gamma_1}
g_1^{\gamma_1\gamma^\prime}.\hskip 0.5cm\eeqa
\beqa
A_6(cccnn,kkkn;t^2\e,t^3\e)&=&\sum_{\gamma_1,\gamma_2}\frac{(-\I
t)^2}{2}\frac{\e^{-\I E_{\gamma}t}
\left|g_1^{\gamma\gamma_1}\right|^4g_1^{\gamma\gamma_2}
g_1^{\gamma_2\gamma^\prime}\eta_{\gamma\gamma^\prime}\eta_{\gamma_1\gamma_2}
\eta_{\gamma_1\gamma^\prime}}{\left(E_\gamma-E_{\gamma_1}\right)^2
\left(E_{\gamma}-E_{\gamma_2}\right)\left(E_{\gamma}-E_{\gamma^\prime}\right)}
.\hskip 0.5cm\eeqa
\beqa
A_6(ccncn,kkc;t^2\e,t^3\e)&=&\sum_{\gamma_1,\gamma_2}\left[\frac{(-\I
t)^2}{2}\frac{\e^{-\I
E_{\gamma}t}}{\left(E_\gamma-E_{\gamma_1}\right)^3
\left(E_{\gamma}-E_{\gamma^\prime}\right)}\right.\nonumber\\
& &\left.-\frac{(-\I t)^2}{2}\frac{\e^{-\I
E_{\gamma_1}t}}{\left(E_\gamma-E_{\gamma_1}\right)^3
\left(E_{\gamma_1}-E_{\gamma_2}\right)}\right]
\left|g_1^{\gamma\gamma_1}\right|^4\left|g_1^{\gamma_1\gamma_2}\right|^2
\eta_{\gamma\gamma_2}\delta_{\gamma\gamma^\prime}.\hskip 1.0cm\eeqa
\beqa
A_6(ccncn,kkn;t^2\e,t^3\e)&=&\sum_{\gamma_1,\gamma_2}\frac{(-\I
t)^2}{2}\frac{\e^{-\I E_{\gamma_1}t}
\left|g_1^{\gamma\gamma_1}\right|^2\left|g_1^{\gamma_1\gamma_2}\right|^2
g_1^{\gamma\gamma_1}g_1^{\gamma_1\gamma^\prime}\eta_{\gamma\gamma_2}\eta_{\gamma\gamma^\prime}
\eta_{\gamma_2\gamma^\prime}}{\left(E_\gamma-E_{\gamma_1}\right)^2
\left(E_{\gamma_1}-E_{\gamma_2}\right)\left(E_{\gamma_1}-E_{\gamma^\prime}\right)}
.\hskip 0.5cm\eeqa
\beqa A_6(ccnnc,kkck;t^2\e,t^3\e)&=&\sum_{\gamma_1}\frac{(-\I
t)^2}{2}\frac{\e^{-\I E_{\gamma}t}
\left|g_1^{\gamma\gamma_1}\right|^2\left|g_1^{\gamma\gamma^\prime}\right|^2g_1^{\gamma\gamma_1}
g_1^{\gamma_1\gamma^\prime}}{\left(E_\gamma-E_{\gamma_1}\right)^2
\left(E_{\gamma}-E_{\gamma^\prime}\right)^2} .\eeqa
\beqa A_6(cnccn,kc;t^2\e,t^3\e)&=&\sum_{\gamma_1}\frac{(-\I
t)^2}{2}\frac{\e^{-\I E_{\gamma}t}
\left|g_1^{\gamma\gamma_1}\right|^2\left|g_1^{\gamma\gamma^\prime}\right|^2g_1^{\gamma\gamma_1}
g_1^{\gamma_1\gamma^\prime}}{\left(E_\gamma-E_{\gamma_1}\right)^2
\left(E_{\gamma}-E_{\gamma^\prime}\right)^2} .\eeqa
\beqa A_6(cnccn,kn;t^2\e,t^3\e)&=&\sum_{\gamma_1,\gamma_2}\frac{(-\I
t)^2}{2}\frac{\e^{-\I E_{\gamma}t}
\left|g_1^{\gamma\gamma_1}\right|^2\left|g_1^{\gamma\gamma_2}\right|^2
g_1^{\gamma\gamma_2}g_1^{\gamma_2\gamma^\prime}\eta_{\gamma\gamma^\prime}\eta_{\gamma_1\gamma_2}
\eta_{\gamma_1\gamma^\prime}}{\left(E_\gamma-E_{\gamma_1}\right)
\left(E_{\gamma}-E_{\gamma_2}\right)^2\left(E_{\gamma}-E_{\gamma^\prime}\right)}
.\hskip 0.5cm\eeqa
\beqa
A_6(cncnc,kck;t^2\e,t^3\e)&=&\sum_{\gamma_1,\gamma_2}\left[\frac{(-\I
t)^3}{3!}\frac{\e^{-\I
E_{\gamma}t}}{\left(E_\gamma-E_{\gamma_1}\right)^2
\left(E_\gamma-E_{\gamma_2}\right)}\right.\nonumber\\
& &-\frac{(-\I t)^2}{2}\frac{\e^{-\I
E_{\gamma}t}}{\left(E_\gamma-E_{\gamma_1}\right)^2
\left(E_{\gamma}-E_{\gamma_2}\right)^2}\nonumber\\
& &\left.-\frac{(-\I t)^2}{2}\frac{\e^{-\I
E_{\gamma}t}}{\left(E_\gamma-E_{\gamma_1}\right)^3
\left(E_{\gamma}-E_{\gamma_2}\right)}\right]\nonumber\\
& &\times
\left|g_1^{\gamma\gamma_1}\right|^4\left|g_1^{\gamma\gamma_2}\right|^2
\eta_{\gamma_1\gamma_2}\delta_{\gamma\gamma^\prime}.\eeqa
\beqa
A_6(cncnc,knk;t^2\e,t^3\e)&=&\sum_{\gamma_1,\gamma_2,\gamma_3}\left[\frac{(-\I
t)^3}{3!}\frac{\e^{-\I
E_{\gamma}t}}{\left(E_\gamma-E_{\gamma_1}\right)
\left(E_\gamma-E_{\gamma_2}\right)\left(E_\gamma-E_{\gamma_3}\right)}\right.\nonumber\\
& &-\frac{(-\I t)^2}{2}\frac{\e^{-\I
E_{\gamma}t}}{\left(E_\gamma-E_{\gamma_1}\right)
\left(E_{\gamma}-E_{\gamma_2}\right)\left(E_\gamma-E_{\gamma_3}\right)^2}\nonumber\\
& & -\frac{(-\I t)^2}{2}\frac{\e^{-\I
E_{\gamma}t}}{\left(E_\gamma-E_{\gamma_1}\right)
\left(E_{\gamma}-E_{\gamma_2}\right)^2\left(E_\gamma-E_{\gamma_3}\right)}\nonumber\\
& &\left.-\frac{(-\I t)^2}{2}\frac{\e^{-\I
E_{\gamma}t}}{\left(E_\gamma-E_{\gamma_1}\right)^2
\left(E_{\gamma}-E_{\gamma_2}\right)\left(E_\gamma-E_{\gamma_3}\right)}\right]\nonumber\\
& &\times
\left|g_1^{\gamma\gamma_1}\right|^2\left|g_1^{\gamma\gamma_2}\right|^2
\left|g_1^{\gamma\gamma_3}\right|^2\eta_{\gamma_1\gamma_2}\eta_{\gamma_1\gamma_3}
\eta_{\gamma_2\gamma_3}\delta_{\gamma\gamma^\prime}.\eeqa
\beqa A_6(cnncc,kckk;t^2\e,t^3\e)&=&\sum_{\gamma_1}\frac{(-\I
t)^2}{2}\frac{\e^{-\I E_{\gamma^\prime}t}
\left|g_1^{\gamma_1\gamma^\prime}\right|^2\left|g_1^{\gamma\gamma^\prime}\right|^2g_1^{\gamma\gamma_1}
g_1^{\gamma_1\gamma^\prime}}{\left(E_\gamma-E_{\gamma^\prime}\right)^2
\left(E_{\gamma_1}-E_{\gamma^\prime}\right)^2} .\eeqa
\beqa A_6(ncccn,c;t^2\e,t^3\e)&=&\sum_{\gamma_1,\gamma_2}\frac{(-\I
t)^2}{2}\frac{\e^{-\I E_{\gamma_1}t}
\left|g_1^{\gamma\gamma_1}\right|^2\left|g_1^{\gamma_1\gamma_2}\right|^4
\eta_{\gamma\gamma_2}\delta_{\gamma\gamma^\prime}}
{\left(E_\gamma-E_{\gamma_1}\right)^2
\left(E_{\gamma_1}-E_{\gamma_2}\right)^2} .\eeqa
\beqa A_6(ncccn,n;t^2\e,t^3\e)&=&-\sum_{\gamma_1,\gamma_2}\frac{(-\I
t)^2}{2}\frac{\e^{-\I E_{\gamma_1}t}
\left|g_1^{\gamma_1\gamma_2}\right|^4
g_1^{\gamma\gamma_1}g_1^{\gamma_1\gamma^\prime}\eta_{\gamma\gamma^\prime}\eta_{\gamma\gamma_2}
\eta_{\gamma_2\gamma^\prime}}{\left(E_\gamma-E_{\gamma_1}\right)
\left(E_{\gamma_1}-E_{\gamma_2}\right)^2\left(E_{\gamma_1}-E_{\gamma^\prime}\right)}
.\hskip 0.5cm\eeqa
\beqa A_6(nccnc,ck;t^2\e,t^3\e)&=&\sum_{\gamma_1}\frac{(-\I
t)^2}{2}\frac{\e^{-\I E_{\gamma^\prime}t}
\left|g_1^{\gamma\gamma^\prime}\right|^2\left|g_1^{\gamma_1\gamma^\prime}\right|^2
g_1^{\gamma\gamma_1}g_1^{\gamma_1\gamma^\prime}}
{\left(E_\gamma-E_{\gamma^\prime}\right)^2
\left(E_{\gamma_1}-E_{\gamma^\prime}\right)^2}.\eeqa
\beqa A_6(nccnc,nk;t^2\e,t^3\e)&=&\sum_{\gamma_1,\gamma_2}\frac{(-\I
t)^2}{2}\frac{\e^{-\I E_{\gamma^\prime}t}
\left|g_1^{\gamma_1\gamma^\prime}\right|^2
\left|g_1^{\gamma_2\gamma^\prime}\right|^2
g_1^{\gamma\gamma_1}g_1^{\gamma_1\gamma^\prime}\eta_{\gamma\gamma^\prime}\eta_{\gamma\gamma_2}
\eta_{\gamma_1\gamma_2}}{\left(E_\gamma-E_{\gamma^\prime}\right)
\left(E_{\gamma_1}-E_{\gamma^\prime}\right)^2\left(E_{\gamma_2}-E_{\gamma^\prime}\right)}
.\hskip 0.8cm\eeqa
\beqa
A_6(ncncc,ckk;t^2\e,t^3\e)&=&\sum_{\gamma_1,\gamma_2}\left[\frac{(-\I
t)^2}{2}\frac{\e^{-\I
E_{\gamma}t}}{\left(E_\gamma-E_{\gamma_1}\right)^3
\left(E_\gamma-E_{\gamma_2}\right)}\right.\nonumber\\
& &\left.-\frac{(-\I t)^2}{2}\frac{\e^{-\I
E_{\gamma_1}t}}{\left(E_\gamma-E_{\gamma_1}\right)^3
\left(E_{\gamma_1}-E_{\gamma_2}\right)}\right]
\left|g_1^{\gamma\gamma_1}\right|^4\left|g_1^{\gamma_1\gamma_2}\right|^2
\eta_{\gamma\gamma_2}\delta_{\gamma\gamma^\prime}.\hskip 0.8cm\eeqa
\beqa
A_6(ncncc,nkk;t^2\e,t^3\e)&=&-\sum_{\gamma_1,\gamma_2}\frac{(-\I
t)^2}{2}\frac{\e^{-\I E_{\gamma_1}t}
\left|g_1^{\gamma_1\gamma_2}\right|^2
\left|g_1^{\gamma_1\gamma^\prime}\right|^2
g_1^{\gamma\gamma_1}g_1^{\gamma_1\gamma^\prime}\eta_{\gamma\gamma^\prime}\eta_{\gamma\gamma_2}
\eta_{\gamma_2\gamma^\prime}}{\left(E_\gamma-E_{\gamma_1}\right)
\left(E_{\gamma_1}-E_{\gamma_2}\right)\left(E_{\gamma_1}-E_{\gamma^\prime}\right)^2}
.\hskip 0.8cm\eeqa
\beqa A_6(nnccc,ckkk;t^2\e,t^3\e)&=&\sum_{\gamma_1}\left[\frac{(-\I
t)^2}{2}\frac{\e^{-\I
E_{\gamma}t}}{\left(E_\gamma-E_{\gamma_1}\right)
\left(E_\gamma-E_{\gamma^\prime}\right)^3}\right.\nonumber\\
& &\left.+\frac{(-\I t)^2}{2}\frac{\e^{-\I
E_{\gamma^\prime}t}}{\left(E_{\gamma}-E_{\gamma^\prime}\right)^3
\left(E_{\gamma_1}-E_{\gamma^\prime}\right)}\right]
\left|g_1^{\gamma\gamma^\prime}\right|^4g_1^{\gamma\gamma_1}g_1^{\gamma_1\gamma^\prime}
.\hskip 0.8cm\eeqa
\beqa
A_6(nnccc,nkkk;t^2\e,t^3\e)&=&\sum_{\gamma_1,\gamma_2}\frac{(-\I
t)^2}{2}\frac{\e^{-\I E_{\gamma^\prime}t}
\left|g_1^{\gamma_2\gamma^\prime}\right|^4
g_1^{\gamma\gamma_1}g_1^{\gamma_1\gamma^\prime}\eta_{\gamma\gamma^\prime}\eta_{\gamma\gamma_2}
\eta_{\gamma_1\gamma_2}}{\left(E_\gamma-E_{\gamma^\prime}\right)
\left(E_{\gamma_1}-E_{\gamma^\prime}\right)\left(E_{\gamma_2}-E_{\gamma^\prime}\right)^2}
.\hskip 0.8cm\eeqa
\beqa A_6(ccnnn,kkcc;t^2\e,t^3\e)&=&\sum_{\gamma_1}\left[\frac{(-\I
t)^2}{2}\frac{\e^{-\I
E_{\gamma}t}}{\left(E_\gamma-E_{\gamma_1}\right)
\left(E_\gamma-E_{\gamma^\prime}\right)^3}\right.\nonumber\\
& &\left.+\frac{(-\I t)^2}{2}\frac{\e^{-\I
E_{\gamma^\prime}t}}{\left(E_{\gamma}-E_{\gamma^\prime}\right)^3
\left(E_{\gamma_1}-E_{\gamma^\prime}\right)}\right]
\left|g_1^{\gamma\gamma^\prime}\right|^2g_1^{\gamma\gamma^\prime}g_1^{\gamma^\prime\gamma_1}
g_1^{\gamma_1\gamma}g_1^{\gamma\gamma^\prime} .\hskip 1.2cm\eeqa
\beqa
A_6(ccnnn,kkcn;t^2\e,t^3\e)&=&\sum_{\gamma_1,\gamma_2}\frac{(-\I
t)^2}{2}\frac{\e^{-\I E_{\gamma}t}
\left|g_1^{\gamma\gamma_1}\right|^2
g_1^{\gamma\gamma_1}g_1^{\gamma_1\gamma_2}g_1^{\gamma_2\gamma}g_1^{\gamma\gamma^\prime}
\eta_{\gamma_1\gamma^\prime}\eta_{\gamma_2\gamma^\prime}}
{\left(E_\gamma-E_{\gamma_1}\right)^2
\left(E_{\gamma}-E_{\gamma_2}\right)\left(E_{\gamma}-E_{\gamma^\prime}\right)}
.\hskip 0.8cm\eeqa
\beqa
A_6(ccnnn,kknc;t^2\e,t^3\e)&=&\sum_{\gamma_1,\gamma_2}\frac{(-\I
t)^2}{2}\frac{\e^{-\I E_{\gamma^\prime}t}
\left|g_1^{\gamma\gamma^\prime}\right|^2
g_1^{\gamma\gamma^\prime}g_1^{\gamma^\prime\gamma_1}g_1^{\gamma_1\gamma_2}g_1^{\gamma_2\gamma^\prime}
\eta_{\gamma\gamma_1}
\eta_{\gamma\gamma_2}}{\left(E_\gamma-E_{\gamma^\prime}\right)^2
\left(E_{\gamma_1}-E_{\gamma^\prime}\right)\left(E_{\gamma_2}-E_{\gamma^\prime}\right)}
.\hskip 0.8cm\eeqa
\beqa
A_6(ccnnn,kknn,kkc;t^2\e,t^3\e)&=&\sum_{\gamma_1,\gamma_2,\gamma_3}\frac{(-\I
t)^2}{2}\frac{\e^{-\I E_{\gamma}t}
\left|g_1^{\gamma\gamma_1}\right|^2
g_1^{\gamma\gamma_1}g_1^{\gamma_1\gamma_2}g_1^{\gamma_2\gamma_3}g_1^{\gamma_3\gamma}
\eta_{\gamma\gamma_2}
\eta_{\gamma_1\gamma_3}\delta_{\gamma\gamma^\prime}}{\left(E_\gamma-E_{\gamma_1}\right)^2
\left(E_{\gamma}-E_{\gamma_2}\right)\left(E_{\gamma}-E_{\gamma_3}\right)}
.\hskip 1.2cm\eeqa
\beqa A_6(cncnn,kkkc,kck;t^2\e,t^3\e)&=&\sum_{\gamma_1}\frac{(-\I
t)^2}{2}\frac{\e^{-\I E_{\gamma}t}
\left|g_1^{\gamma\gamma_1}\right|^2\left|g_1^{\gamma\gamma^\prime}\right|^2
g_1^{\gamma\gamma_1}g_1^{\gamma_1\gamma^\prime}}
{\left(E_\gamma-E_{\gamma_1}\right)^2
\left(E_{\gamma}-E_{\gamma^\prime}\right)^2}.\eeqa
\beqa
A_6(cncnn,kkkc,knk;t^2\e,t^3\e)&=&\sum_{\gamma_1,\gamma_2}\frac{(-\I
t)^2}{2}\frac{\e^{-\I E_{\gamma}t}
\left|g_1^{\gamma\gamma_1}\right|^2\left|g_1^{\gamma\gamma^\prime}\right|^2
g_1^{\gamma\gamma_2}g_1^{\gamma_2\gamma^\prime}
\eta_{\gamma_1\gamma_2}
\eta_{\gamma_1\gamma^\prime}}{\left(E_\gamma-E_{\gamma_1}\right)
\left(E_{\gamma}-E_{\gamma_2}\right)\left(E_{\gamma}-E_{\gamma^\prime}\right)^2}
.\hskip 1.2cm\eeqa
\beqa
A_6(cncnn,kkkn,kck;t^2\e,t^3\e)&=&\sum_{\gamma_1,\gamma_2}\frac{(-\I
t)^2}{2}\frac{\e^{-\I E_{\gamma}t}
\left|g_1^{\gamma\gamma_1}\right|^2\left|g_1^{\gamma\gamma_2}\right|^2
g_1^{\gamma\gamma_1}g_1^{\gamma_1\gamma^\prime}
\eta_{\gamma\gamma^\prime}\eta_{\gamma_1\gamma_2}
\eta_{\gamma_2\gamma^\prime}}{\left(E_\gamma-E_{\gamma_1}\right)^2
\left(E_{\gamma}-E_{\gamma_2}\right)\left(E_{\gamma}-E_{\gamma^\prime}\right)}
.\hskip 1.2cm\eeqa
\beqa
A_6(cncnn,kkkn,knk,kc;t^2\e,t^3\e)&=&\sum_{\gamma_1,\gamma_2}\frac{(-\I
t)^2}{2}\frac{\e^{-\I E_{\gamma}t}
\left|g_1^{\gamma\gamma_1}\right|^2\left|g_1^{\gamma\gamma^\prime}\right|^2
g_1^{\gamma\gamma_2}g_1^{\gamma_2\gamma^\prime}
\eta_{\gamma_1\gamma_2}
\eta_{\gamma_1\gamma^\prime}}{\left(E_\gamma-E_{\gamma_1}\right)
\left(E_{\gamma}-E_{\gamma_2}\right)\left(E_{\gamma}-E_{\gamma^\prime}\right)^2}
.\hskip 1.2cm\eeqa
\beqa & &
A_6(cncnn,kkkn,knk,kn;t^2\e,t^3\e)\nonumber\\
& & \quad =\sum_{\gamma_1,\gamma_2,\gamma_3}\frac{(-\I
t)^2}{2}\frac{\e^{-\I E_{\gamma}t}
\left|g_1^{\gamma\gamma_1}\right|^2\left|g_1^{\gamma\gamma_2}\right|^2
g_1^{\gamma\gamma_3}g_1^{\gamma_3\gamma^\prime}
\eta_{\gamma\gamma^\prime}
\eta_{\gamma_1\gamma_2}\eta_{\gamma_1\gamma_3}
\eta_{\gamma_1\gamma^\prime}\eta_{\gamma_2\gamma_3}
\eta_{\gamma_2\gamma^\prime}}{\left(E_\gamma-E_{\gamma_1}\right)
\left(E_{\gamma}-E_{\gamma_2}\right)\left(E_{\gamma}-E_{\gamma_3}\right)
\left(E_{\gamma}-E_{\gamma^\prime}\right)} .\hskip 1.2cm\eeqa
\beqa
A_6(cnncn,kckk,kkc;t^2\e,t^3\e)&=&\sum_{\gamma_1,\gamma_2}\frac{(-\I
t)^2}{2}\frac{\e^{-\I E_{\gamma}t}
\left|g_1^{\gamma\gamma_1}\right|^2\left|g_1^{\gamma_1\gamma_2}\right|^2
\left|g_1^{\gamma\gamma_2}\right|^2 \delta_{\gamma\gamma^\prime}
}{\left(E_\gamma-E_{\gamma_1}\right)^2
\left(E_{\gamma}-E_{\gamma_2}\right)^2} \eeqa
\beqa
A_6(cnncn,knkk,kkc;t^2\e,t^3\e)&=&\sum_{\gamma_1,\gamma_2,\gamma_3}\frac{(-\I
t)^2}{2}\frac{\e^{-\I E_{\gamma}t}
\left|g_1^{\gamma\gamma_1}\right|^2\left|g_1^{\gamma\gamma_2}\right|^2
\left|g_1^{\gamma_2\gamma_3}\right|^2
\eta_{\gamma\gamma_3}\eta_{\gamma_1\gamma_2}\eta_{\gamma_1\gamma_3}\delta_{\gamma\gamma^\prime}
}{\left(E_\gamma-E_{\gamma_1}\right)
\left(E_{\gamma}-E_{\gamma_2}\right)^2\left(E_{\gamma}-E_{\gamma_3}\right)}.\hskip
1.3cm \eeqa
\beqa A_6(cnnnc,kcck;t^2\e,t^3\e)&=&\sum_{\gamma_1}\left[\frac{(-\I
t)^2}{2}\frac{\e^{-\I
E_{\gamma}t}}{\left(E_\gamma-E_{\gamma_1}\right)
\left(E_{\gamma}-E_{\gamma^\prime}\right)^3}\right.\nonumber\\
& &\left.+\frac{(-\I t)^2}{2}\frac{\e^{-\I
E_{\gamma^\prime}t}}{\left(E_\gamma-E_{\gamma^\prime}\right)^3
\left(E_{\gamma_1}-E_{\gamma^\prime}\right)}\right]
\left|g_1^{\gamma\gamma^\prime}\right|^4g_1^{\gamma\gamma_1}
g_1^{\gamma_1\gamma^\prime}.\eeqa
\beqa
A_6(cnnnc,kcnk;t^2\e,t^3\e)&=&\sum_{\gamma_1,\gamma_2}\frac{(-\I
t)^2}{2}\frac{\e^{-\I E_{\gamma^\prime}t}
\left|g_1^{\gamma\gamma^\prime}\right|^2\left|g_1^{\gamma_2\gamma^\prime}\right|^2
g_1^{\gamma\gamma_1}g_1^{\gamma_1\gamma^\prime}
\eta_{\gamma\gamma_2}\eta_{\gamma_1\gamma_2}
}{\left(E_\gamma-E_{\gamma^\prime}\right)^2
\left(E_{\gamma_1}-E_{\gamma^\prime}\right)\left(E_{\gamma_2}-E_{\gamma^\prime}\right)}.\hskip
1.3cm \eeqa
\beqa
A_6(cnnnc,knck;t^2\e,t^3\e)&=&\sum_{\gamma_1,\gamma_2}\frac{(-\I
t)^2}{2}\frac{\e^{-\I E_{\gamma}t}
\left|g_1^{\gamma\gamma_1}\right|^2\left|g_1^{\gamma\gamma^\prime}\right|^2
g_1^{\gamma\gamma_2}g_1^{\gamma_2\gamma^\prime}
\eta_{\gamma_1\gamma_2}\eta_{\gamma_1\gamma^\prime}
}{\left(E_\gamma-E_{\gamma_1}\right)
\left(E_{\gamma}-E_{\gamma_2}\right)\left(E_{\gamma_2}-E_{\gamma^\prime}\right)^2}.\hskip
1.3cm \eeqa
\beqa A_6(nccnn,kkkc,ck;t^2\e,t^3\e)&=&\sum_{\gamma_1}\frac{(-\I
t)^2}{2}\frac{\e^{-\I E_{\gamma^\prime}t}
\left|g_1^{\gamma_1\gamma^\prime}\right|^2 g_1^{\gamma\gamma^\prime}
g_1^{\gamma^\prime\gamma_1}g_1^{\gamma_1\gamma}g_1^{\gamma\gamma^\prime}}
{\left(E_\gamma-E_{\gamma^\prime}\right)^2
\left(E_{\gamma_1}-E_{\gamma^\prime}\right)^2}.\eeqa
\beqa
A_6(nccnn,kkkc,nk;t^2\e,t^3\e)&=&\sum_{\gamma_1,\gamma_2}\frac{(-\I
t)^2}{2}\frac{\e^{-\I E_{\gamma^\prime}t}
\left|g_1^{\gamma_1\gamma^\prime}\right|^2
g_1^{\gamma\gamma^\prime}g_1^{\gamma^\prime\gamma_1}
g_1^{\gamma_1\gamma_2}g_1^{\gamma_2\gamma^\prime}
\eta_{\gamma\gamma_1}\eta_{\gamma\gamma_2}
}{\left(E_\gamma-E_{\gamma^\prime}\right)
\left(E_{\gamma_1}-E_{\gamma^\prime}\right)^2\left(E_{\gamma_2}-E_{\gamma^\prime}\right)}.\hskip
1.3cm \eeqa
\beqa A_6(ncncn,ckc;t^2\e,t^3\e)&=&\sum_{\gamma_1}\frac{(-\I
t)^2}{2}\frac{\e^{-\I E_{\gamma_1}t}
\left|g_1^{\gamma\gamma_1}\right|^2
\left|g_1^{\gamma_1\gamma^\prime}\right|^2
g_1^{\gamma\gamma_1}g_1^{\gamma_1\gamma^\prime}\eta_{\gamma\gamma^\prime}}
{\left(E_\gamma-E_{\gamma_1}\right)^2
\left(E_{\gamma_1}-E_{\gamma^\prime}\right)^2}.\eeqa
\beqa
A_6(ncncn,ckn;t^2\e,t^3\e)&=&\sum_{\gamma_1,\gamma_2}\frac{(-\I
t)^2}{2}\frac{\e^{-\I E_{\gamma_1}t}
\left|g_1^{\gamma\gamma_1}\right|^2
\left|g_1^{\gamma_1\gamma_2}\right|^2
g_1^{\gamma\gamma_1}g_1^{\gamma_1\gamma^\prime}\eta_{\gamma\gamma_2}
\eta_{\gamma\gamma^\prime}\eta_{\gamma_2\gamma^\prime}}
{\left(E_\gamma-E_{\gamma_1}\right)^2\left(E_{\gamma_1}-E_{\gamma_2}\right)
\left(E_{\gamma_1}-E_{\gamma^\prime}\right)}.\hskip 0.5cm\eeqa
\beqa
A_6(ncncn,nkc;t^2\e,t^3\e)&=&-\sum_{\gamma_1,\gamma_2}\frac{(-\I
t)^2}{2}\frac{\e^{-\I E_{\gamma_1}t}
\left|g_1^{\gamma_1\gamma^\prime}\right|^2
\left|g_1^{\gamma_1\gamma_2}\right|^2
g_1^{\gamma\gamma_1}g_1^{\gamma_1\gamma^\prime}
\eta_{\gamma\gamma^\prime}\eta_{\gamma\gamma_2}
\eta_{\gamma\gamma^\prime}\eta_{\gamma_2\gamma^\prime}}
{\left(E_\gamma-E_{\gamma_1}\right)\left(E_{\gamma_1}-E_{\gamma_2}\right)
\left(E_{\gamma_1}-E_{\gamma^\prime}\right)^2}.\hskip 1.2cm\eeqa
\beqa
A_6(ncncn,nkn,c;t^2\e,t^3\e)&=&\sum_{\gamma_1,\gamma_2,\gamma_3}\frac{(-\I
t)^2}{2}\frac{\e^{-\I E_{\gamma_1}t}
\left|g_1^{\gamma\gamma_1}\right|^2
\left|g_1^{\gamma_1\gamma_2}\right|^2
\left|g_1^{\gamma_1\gamma_3}\right|^2
\eta_{\gamma\gamma_2}\eta_{\gamma\gamma_3}
\eta_{\gamma_2\gamma_3}\delta_{\gamma\gamma^\prime}}
{\left(E_\gamma-E_{\gamma_1}\right)^2\left(E_{\gamma_1}-E_{\gamma_2}\right)
\left(E_{\gamma_1}-E_{\gamma_3}\right)}.\hskip 1.2cm\eeqa
\beqa & & A_6(ncncn,nkn,n;t^2\e,t^3\e) \nonumber\\ & &\quad
=-\sum_{\gamma_1,\gamma_2,\gamma_3}\frac{(-\I t)^2}{2}\frac{\e^{-\I
E_{\gamma_1}t} \left|g_1^{\gamma_1\gamma_2}\right|^2
\left|g_1^{\gamma_1\gamma_3}\right|^2
g_1^{\gamma\gamma_1}g_1^{\gamma_1\gamma^\prime}\eta_{\gamma\gamma_2}
\eta_{\gamma\gamma_3}
\eta_{\gamma\gamma^\prime}\eta_{\gamma_2\gamma_3}
\eta_{\gamma_2\gamma^\prime}\eta_{\gamma_3\gamma^\prime}}
{\left(E_\gamma-E_{\gamma_1}\right)\left(E_{\gamma_1}-E_{\gamma_2}\right)
\left(E_{\gamma_1}-E_{\gamma_3}\right)\left(E_{\gamma_1}-E_{\gamma^\prime}\right)}.\hskip
1.2cm\eeqa
\beqa
A_6(ncnnc,kkck,ckk;t^2\e,t^3\e)&=&\sum_{\gamma_1,\gamma_2}\frac{(-\I
t)^2}{2}\frac{\e^{-\I E_{\gamma}t}
\left|g_1^{\gamma\gamma_1}\right|^2\left|g_1^{\gamma\gamma_2}\right|^2
\left|g_1^{\gamma_1\gamma_2}\right|^2 \delta_{\gamma\gamma^\prime}}
{\left(E_\gamma-E_{\gamma_1}\right)^2
\left(E_{\gamma}-E_{\gamma_2}\right)^2}.\eeqa
\beqa
A_6(ncnnc,kknk,ckk;t^2\e,t^3\e)&=&\sum_{\gamma_1,\gamma_2,\gamma_3}\frac{(-\I
t)^2}{2}\frac{\e^{-\I E_{\gamma}t}
\left|g_1^{\gamma\gamma_1}\right|^2\left|g_1^{\gamma\gamma_3}\right|^2
\left|g_1^{\gamma_1\gamma_2}\right|^2\eta_{\gamma\gamma_2}
\eta_{\gamma_1\gamma_3}\eta_{\gamma_2\gamma_3}
\delta_{\gamma\gamma^\prime}} {\left(E_\gamma-E_{\gamma_1}\right)^2
\left(E_{\gamma}-E_{\gamma_2}\right)\left(E_{\gamma}-E_{\gamma_3}\right)}.\hskip
1.3cm\eeqa
\beqa A_6(nnccn,cc;t^2\e,t^3\e)&=&\sum_{\gamma_1}\frac{(-\I
t)^2}{2}\frac{\e^{-\I E_{\gamma}t}
\left|g_1^{\gamma\gamma_1}\right|^2g_1^{\gamma\gamma^\prime}g_1^{\gamma^\prime\gamma_1}
g_1^{\gamma_1\gamma}g_1^{\gamma\gamma^\prime}}
{\left(E_\gamma-E_{\gamma_1}\right)^2
\left(E_{\gamma}-E_{\gamma^\prime}\right)^2}.\eeqa
\beqa A_6(nnccn,cn;t^2\e,t^3\e)&=&\sum_{\gamma_1,\gamma_2}\frac{(-\I
t)^2}{2}\frac{\e^{-\I E_{\gamma}t}
\left|g_1^{\gamma\gamma_2}\right|^2g_1^{\gamma\gamma_1}g_1^{\gamma_1\gamma_2}
g_1^{\gamma_2\gamma}g_1^{\gamma\gamma^\prime}\eta_{\gamma_1\gamma^\prime}
\eta_{\gamma_2\gamma^\prime}}{\left(E_\gamma-E_{\gamma_1}\right)
\left(E_{\gamma}-E_{\gamma_2}\right)^2\left(E_{\gamma}-E_{\gamma^\prime}\right)}.
\hskip 1.3cm\eeqa
\beqa A_6(nncnc,ckkk,kck;t^2\e,t^3\e)&=&\sum_{\gamma_1}\frac{(-\I
t)^2}{2}\frac{\e^{-\I E_{\gamma^\prime}t}
\left|g_1^{\gamma\gamma^\prime}\right|^2
\left|g_1^{\gamma_1\gamma^\prime}\right|^2 g_1^{\gamma\gamma_1}
g_1^{\gamma_1\gamma^\prime}}
{\left(E_\gamma-E_{\gamma^\prime}\right)^2
\left(E_{\gamma_1}-E_{\gamma^\prime}\right)^2}.\eeqa
\beqa
A_6(nncnc,ckkk,knk;t^2\e,t^3\e)&=&\sum_{\gamma_1,\gamma_2}\frac{(-\I
t)^2}{2}\frac{\e^{-\I E_{\gamma^\prime}t}
\left|g_1^{\gamma\gamma^\prime}\right|^2
\left|g_1^{\gamma_2\gamma^\prime}\right|^2 g_1^{\gamma\gamma_1}
g_1^{\gamma_1\gamma^\prime}\eta_{\gamma\gamma_2}\eta_{\gamma_1\gamma_2}}
{\left(E_\gamma-E_{\gamma^\prime}\right)^2
\left(E_{\gamma_1}-E_{\gamma^\prime}\right)\left(E_{\gamma_2}-E_{\gamma^\prime}\right)}.\hskip
1.3cm\eeqa
\beqa
A_6(nncnc,nkkk,kck;t^2\e,t^3\e)&=&\sum_{\gamma_1,\gamma_2}\frac{(-\I
t)^2}{2}\frac{\e^{-\I E_{\gamma^\prime}t}
\left|g_1^{\gamma_1\gamma^\prime}\right|^2
\left|g_1^{\gamma_2\gamma^\prime}\right|^2 g_1^{\gamma\gamma_1}
g_1^{\gamma_1\gamma^\prime}\eta_{\gamma\gamma_2}\eta_{\gamma_1\gamma_2}\eta_{\gamma\gamma^\prime}}
{\left(E_\gamma-E_{\gamma^\prime}\right)
\left(E_{\gamma_1}-E_{\gamma^\prime}\right)^2\left(E_{\gamma_2}-E_{\gamma^\prime}\right)}.\hskip
1.3cm\eeqa
\beqa
A_6(nncnc,nkkk,knk,ck;t^2\e,t^3\e)&=&\sum_{\gamma_1,\gamma_2}\frac{(-\I
t)^2}{2}\frac{\e^{-\I E_{\gamma^\prime}t}
\left|g_1^{\gamma\gamma^\prime}\right|^2
\left|g_1^{\gamma_2\gamma^\prime}\right|^2 g_1^{\gamma\gamma_1}
g_1^{\gamma_1\gamma^\prime}\eta_{\gamma\gamma_2}\eta_{\gamma_1\gamma_2}}
{\left(E_\gamma-E_{\gamma^\prime}\right)^2
\left(E_{\gamma_1}-E_{\gamma^\prime}\right)\left(E_{\gamma_2}-E_{\gamma^\prime}\right)}.\hskip
1.3cm\eeqa
\beqa & &
A_6(nncnc,nkkk,knk,nk;t^2\e,t^3\e)\nonumber\\
& &\quad =\sum_{\gamma_1,\gamma_2,\gamma_3}\frac{(-\I
t)^2}{2}\frac{\e^{-\I E_{\gamma^\prime}t}
\left|g_1^{\gamma_2\gamma^\prime}\right|^2
\left|g_1^{\gamma_3\gamma^\prime}\right|^2 g_1^{\gamma\gamma_1}
g_1^{\gamma_1\gamma^\prime}\eta_{\gamma\gamma_2}\eta_{\gamma\gamma_3}
\eta_{\gamma\gamma^\prime}\eta_{\gamma_1\gamma_2}\eta_{\gamma_1\gamma_3}\eta_{\gamma_2\gamma_3}}
{\left(E_\gamma-E_{\gamma^\prime}\right)
\left(E_{\gamma_1}-E_{\gamma^\prime}\right)\left(E_{\gamma_2}-E_{\gamma^\prime}\right)
\left(E_{\gamma_3}-E_{\gamma^\prime}\right)}.\eeqa
\beqa A_6(nnncc,cckk;t^2\e,t^3\e)&=&\sum_{\gamma_1}\left[\frac{(-\I
t)^2}{2}\frac{\e^{-\I
E_{\gamma}t}}{\left(E_\gamma-E_{\gamma_1}\right)
\left(E_\gamma-E_{\gamma^\prime}\right)^3}\right.\nonumber\\
& &\left.+\frac{(-\I t)^2}{2}\frac{\e^{-\I
E_{\gamma^\prime}t}}{\left(E_{\gamma}-E_{\gamma^\prime}\right)^3
\left(E_{\gamma_1}-E_{\gamma^\prime}\right)}\right]
\left|g_1^{\gamma\gamma^\prime}\right|^2g_1^{\gamma\gamma^\prime}g_1^{\gamma^\prime\gamma_1}
g_1^{\gamma_1\gamma}g_1^{\gamma\gamma^\prime} .\hskip 1.2cm\eeqa
\beqa
A_6(nnncc,cnkk;t^2\e,t^3\e)&=&\sum_{\gamma_1,\gamma_2}\frac{(-\I
t)^2}{2}\frac{\e^{-\I E_{\gamma}t}
\left|g_1^{\gamma\gamma^\prime}\right|^2
g_1^{\gamma\gamma_1}g_1^{\gamma_1\gamma_2}
g_1^{\gamma_2\gamma}g_1^{\gamma\gamma^\prime}
\eta_{\gamma_1\gamma^\prime}\eta_{\gamma_2\gamma^\prime}}
{\left(E_\gamma-E_{\gamma_1}\right)
\left(E_{\gamma}-E_{\gamma_2}\right)\left(E_{\gamma}-E_{\gamma^\prime}\right)^2}.\hskip
1.3cm\eeqa
\beqa
A_6(nnncc,nckk;t^2\e,t^3\e)&=&\sum_{\gamma_1,\gamma_2}\frac{(-\I
t)^2}{2}\frac{\e^{-\I E_{\gamma^\prime}t}
\left|g_1^{\gamma_2\gamma^\prime}\right|^2
g_1^{\gamma\gamma^\prime}g_1^{\gamma^\prime\gamma_1}
g_1^{\gamma_1\gamma_2}g_1^{\gamma_2\gamma^\prime}
\eta_{\gamma\gamma_1}\eta_{\gamma\gamma_2}}
{\left(E_\gamma-E_{\gamma^\prime}\right)
\left(E_{\gamma_1}-E_{\gamma^\prime}\right)\left(E_{\gamma_2}-E_{\gamma^\prime}\right)^2}.\hskip
1.3cm\eeqa
\beqa
A_6(nnncc,nnkk,c;t^2\e,t^3\e)&=&\sum_{\gamma_1,\gamma_2,\gamma_3}\frac{(-\I
t)^2}{2}\frac{\e^{-\I E_{\gamma}t}
\left|g_1^{\gamma\gamma_3}\right|^2
g_1^{\gamma\gamma_1}g_1^{\gamma_1\gamma_2}
g_1^{\gamma_2\gamma_3}g_1^{\gamma_3\gamma}
\eta_{\gamma\gamma_2}\eta_{\gamma_1\gamma_3}\delta_{\gamma\gamma^\prime}}
{\left(E_\gamma-E_{\gamma_1}\right)
\left(E_{\gamma}-E_{\gamma_2}\right)\left(E_{\gamma}-E_{\gamma_3}\right)^2}.\hskip
1.3cm\eeqa
\beqa A_6(cnnnn,kccc;t^2\e,t^3\e)&=&\sum_{\gamma_1}\frac{(-\I
t)^2}{2}\frac{\e^{-\I E_{\gamma}t}
\left|g_1^{\gamma\gamma_1}\right|^2
g_1^{\gamma\gamma^\prime}g_1^{\gamma^\prime\gamma_1}
g_1^{\gamma_1\gamma}g_1^{\gamma\gamma^\prime}}
{\left(E_\gamma-E_{\gamma_1}\right)^2
\left(E_{\gamma}-E_{\gamma^\prime}\right)^2}.\hskip 1.3cm\eeqa
\beqa
A_6(cnnnn,kccn;t^2\e,t^3\e)&=&\sum_{\gamma_1,\gamma_2}\frac{(-\I
t)^2}{2}\frac{\e^{-\I E_{\gamma}t}
\left|g_1^{\gamma\gamma_1}\right|^2
g_1^{\gamma\gamma_2}g_1^{\gamma_2\gamma_1}
g_1^{\gamma_1\gamma}g_1^{\gamma\gamma^\prime}
\eta_{\gamma_1\gamma^\prime}\eta_{\gamma_2\gamma^\prime}}
{\left(E_\gamma-E_{\gamma_1}\right)^2
\left(E_{\gamma}-E_{\gamma_2}\right)\left(E_{\gamma}-E_{\gamma^\prime}\right)}.\hskip
1.3cm\eeqa
\beqa
A_6(cnnnn,kncc;t^2\e,t^3\e)&=&\sum_{\gamma_1,\gamma_2}\frac{(-\I
t)^2}{2}\frac{\e^{-\I E_{\gamma}t}
\left|g_1^{\gamma\gamma_1}\right|^2
g_1^{\gamma\gamma^\prime}g_1^{\gamma^\prime\gamma_2}
g_1^{\gamma_2\gamma}g_1^{\gamma\gamma^\prime}
\eta_{\gamma_1\gamma^\prime}\eta_{\gamma_1\gamma_2}}
{\left(E_\gamma-E_{\gamma_1}\right)
\left(E_{\gamma}-E_{\gamma_2}\right)\left(E_{\gamma}-E_{\gamma^\prime}\right)^2}.\hskip
1.3cm\eeqa
\beqa
A_6(cnnnn,kcnn,kkc;t^2\e,t^3\e)&=&\sum_{\gamma_1,\gamma_2,\gamma_3}\frac{(-\I
t)^2}{2}\frac{\e^{-\I E_{\gamma}t}
\left|g_1^{\gamma\gamma_1}\right|^2
g_1^{\gamma\gamma_2}g_1^{\gamma_2\gamma_1}
g_1^{\gamma_1\gamma_3}g_1^{\gamma_3\gamma}
\eta_{\gamma_2\gamma_3}\delta_{\gamma\gamma^\prime}}
{\left(E_\gamma-E_{\gamma_1}\right)^2
\left(E_{\gamma}-E_{\gamma_2}\right)\left(E_{\gamma}-E_{\gamma_3}\right)}.\hskip
1.3cm\eeqa
\beqa
A_6(cnnnn,kncn,kc;t^2\e,t^3\e)&=&\sum_{\gamma_1,\gamma_2}\frac{(-\I
t)^2}{2}\frac{\e^{-\I E_{\gamma}t}
\left|g_1^{\gamma\gamma^\prime}\right|^2
g_1^{\gamma\gamma_1}g_1^{\gamma_1\gamma_2}
g_1^{\gamma_2\gamma}g_1^{\gamma\gamma^\prime}
\eta_{\gamma_1\gamma^\prime}\eta_{\gamma_2\gamma^\prime}}
{\left(E_\gamma-E_{\gamma_1}\right)
\left(E_{\gamma}-E_{\gamma_2}\right)\left(E_{\gamma}-E_{\gamma^\prime}\right)^2}.\hskip
1.3cm\eeqa
\beqa & &
A_6(cnnnn,kncn,kn;t^2\e,t^3\e)\nonumber\\
& &\quad =\sum_{\gamma_1,\gamma_2,\gamma_3}\frac{(-\I
t)^2}{2}\frac{\e^{-\I E_{\gamma}t}
\left|g_1^{\gamma\gamma_1}\right|^2 g_1^{\gamma\gamma_2}
g_1^{\gamma_2\gamma_3}g_1^{\gamma_3\gamma}g_1^{\gamma\gamma^\prime}
\eta_{\gamma_1\gamma_2}\eta_{\gamma_1\gamma_3}
\eta_{\gamma_1\gamma^\prime}\eta_{\gamma_2\gamma^\prime}\eta_{\gamma_3\gamma^\prime}}
{\left(E_\gamma-E_{\gamma_1}\right)
\left(E_{\gamma}-E_{\gamma_2}\right)\left(E_{\gamma}-E_{\gamma_3}\right)
\left(E_{\gamma}-E_{\gamma^\prime}\right)}.\eeqa
\beqa A_6(cnnnn,knnn,kcc;t^2\e,t^3\e)
&=&\sum_{\gamma_1,\gamma_2,\gamma_3}\frac{(-\I t)^2}{2}\frac{\e^{-\I
E_{\gamma}t} \left|g_1^{\gamma\gamma_1}\right|^2
g_1^{\gamma\gamma_2}
g_1^{\gamma_2\gamma_3}g_1^{\gamma_3\gamma_1}g_1^{\gamma_1\gamma}
\eta_{\gamma\gamma_3}\eta_{\gamma_1\gamma_2}\delta_{\gamma\gamma^\prime}}
{\left(E_\gamma-E_{\gamma_1}\right)^2
\left(E_{\gamma}-E_{\gamma_2}\right)\left(E_{\gamma}-E_{\gamma_3}\right)}.
\hskip 1.3cm \eeqa
\beqa & &
A_6(cnnnn,knnn,knc;t^2\e,t^3\e)\nonumber\\
& &\quad =\sum_{\gamma_1,\gamma_2,\gamma_3,\gamma_4}\frac{(-\I
t)^2}{2}\frac{\e^{-\I E_{\gamma}t}
\left|g_1^{\gamma\gamma_1}\right|^2 g_1^{\gamma\gamma_2}
g_1^{\gamma_2\gamma_3}g_1^{\gamma_3\gamma_4}g_1^{\gamma_4\gamma}
\eta_{\gamma\gamma_3}\eta_{\gamma_1\gamma_2}\eta_{\gamma_1\gamma_3}
\eta_{\gamma_1\gamma_4}\eta_{\gamma_2\gamma_4}\delta_{\gamma\gamma^\prime}}
{\left(E_\gamma-E_{\gamma_1}\right)
\left(E_{\gamma}-E_{\gamma_2}\right)\left(E_{\gamma}-E_{\gamma_3}\right)
\left(E_{\gamma}-E_{\gamma_4}\right)}.\eeqa
\beqa A_6(ncnnn,kkcc,ckk;t^2\e,t^3\e)&=&\sum_{\gamma_1}\frac{(-\I
t)^2}{2}\frac{\e^{-\I E_{\gamma^\prime}t}
\left|g_1^{\gamma\gamma^\prime}\right|^2\left|g_1^{\gamma_1\gamma^\prime}\right|^2
g_1^{\gamma\gamma_1}g_1^{\gamma_1\gamma^\prime}}
{\left(E_\gamma-E_{\gamma^\prime}\right)^2
\left(E_{\gamma_1}-E_{\gamma^\prime}\right)^2}.\hskip 1.3cm\eeqa
\beqa A_6(ncnnn,kkcc,nkk;t^2\e,t^3\e)
&=&\sum_{\gamma_1,\gamma_2}\frac{(-\I t)^2}{2}\frac{\e^{-\I
E_{\gamma^\prime}t}\left|g_1^{\gamma_1\gamma^\prime}\right|^2
g_1^{\gamma\gamma^\prime}
g_1^{\gamma^\prime\gamma_2}g_1^{\gamma_2\gamma_1}g_1^{\gamma_1\gamma^\prime}
\eta_{\gamma\gamma_1}\eta_{\gamma\gamma_2}}
{\left(E_\gamma-E_{\gamma^\prime}\right)
\left(E_{\gamma_1}-E_{\gamma^\prime}\right)^2\left(E_{\gamma_2}-E_{\gamma^\prime}\right)}.
\hskip 1.3cm \eeqa
\beqa A_6(ncnnn,kknc,ckk;t^2\e,t^3\e)
&=&\sum_{\gamma_1,\gamma_2}\frac{(-\I t)^2}{2}\frac{\e^{-\I
E_{\gamma^\prime}t}\left|g_1^{\gamma\gamma^\prime}\right|^2
\left|g_1^{\gamma_1\gamma^\prime}\right|^2 g_1^{\gamma\gamma_2}
g_1^{\gamma_2\gamma^\prime}
\eta_{\gamma\gamma_1}\eta_{\gamma_1\gamma_2}}
{\left(E_\gamma-E_{\gamma^\prime}\right)^2
\left(E_{\gamma_1}-E_{\gamma^\prime}\right)\left(E_{\gamma_2}-E_{\gamma^\prime}\right)}.
\hskip 1.3cm \eeqa
\beqa A_6(ncnnn,kknc,nkk,ck;t^2\e,t^3\e)
&=&\sum_{\gamma_1,\gamma_2}\frac{(-\I t)^2}{2}\frac{\e^{-\I
E_{\gamma^\prime}t}\left|g_1^{\gamma_1\gamma^\prime}\right|^2
g_1^{\gamma\gamma^\prime} g_1^{\gamma^\prime\gamma_2}
g_1^{\gamma_2\gamma} g_1^{\gamma\gamma^\prime}
\eta_{\gamma\gamma_1}\eta_{\gamma_1\gamma_2}}
{\left(E_\gamma-E_{\gamma^\prime}\right)^2
\left(E_{\gamma_1}-E_{\gamma^\prime}\right)\left(E_{\gamma_2}-E_{\gamma^\prime}\right)}.
\hskip 1.3cm \eeqa
\beqa & &
A_6(ncnnn,kknc,nkk,nk;t^2\e,t^3\e)\nonumber\\
& &\quad =\sum_{\gamma_1,\gamma_2,\gamma_3}\frac{(-\I
t)^2}{2}\frac{\e^{-\I E_{\gamma^\prime}t}
\left|g_1^{\gamma_1\gamma^\prime}\right|^2 g_1^{\gamma\gamma^\prime}
g_1^{\gamma^\prime\gamma_2}g_1^{\gamma_2\gamma_3}g_1^{\gamma_3\gamma^\prime}
\eta_{\gamma\gamma_1}\eta_{\gamma\gamma_2}\eta_{\gamma\gamma_3}
\eta_{\gamma_1\gamma_2}\eta_{\gamma_1\gamma_3}}
{\left(E_\gamma-E_{\gamma^\prime}\right)
\left(E_{\gamma_1}-E_{\gamma^\prime}\right)\left(E_{\gamma_2}-E_{\gamma^\prime}\right)
\left(E_{\gamma_3}-E_{\gamma^\prime}\right)}.\eeqa
\beqa
A_6(nncnn,ckkc,kck;t^2\e,t^3\e)&=&\sum_{\gamma_1,\gamma_2}\frac{(-\I
t)^2}{2}\frac{\e^{-\I E_{\gamma}t}
\left|g_1^{\gamma\gamma_1}\right|^2\left|g_1^{\gamma\gamma_2}\right|^2
\left|g_1^{\gamma_1\gamma_2}\right|^2\delta_{\gamma\gamma^\prime}}
{\left(E_\gamma-E_{\gamma_1}\right)^2
\left(E_{\gamma}-E_{\gamma_2}\right)^2}.\eeqa
\beqa A_6(nncnn,ckkc,knk;t^2\e,t^3\e)
&=&\sum_{\gamma_1,\gamma_2,\gamma_3}\frac{(-\I t)^2}{2}\frac{\e^{-\I
E_{\gamma}t}\left|g_1^{\gamma\gamma_2}\right|^2 g_1^{\gamma\gamma_1}
g_1^{\gamma_1\gamma_2} g_1^{\gamma_2\gamma_3} g_1^{\gamma_3\gamma}
\eta_{\gamma_1\gamma_3}\delta_{\gamma\gamma^\prime}}
{\left(E_\gamma-E_{\gamma_1}\right)
\left(E_{\gamma}-E_{\gamma_2}\right)^2\left(E_{\gamma}-E_{\gamma_3}\right)}.
\hskip 1.3cm \eeqa
\beqa A_6(nnncn,cckk,kkc;t^2\e,t^3\e)&=&\sum_{\gamma_1}\frac{(-\I
t)^2}{2}\frac{\e^{-\I E_{\gamma}t}
\left|g_1^{\gamma\gamma_1}\right|^2\left|g_1^{\gamma\gamma^\prime}\right|^2
g_1^{\gamma\gamma_1}g_1^{\gamma_1\gamma^\prime}}
{\left(E_\gamma-E_{\gamma_1}\right)^2
\left(E_{\gamma}-E_{\gamma^\prime}\right)^2}.\eeqa
\beqa A_6(nnncn,cckk,kkn;t^2\e,t^3\e)
&=&\sum_{\gamma_1,\gamma_2}\frac{(-\I t)^2}{2}\frac{\e^{-\I
E_{\gamma}t}\left|g_1^{\gamma\gamma_1}\right|^2 g_1^{\gamma\gamma_1}
g_1^{\gamma_1\gamma_2} g_1^{\gamma_2\gamma}
g_1^{\gamma\gamma^\prime}
\eta_{\gamma_1\gamma^\prime}\eta_{\gamma_2\gamma^\prime}}
{\left(E_\gamma-E_{\gamma_1}\right)^2
\left(E_{\gamma}-E_{\gamma_2}\right)\left(E_{\gamma}-E_{\gamma^\prime}\right)}.
\hskip 1.3cm \eeqa
\beqa A_6(nnncn,cnkk,kkc;t^2\e,t^3\e)
&=&\sum_{\gamma_1,\gamma_2}\frac{(-\I t)^2}{2}\frac{\e^{-\I
E_{\gamma}t}\left|g_1^{\gamma\gamma_2}\right|^2
\left|g_1^{\gamma\gamma^\prime}\right|^2 g_1^{\gamma\gamma_1}
g_1^{\gamma_1\gamma^\prime}
\eta_{\gamma_1\gamma_2}\eta_{\gamma_2\gamma^\prime}}
{\left(E_\gamma-E_{\gamma_1}\right)
\left(E_{\gamma}-E_{\gamma_2}\right)\left(E_{\gamma}-E_{\gamma^\prime}\right)^2}.
\hskip 1.3cm \eeqa
\beqa A_6(nnncn,cnkk,kkn,kc;t^2\e,t^3\e)
&=&\sum_{\gamma_1,\gamma_2}\frac{(-\I t)^2}{2}\frac{\e^{-\I
E_{\gamma}t}\left|g_1^{\gamma\gamma_2}\right|^2
g_1^{\gamma\gamma^\prime} g_1^{\gamma^\prime\gamma_1}
g_1^{\gamma_1\gamma} g_1^{\gamma\gamma^\prime}
\eta_{\gamma_1\gamma_2}\eta_{\gamma_2\gamma^\prime}}
{\left(E_\gamma-E_{\gamma_1}\right)
\left(E_{\gamma}-E_{\gamma_2}\right)\left(E_{\gamma}-E_{\gamma^\prime}\right)^2}.
\hskip 1.3cm \eeqa
\beqa & &
A_6(nnncn,cnkk,kkn,kn;t^2\e,t^3\e)\nonumber\\
& &\quad =\sum_{\gamma_1,\gamma_2,\gamma_3}\frac{(-\I
t)^2}{2}\frac{\e^{-\I E_{\gamma}t}
\left|g_1^{\gamma\gamma_3}\right|^2 g_1^{\gamma\gamma_1}
g_1^{\gamma_1\gamma_2}g_1^{\gamma_2\gamma}g_1^{\gamma\gamma^\prime}
\eta_{\gamma_1\gamma_3}\eta_{\gamma_1\gamma^\prime}\eta_{\gamma_2\gamma_3}
\eta_{\gamma_2\gamma^\prime}\eta_{\gamma_3\gamma^\prime}}
{\left(E_\gamma-E_{\gamma_1}\right)
\left(E_{\gamma}-E_{\gamma_2}\right)\left(E_{\gamma}-E_{\gamma_3}\right)
\left(E_{\gamma}-E_{\gamma^\prime}\right)}.\eeqa
\beqa A_6(nnnnc,ccck;t^2\e,t^3\e)&=&\sum_{\gamma_1}\frac{(-\I
t)^2}{2}\frac{\e^{-\I E_{\gamma^\prime}t}
\left|g_1^{\gamma_1\gamma^\prime}\right|^2
g_1^{\gamma\gamma^\prime}g_1^{\gamma^\prime\gamma_1}g_1^{\gamma_1\gamma}g_1^{\gamma\gamma^\prime}}
{\left(E_\gamma-E_{\gamma^\prime}\right)^2
\left(E_{\gamma_1}-E_{\gamma^\prime}\right)^2}.\eeqa
\beqa
A_6(nnnnc,ccnk;t^2\e,t^3\e)&=&\sum_{\gamma_1,\gamma_2}\frac{(-\I
t)^2}{2}\frac{\e^{-\I E_{\gamma^\prime}t}
\left|g_1^{\gamma_2\gamma^\prime}\right|^2
g_1^{\gamma\gamma^\prime}g_1^{\gamma^\prime\gamma_1}
g_1^{\gamma_1\gamma}g_1^{\gamma\gamma^\prime}\eta_{\gamma\gamma_2}\eta_{\gamma_1\gamma_2}}
{\left(E_\gamma-E_{\gamma^\prime}\right)^2
\left(E_{\gamma_1}-E_{\gamma^\prime}\right)\left(E_{\gamma_2}-E_{\gamma^\prime}\right)}.
\hskip 1.0cm\eeqa
\beqa
A_6(nnnnc,ncck;t^2\e,t^3\e)&=&\sum_{\gamma_1,\gamma_2}\frac{(-\I
t)^2}{2}\frac{\e^{-\I E_{\gamma^\prime}t}
\left|g_1^{\gamma_1\gamma^\prime}\right|^2
g_1^{\gamma\gamma^\prime}g_1^{\gamma^\prime\gamma_1}
g_1^{\gamma_1\gamma_2}g_1^{\gamma_2\gamma^\prime}\eta_{\gamma\gamma_1}\eta_{\gamma\gamma_2}}
{\left(E_\gamma-E_{\gamma^\prime}\right)
\left(E_{\gamma_1}-E_{\gamma^\prime}\right)^2\left(E_{\gamma_2}-E_{\gamma^\prime}\right)}.
\hskip 1.0cm\eeqa
\beqa
A_6(nnnnc,ncnk,ck;t^2\e,t^3\e)&=&\sum_{\gamma_1,\gamma_2}\frac{(-\I
t)^2}{2}\frac{\e^{-\I E_{\gamma^\prime}t}
\left|g_1^{\gamma\gamma^\prime}\right|^2
g_1^{\gamma\gamma^\prime}g_1^{\gamma^\prime\gamma_1}
g_1^{\gamma_1\gamma_2}g_1^{\gamma_2\gamma^\prime}\eta_{\gamma\gamma_1}\eta_{\gamma\gamma_2}}
{\left(E_\gamma-E_{\gamma^\prime}\right)^2
\left(E_{\gamma_1}-E_{\gamma^\prime}\right)\left(E_{\gamma_2}-E_{\gamma^\prime}\right)}.
\hskip 1.0cm\eeqa
\beqa & &
A_6(nnnnc,ncnk,nk;t^2\e,t^3\e)\nonumber\\
& &\quad =\sum_{\gamma_1,\gamma_2,\gamma_3}\frac{(-\I
t)^2}{2}\frac{\e^{-\I E_{\gamma^\prime}t}
\left|g_1^{\gamma_3\gamma^\prime}\right|^2 g_1^{\gamma\gamma^\prime}
g_1^{\gamma^\prime\gamma_1}g_1^{\gamma_1\gamma_2}g_1^{\gamma_2\gamma^\prime}
\eta_{\gamma\gamma_1}\eta_{\gamma\gamma_2}\eta_{\gamma\gamma_3}
\eta_{\gamma_1\gamma_3}\eta_{\gamma_2\gamma_3}}
{\left(E_\gamma-E_{\gamma^\prime}\right)
\left(E_{\gamma_1}-E_{\gamma^\prime}\right)\left(E_{\gamma_2}-E_{\gamma^\prime}\right)
\left(E_{\gamma_3}-E_{\gamma^\prime}\right)}.\eeqa
\beqa
A_6(nnnnc,nnck,c;t^2\e,t^3\e)&=&\sum_{\gamma_1,\gamma_2,\gamma_3}\frac{(-\I
t)^2}{2}\frac{\e^{-\I E_{\gamma}t}
\left|g_1^{\gamma\gamma_2}\right|^2
g_1^{\gamma\gamma_1}g_1^{\gamma_1\gamma_2}
g_1^{\gamma_2\gamma_3}g_1^{\gamma_3\gamma}\eta_{\gamma_1\gamma_3}\delta_{\gamma\gamma^\prime}}
{\left(E_\gamma-E_{\gamma_1}\right)
\left(E_{\gamma}-E_{\gamma_2}\right)^2\left(E_{\gamma}-E_{\gamma_3}\right)}.
\hskip 1.0cm\eeqa
\beqa A_6(nnnnc,nnnk,cck;t^2\e,t^3\e)
&=&\sum_{\gamma_1,\gamma_2,\gamma_3}\frac{(-\I t)^2}{2}\frac{\e^{-\I
E_{\gamma}t} \left|g_1^{\gamma\gamma_1}\right|^2
g_1^{\gamma\gamma_1}
g_1^{\gamma_1\gamma_2}g_1^{\gamma_2\gamma_3}g_1^{\gamma_3\gamma}
\eta_{\gamma\gamma_2}\eta_{\gamma_1\gamma_3}
\delta_{\gamma\gamma^\prime}} {\left(E_\gamma-E_{\gamma_1}\right)^2
\left(E_{\gamma}-E_{\gamma_2}\right)\left(E_{\gamma}-E_{\gamma_3}\right)
}.\hskip 1.3cm\eeqa
\beqa & &
A_6(nnnnc,nnnk,cnk;t^2\e,t^3\e)\nonumber\\
& &\quad =\sum_{\gamma_1,\gamma_2,\gamma_3,\gamma_4}\frac{(-\I
t)^2}{2}\frac{\e^{-\I E_{\gamma}t}
\left|g_1^{\gamma\gamma_4}\right|^2 g_1^{\gamma\gamma_1}
g_1^{\gamma_1\gamma_2}g_1^{\gamma_2\gamma_3}g_1^{\gamma_3\gamma}
\eta_{\gamma\gamma_2}\eta_{\gamma\gamma_3}\eta_{\gamma_1\gamma_3}
\eta_{\gamma_1\gamma_4}\eta_{\gamma_2\gamma_4}\eta_{\gamma_3\gamma_4}\delta_{\gamma\gamma^\prime}}
{\left(E_\gamma-E_{\gamma_1}\right)
\left(E_{\gamma}-E_{\gamma_2}\right)\left(E_{\gamma}-E_{\gamma_3}\right)
\left(E_{\gamma}-E_{\gamma_4}\right)}.\hskip 1.3cm\eeqa
\beqa
A_6(nnnnn,cccc;t^2\e,t^3\e)&=&\sum_{\gamma_1,\gamma_2}\frac{(-\I
t)^2}{2}\frac{\e^{-\I E_{\gamma}t}\left(
g_1^{\gamma\gamma_1}g_1^{\gamma_1\gamma_2}
g_1^{\gamma_2\gamma}\right)^2\delta_{\gamma\gamma^\prime}}
{\left(E_\gamma-E_{\gamma_1}\right)^2
\left(E_{\gamma}-E_{\gamma_2}\right)^2}. \eeqa
\beqa
A_6(nnnnn,ccnc;t^2\e,t^3\e)&=&\sum_{\gamma_1,\gamma_2,\gamma_3}\frac{(-\I
t)^2}{2}\frac{\e^{-\I E_{\gamma}t}
g_1^{\gamma\gamma_1}g_1^{\gamma_1\gamma_2}
g_1^{\gamma_2\gamma}g_1^{\gamma\gamma_1}g_1^{\gamma_1\gamma_3}
g_1^{\gamma_3\gamma}\eta_{\gamma_2\gamma_3}\delta_{\gamma\gamma^\prime}}
{\left(E_\gamma-E_{\gamma_1}\right)^2
\left(E_{\gamma}-E_{\gamma_2}\right)\left(E_{\gamma}-E_{\gamma_3}\right)}.
\hskip 1.3cm\eeqa
\beqa
A_6(nnnnn,cncc;t^2\e,t^3\e)&=&\sum_{\gamma_1,\gamma_2,\gamma_3}\frac{(-\I
t)^2}{2}\frac{\e^{-\I E_{\gamma}t}
g_1^{\gamma\gamma_1}g_1^{\gamma_1\gamma_2}
g_1^{\gamma_2\gamma}g_1^{\gamma\gamma_3}g_1^{\gamma_3\gamma_2}
g_1^{\gamma_2\gamma}\eta_{\gamma_1\gamma_3}\delta_{\gamma\gamma^\prime}}
{\left(E_\gamma-E_{\gamma_1}\right)
\left(E_{\gamma}-E_{\gamma_2}\right)^2\left(E_{\gamma}-E_{\gamma_3}\right)}.
\hskip 1.3cm\eeqa
\beqa
A_6(nnnnn,cnnc,kck;t^2\e,t^3\e)&=&\sum_{\gamma_1,\gamma_2,\gamma_3}\frac{(-\I
t)^2}{2}\frac{\e^{-\I E_{\gamma}t}
g_1^{\gamma\gamma_1}g_1^{\gamma_1\gamma_2}
g_1^{\gamma_2\gamma}g_1^{\gamma\gamma_3}g_1^{\gamma_3\gamma_1}
g_1^{\gamma_1\gamma}\eta_{\gamma_2\gamma_3}\delta_{\gamma\gamma^\prime}}
{\left(E_\gamma-E_{\gamma_1}\right)^2
\left(E_{\gamma}-E_{\gamma_2}\right)\left(E_{\gamma}-E_{\gamma_3}\right)}.
\hskip 1.3cm\eeqa
\beqa & &
A_6(nnnnn,cnnc,knk;t^2\e,t^3\e)\nonumber\\
& & \quad =\sum_{\gamma_1,\gamma_2,\gamma_3,\gamma_4}\frac{(-\I
t)^2}{2}\frac{\e^{-\I E_{\gamma}t}
g_1^{\gamma\gamma_1}g_1^{\gamma_1\gamma_2}
g_1^{\gamma_2\gamma}g_1^{\gamma\gamma_3}g_1^{\gamma_3\gamma_4}
g_1^{\gamma_4\gamma}\eta_{\gamma_1\gamma_3}\eta_{\gamma_1\gamma_4}
\eta_{\gamma_2\gamma_3}\eta_{\gamma_2\gamma_4}\delta_{\gamma\gamma^\prime}}
{\left(E_\gamma-E_{\gamma_1}\right)
\left(E_{\gamma}-E_{\gamma_2}\right)\left(E_{\gamma}-E_{\gamma_3}\right)
\left(E_{\gamma}-E_{\gamma_4}\right)}. \eeqa

\end{document}